\documentclass[12pt]{article}
\usepackage{amsmath,amssymb,amsthm}
\usepackage{txfonts,bm,pifont}
\usepackage{graphicx}
\usepackage[pdfencoding=auto,bookmarks=true,bookmarksnumbered=true,colorlinks=true]{hyperref}
\usepackage{color}
\usepackage{float}
\usepackage{multirow,bigdelim}
\theoremstyle{plain}
\newtheorem{thm}{Theorem}[section]
\newtheorem{prop}[thm]{Proposition}

\newtheorem{lemma}[thm]{Lemma}

\newtheorem{rem}[thm]{Remark}

\newcommand{\R}{\mathbb{R}}

\newcommand{\col}[2]{{\scriptstyle {#1\atop #2}}}
\newcommand{\colu}[3]{\begin{array}{c}{\scriptscriptstyle #1}\\[-3mm]
{\scriptscriptstyle #2}\\[-3mm]{\scriptscriptstyle #3}\end{array}}
\makeatletter
\newdimen\LENB \newdimen\LENW \newdimen\THI 
\newdimen\LENWH \newdimen\LENTOT \newcount\N 
\def\vbrknlnele#1#2#3{
  \LENB=#1pt \LENW=#2pt \THI=#3pt
  \LENWH=\LENW \divide\LENWH by 2
  \LENTOT=\LENB \advance\LENTOT by \LENW
  \vbox to \LENTOT{
    \vbox to \LENWH{}
    \nointerlineskip
    \vbox to \LENB{\hbox to \THI{\vrule width \THI height \LENB}}
    \nointerlineskip
    \vbox to \LENWH{}
  }}

\def\vbrknln#1{
  \N=#1
  \vcenter{
    \vbox{
      \loop\ifnum\N>0
        \vbox to 4pt{\vbrknlnele{2}{2}{0.1}}
        \nointerlineskip
        \advance\N by -1
      \repeat
  }}}

\def\vbl#1{\hskip-5pt \vbrknln{#1} \hskip-5pt}

\def\hbrknlnele#1#2#3{
  \LENB=#1pt \LENW=#2pt \THI=#3pt
  \LENTOT=\LENB \advance\LENTOT by \LENW
  \vcenter{
    \vbox to \THI{
      \hbox to \LENTOT{
        \hfil
        \vrule width \LENB height \THI
        \hfil}
  }}}

\def\hblele{\hbrknlnele{2}{2.2}{0.1}}

\def\hblfil{\cleaders\hbox{$ \m@th \mkern1mu \hblele \mkern1mu
$}\hfill} 
\makeatother
\makeatletter
\@addtoreset{equation}{section} 

\makeatother
\begin{document}
\begin{center}
\begin{Large}
Discrete Local Induction Equation\\[5mm]
\end{Large}
\begin{normalsize}
Sampei {\sc Hirose}\\
Department of Engineering and Design,
Shibaura Institute of Technology\\
307 Fukasaku, Minuma-ku, Saitama 337-8570, Japan\\
e-mail: hirose3@shibaura-it.ac.jp\\[2mm]
Jun-ichi {\sc Inoguchi}\\
Institute of Mathematics, University of Tsukuba\\
Tsukuba 305-8571, Japan\\
e-mail: inoguchi@math.tsukuba.ac.jp\\[2mm]
Kenji {\sc Kajiwara}\\
Institute of Mathematics for Industry, Kyushu University\\
744 Motooka, Fukuoka 819-0395, Japan\\
e-mail: kaji@imi.kyushu-u.ac.jp\\[2mm]
Nozomu {\sc Matsuura}\\
Department of Education and Creation Engineering,
Kurume Institute of Technology\\
2228-66 Kamitsu, Kurume 830-0052, Japan\\
e-mail: nozomu@kurume-it.ac.jp\\[2mm]
Yasuhiro {\sc Ohta}\\
Department of Mathematics, Kobe University\\
Rokko, Kobe 657-8501, Japan\\
e-mail: ohta@math.sci.kobe-u.ac.jp
\end{normalsize}
\end{center}
\begin{abstract}
The local induction equation, or the binormal flow on space curves is a well-known model of
deformation of space curves as it describes the dynamics of vortex filaments, and the complex
curvature is governed by the nonlinear Schr\"odinger equation. In this paper, we present its
discrete analogue, namely, a model of deformation of discrete space curves by the discrete nonlinear
Schr\"odinger equation. We also present explicit formulas for both smooth and discrete curves
in terms of $\tau$ functions of the two-component KP hierarchy.
\end{abstract}
\section{Introduction}
The {\em local induction equation} (LIE)
\begin{equation} \label{eqn:local_induction}
\frac{\partial\gamma}{\partial t} = \frac{\partial\gamma}{\partial x}\times \frac{\partial^2\gamma}{\partial x^2},
\end{equation}
is one of the most important models of deformation of space curves, where
$\gamma(x,t)\in\mathbb{R}^3$ is a smooth space curve parametrized by the arc-length $x$ and $t$ is a
deformation parameter.  In a physical setting, it describes the dynamics of vortex filaments driven
by the self-induction in the inviscid fluid under the local induction approximation
\cite{Hasimoto:JFM1972}. It is well-known that if $\gamma$ obeys LIE, then the curvature and the
torsion, or equivalently, the complex curvature of $\gamma$ solves the {\em nonlinear Schr\"odinger
equation} (NLS)
\begin{equation}\label{eqn:NLS}
\sqrt{-1}\frac{\partial u}{\partial t} + \frac{\partial^2u}{\partial x^2} + \frac{1}{2}|u|^2 u = 0.
\end{equation}
Here, $u=u(x,t)\in\mathbb{C}$ is the {\em complex curvature} defined by the
{\em Hasimoto transformation}
\begin{equation}\label{eqn:Hasimoto_transform}
 u = \kappa e^{\sqrt{-1}\Lambda},\quad \Lambda = \int^x \lambda\,dx,
\end{equation}
where $\kappa$ and $\lambda$ are the curvature and the torsion of $\gamma$, respectively.

In this paper, we consider the integrable discretization of LIE in both space and time; we construct
the discrete deformation of discrete space curves which is described by the {\em discrete nonlinear
Schr\"odinger equation}
(dNLS)\cite{Ablowitz-Ladik:SIAM1976,Ablowitz-Ladik:SIAM1977,Hirota-Ohta:PSJ_91Spring,
Hirota-Tsujimoto-Ohta:JSIAM1993,Tsujimoto:App_book}
\begin{equation}\label{eqn:dNLS}
\begin{split}
&  {\textstyle \left({\scriptstyle  \sqrt{-1}}\frac{\epsilon^2}{\delta} - 1\right)}u_{n}^{m+1} 
-  {\textstyle \left({\scriptstyle\sqrt{-1}}\frac{\epsilon^2}{\delta} + 1\right)}u_n^m
+ (u_{n+1}^m+u_{n-1}^{m+1})\Big({\textstyle 1+\frac{\epsilon^2}{4}|u_n^m|^2}\Big)\Gamma_n^m=0 ,\\[2mm]
&\hskip60pt  \frac{\Gamma_{n+1}^m}{\Gamma_n^m} = \frac{1+\frac{\epsilon^2}{4}|u_n^m|^2}
{1+\frac{\epsilon^2}{4}|u_n^{m+1}|^2},
\end{split}
\end{equation}
where $u_n^m\in\mathbb{C}$, $\Gamma_n^m\in\mathbb{R}$, $n$, $m$ are discrete independent variables
corresponding to space and time, respectively, and $\epsilon, \delta\in\mathbb{R}$ are
constants corresponding to the lattice intervals of space and time, respectively.
dNLS \eqref{eqn:dNLS} reduces to NLS \eqref{eqn:NLS} by taking the simple continuous limit
$x=n\epsilon$, $t=m\delta$, $\epsilon,\delta\to 0$.  

Let us explain the relationship between LIE \eqref{eqn:local_induction} and NLS more concretely.  We
use the Frenet frame $\Phi=\Phi(x,t)=[T(x,t), N(x,t),B(x,t)]\in\mathrm{SO}(3)$, where $T$, $N$, $B$
are the tangent, the normal, and the binormal vectors defined by
\begin{equation}
T=\gamma',\quad 
N=\frac{\gamma''}{\left|\gamma''\right|},\quad
B=T\times N,\quad ' = \frac{\partial}{\partial x},
\end{equation}
respectively. Note that it follows that $ |T|=\left|\gamma'\right|=1$ since $x$ is the
arc-length. Then we have the {\em Frenet-Serret formula}
\begin{equation}\label{eqn:Frenet-Serret}
\frac{\partial \Phi}{\partial x} 
= \Phi L,\quad L=\left[\begin{array}{ccc} 0& -\kappa& 0\\ \kappa & 0 & -\lambda \\ 0 &\lambda &0\end{array}\right],
\end{equation}
where $\kappa = \left|\gamma''\right|$ and $\lambda = -\langle B',N\rangle$ are the curvature and
the torsion, respectively.  In this setting, LIE \eqref{eqn:local_induction} is expressed as the
deformation by the binormal flow
\begin{equation}\label{eqn:continuous_binormal_flow}
 \frac{\partial\gamma}{\partial t} = \kappa B,
\end{equation}
and the corresponding deformation equation of the Frenet frame is given by
\begin{equation}\label{eqn:continuous_deformation_Frenet_frame}
 \frac{\partial\Phi}{\partial t} 
= \Phi M,\quad M=
\begin{bmatrix}
0 & \kappa \lambda & - \kappa'\\
- \kappa \lambda & 0 & - \frac{\kappa''}{\kappa} + \lambda^2\\
\kappa' & \frac{\kappa''}{\kappa} - \lambda^2 & 0
\end{bmatrix}.
\end{equation}
The compatibility condition of the system of linear partial differential equations
\eqref{eqn:Frenet-Serret} and \eqref{eqn:continuous_deformation_Frenet_frame}
\begin{equation}
 \frac{\partial L}{\partial t} -  \frac{\partial M}{\partial x} = LM - ML,
\end{equation}
yields
\begin{equation}\label{eqn:NLS_kappa}
\frac{\partial\kappa}{\partial t}
= - 2 \frac{\partial \kappa}{\partial x} \lambda - \kappa \frac{\partial\lambda}{\partial x},\quad
\frac{\partial \lambda}{\partial t}
= \frac{\partial }{\partial x}\left(\frac{\kappa''}{\kappa}+ \frac{\kappa^2}{2} - \lambda^2\right).
\end{equation}

Discretization of curves and their deformations preserving underlying integrable structures is an
important problem in the discrete differential geometry. For example, continuous deformations of the
discrete plane and space curves have been considered in
\cite{Doliwa-Santini:JMP,Hisakado-Nakayama-Wadati,Hisakado-Wadati,
Hoffmann:LN,Hoffmann-Kutz,IKMO:JPA, Nakayama:JPSJ2007,Nishinari} where the deformations described by
the differential-difference analogue of the mKdV and the NLS equations are formulated. In
particular, Doliwa and Santini formulated the binormal deformation of the discrete space curves
\cite{Doliwa-Santini:JMP} where the discrete analogue of complex curvature is governed by the {\em
semi-discrete NLS equation} (sdNLS) or the {\em Ablowitz-Ladik equation}
\cite{Ablowitz-Ladik:SIAM1976,Ablowitz-Ladik:SIAM1977}
\begin{equation}\label{eqn:sdNLS}
\sqrt{-1}\frac{du_n}{dt} + \frac{u_{n+1} -2u_n + u_{n-1}}{\epsilon^2} + \frac{1}{4}(u_{n+1} + u_{n-1})|u_n|^2=0.
\end{equation}
Nakayama formulated this deformation of discrete space curves as the dynamics of discrete vortex
filaments driven by the self-induction caused by the Biot-Savart law with the discrete analogue of
the local induction approximation \cite{Nakayama:JPSJ2007}.

As to the discrete deformations of discrete curves, the isoperimetric deformation of discrete plane 
curves described by the discrete mKdV equation (dmKdV) has been studied in
\cite{IKMO:KJM,IKMO:MEIS2013,Matsuura:IMRN}. For discrete space curves, the deformations by the
discrete sine-Gordon equation (dsG) and dmKdV has been studied in
\cite{Doliwa-Santini:dsG,IKMO:dmKdV_space_curve,IKMO:MEIS2013}, and the deformation by dNLS is
formulated in \cite{Hoffmann:dNLS,Pinkall:dNLS}.

The dsG and dmKdV describe torsion-preserving isoperimetric and equidistant deformation of the space
discrete curves with constant torsion. However, formulation of discrete deformation of space
discrete curves with varying torsion is a difficult problem. The only example so far is presented by
Hoffmann \cite{Hoffmann:dNLS_HM,Hoffmann:dNLS}, where it has been claimed that composition of
certain two isoperimetric equidistant deformations, which is called the {\em doubly discrete
isotropic Heisenberg magnet} (ddIHM), can be regarded as a discrete analogue of LIE.  This
formulation uses quaternions and its geometric meaning is clear, and it was used for numerical
simulation of fluid flow \cite{Pinkall:dNLS,Weissmann-Pinkall:dNLS_simulation}.  However, in this
formulation, the role of complex curvature is not clear.  In fact, the deformation is expressed as a
coupled system of tangent vectors and displacement vectors of the discrete curves, but the complex
curvature does not appear. Moreover, although it is shown in \cite{Hoffmann:dNLS_HM} that it is
gauge equivalent to the well-known discrete analogue of NLS \eqref{eqn:NLS} proposed by
Ablowitz-Ladik \cite{Ablowitz-Ladik:SIAM1976,Ablowitz-Ladik:SIAM1977}
\begin{equation}\label{eqn:Ablowitz-Ladik}
\begin{split}
&\left(1 + c_0\right) u^m_n
- \left(1 + c_0^\ast\right) u^{m + 1}_n
+ \left(-c_{- 2}^\ast u^{m + 1}_{n - 1}
+ c_{- 2} u^m_{n + 1}\right)
\left(1 + \frac{\epsilon^2}{4} \left|u^m_n\right|^2\right)
\Gamma_n^m\\
&\hskip40pt + c_2 \left(u^m_{n - 1}
+ \epsilon^2 u^m_n \Delta_n^m{}^\ast\right)
- c_2^\ast \left(u^{m + 1}_{n + 1}
+ \epsilon^2 u^{m + 1}_n \Delta_{n+1}^m\right)=0,\\[2mm]
&\hskip20pt \frac{\Gamma_{n+1}^m}{\Gamma_n^m} = \frac{1+\frac{\epsilon^2}{4} |u_n^{m}|^2}{1+\frac{\epsilon^2}{4} |u_n^{m+1}|^2},\\[2mm]
&\hskip20pt \Delta_{n+1}^{m} - \Delta_n^m = \frac{1}{4}(u_n^{m+1\,*}u_{n+1}^{m+1} - u_{n-1}^{m\,*}u_n^m),
\end{split}
\end{equation}
it is not clear that $u_n^m$ actually plays a role of complex curvature of the corresponding
discrete space curves. In \eqref{eqn:Ablowitz-Ladik}, $*$ means complex conjugate, $c_0$, $c_2$,
$c_{-2}$ are complex constants, $\epsilon$ is a real constant which plays a role of lattice interval
of space, $u_n^m,\ \Delta_n^m\in\mathbb{C}$, $\Gamma_n^m\in\mathbb{R}$, and $n$, $m$ are discrete
variables corresponding to space and time, respectively.

On the other hand, from the viewpoint of discretization of LIE \eqref{eqn:local_induction}, we may
consider dNLS \eqref{eqn:dNLS} as a simpler discrete analogue of NLS, 
which is obtained from \eqref{eqn:Ablowitz-Ladik} by putting the parameters as
\begin{equation}
 c_0^* = c_{-2} = \sqrt{-1}\frac{\delta}{\epsilon^2},\quad c_2 = 0.
\end{equation}
The purpose of this paper is to formulate the discrete model of LIE where the complex curvature of
the discrete curves is governed by dNLS \eqref{eqn:dNLS}. In our formulation, the deformation of
discrete curves is expressed in terms of the discrete Frenet frame with the coefficients given by
the curvature and torsion of the curves explicitly. dNLS arises as the governing equation of the
complex curvature, which is the same as the case of smooth curves. Based on this formulation, we
present explicit formulas for the deformations of both smooth and discrete curves in terms of $\tau$
functions of the two-component KP hierarchy. It is well-known that the matrix of the
Frenet-Serret formula gives the Lax matrix of AKNS type by $\mbox{SO}(3)$--$\mbox{SU}(2)$
correspondence \cite{Lamb:JMP1977} and that the AKNS hiearchy arises as a reduction of the
two-component KP hierarchy \cite{JM:RIMS1983,Ueno-Takasaki:book}.  These formulas are consistent
with this fact.  For completeness, we also discuss the case of continuous deformation of discrete
curves described by sdNLS \eqref{eqn:sdNLS}.

The paper is organized as follows: In Section \ref{sec:main}, we study the
continuous binormal flow and its discretization which are governed by sdNLS and
dNLS, respectively. We recall some basic definitions for discrete space curves in Section
\ref{subsec:discrete_curves} and the continuous binormal flow for discrete space curves in Section
\ref{subsec:continuous_binormalflow}. The main theorem, that is, the formulation
of a discrete LIE is presented together with its properties in Section \ref{subsec:dLIE}.
The proofs of the main theorem and its properties are given in Section
\ref{subsec:proof}. Some numerical results for the discrete LIE are
presented in Appendix \ref{subsec:numerical}. In Section \ref{sec:explicit_formulas},
we study explicit formulas for smooth and discrete space curves
by the $\tau$ functions of the two-components KP hierarchy.  In particular, we construct regular soliton
type solutions for the continuous, semi-discrete and discrete LIE. We present
the formulas in terms of the $\tau$ functions satisfying certain bilinear equations
in Section \ref{subset:explicit_formulas_for_space_curves}, which are proved in Section
\ref{proof_of_explicit_formulas}.  We show that the bilinear equations can be solved by the $\tau$
functions of the two-component KP hierarchy in Section \ref{subset:parametrization}. The soliton type
solutions for the continuous, semi-discrete and discrete LIE are constructed in Section
\ref{subsec:formula_deformation}, and their regularity is established in Section
\ref{subset:regularity}.  The proofs of propositions in Section \ref{subset:parametrization} and
\ref{subsec:formula_deformation} are given in Appendix \ref{proofs_of_parametrization} and
\ref{proofs_of_formula_deformation}, respectively.  Finally, concluding remarks are given in Section
\ref{sec:concluding_remarks}.

\section{Discrete models of LIE}\label{sec:main}
\subsection{Discrete space curves}\label{subsec:discrete_curves}
Let $\gamma_n\in\mathbb{R}^3$ be a discrete space curve with
\begin{equation}
 |\gamma_{n+1}-\gamma_n| = \epsilon,
\end{equation}
where $\epsilon$ is a constant. We introduce the {\em discrete Frenet frame}
$\Phi_n=[T_n,N_n,B_n]\in\mathrm{SO}(3)$ by
\begin{equation}\label{eqn:discrete_Frenet_frame}
T_n = \frac{\gamma_{n+1}-\gamma_n}{\epsilon},\quad
B_n = \frac{T_{n-1}\times T_n}{\left|T_{n-1}\times T_n\right|},\quad
N_n = B_n\times T_n.
\end{equation}
Then it follows that the discrete Frenet frame satisfies the {\em discrete Frenet-Serret formula}
\begin{equation}\label{eqn:discrete_Frenet-Serret}
 \Phi_{n+1} = \Phi_n L_n,\quad L_n = R_1(-\nu_{n+1}) R_3(\kappa_{n+1}),
\end{equation}
where
\begin{equation}
 R_1(\theta) = 
\left[\begin{array}{ccc}
1 & 0          & 0          \\
0 & \cos\theta  & -\sin\theta\\
0 & \sin\theta  &  \cos\theta\\
\end{array}\right],\quad
 R_3(\theta) = 
\left[\begin{array}{ccc}
 \cos\theta  & -\sin\theta & 0\\
 \sin\theta  &  \cos\theta & 0\\
     0      &       0     & 1
\end{array}\right],
\end{equation}
and $\nu_n$, $\kappa_n$ are defined by
\begin{equation}\label{eqn:kappa_nu}
\begin{split}
& \langle T_{n-1},T_n\rangle =\cos\kappa_n,\quad  \langle B_{n},B_{n-1}\rangle =\cos\nu_n,\quad 
\langle B_{n},N_{n-1}\rangle =\sin\nu_n,\\
&\hskip60pt -\pi\leq \nu_n< \pi,\quad 0<\kappa_n<\pi. 
\end{split}
\end{equation}
We introduce the {\em discrete complex curvature} $u_n$ by
\begin{equation}\label{eqn:u_discrete}
 u_n = \frac{2}{\epsilon}\tan\frac{\kappa_n}{2} e^{\sqrt{-1}\Lambda_n},\quad 
\Lambda_{n} - \Lambda_{n-1} = -\nu_n,
\end{equation}
following \cite{Doliwa-Santini:JMP}.
%
\subsection{Continuous binormal flow}\label{subsec:continuous_binormalflow}
The binormal flow on the discrete space curves can be introduced by
\begin{equation}\label{eqn:discrete_binormal_flow}
 \frac{d\gamma_n}{dt} = \frac{2}{\epsilon}\tan\frac{\kappa_n}{2} B_n.
\end{equation}
%
\begin{thm}\label{thm:continuouse_deformation}
If $\gamma_n$ is deformed according to \eqref{eqn:discrete_binormal_flow}, the discrete complex
curvature $u_n$ satisfies
\begin{equation}\label{eqn:sdNLS2}
 \frac{1}{\sqrt{-1}}\frac{du_n}{dt} = \frac{1}{\epsilon^2}
\Big(1+\frac{\epsilon^2}{4}|u_n|^2\Big)(u_{n+1}+u_{n-1})
+\frac{c(t)}{\epsilon^2} u_n,
\end{equation}
where $c(t)$ is an arbitrary function in $t$ depending on the boundary condition.
\end{thm}
%
\noindent
Theorem \ref{thm:continuouse_deformation} or the equivalent statement is given in
\cite{Doliwa-Santini:JMP,Hisakado-Wadati,Hoffmann:dNLS_HM}. We provide a proof of Theorem
\ref{thm:continuouse_deformation} for completeness.

\noindent{\bfseries{\it Proof.}}\quad From \eqref{eqn:discrete_Frenet_frame} and
\eqref{eqn:discrete_binormal_flow}, the Frenet frame $\Phi_n$ satisfies
\begin{equation}\label{eqn:continuous_M}
\frac{d\Phi_n}{dt} = \Phi_nM_n,\quad M_n = 
\left[\begin{array}{ccc}
0&-M_{21} & -M_{31} \\
M_{21} & 0 & -M_{32}\\
M_{31} &M_{32} & 0 
  \end{array}\right],
\end{equation}
where
\begin{equation}\label{eqn:continous_M_entries}
\begin{split}
 M_{21}&= \frac{2}{\epsilon^2}\tan\frac{\kappa_{n+1}}{2} \sin\nu_{n+1},\\
M_{31} &= \frac{2}{\epsilon^2}\left(\tan\frac{\kappa_{n+1}}{2}\cos\nu_{n+1} 
- \tan\frac{\kappa_{n}}{2}\right),\\
M_{32} &= \frac{2}{\epsilon^2\sin\kappa_n}\left(\cos\kappa_n\cos\nu_{n+1}\tan\frac{\kappa_{n+1}}{2}
-(1+\cos\kappa_n)\tan\frac{\kappa_n}{2}+\cos\nu_n\tan\frac{\kappa_{n-1}}{2}
\right).
\end{split}
\end{equation}
It follows that $\kappa_n$ and $\nu_n$ should satisfy
\begin{equation}\label{eqn:kappa_t}
 \frac{d\kappa_n }{dt}
= \frac{2}{\epsilon^2}\left(\sin\nu_{n+1}\tan\frac{\kappa_{n+1}}{2}-\sin\nu_n\tan\frac{\kappa_{n-1}}{2}\right),
\end{equation}
so that the diagonal elements in $M_n$ are zero and that $M_n\in\mathfrak{so}(3)$.
Then the compatibility condition of the Frenet-Serret formula\eqref{eqn:discrete_Frenet-Serret}
and the deformation equation \eqref{eqn:continuous_M} 
\begin{equation}
 \frac{dL_n}{dt}=L_nM_{n+1}-M_nL_n
\end{equation}
yields 
\begin{equation}
\begin{split}
& \frac{d\nu_n}{dt} = -(\chi_{n}-\chi_{n-1}),\\
&\chi_n 
= \frac{1}{\epsilon^2}\left(\cot\frac{\kappa_n}{2}+\tan\frac{\kappa_n}{2}\right)
\left(\cos\nu_{n+1}\tan\frac{\kappa_{n+1}}{2}+\cos\nu_n\tan\frac{\kappa_{n-1}}{2}\right),
\end{split}
 \end{equation}
which is summed up as
\begin{equation}\label{eqn:Lambda_t}
\frac{d\Lambda_n}{dt} = \frac{1}{\epsilon^2}\left(\cot\frac{\kappa_n}{2}+\tan\frac{\kappa_n}{2}\right)
\left(\cos\nu_{n+1}\tan\frac{\kappa_{n+1}}{2}+\cos\nu_n\tan\frac{\kappa_{n-1}}{2}\right) + c(t).
\end{equation}
Here, $c(t)$ is an arbitrary function in $t$ depending on the boundary condition. Then \eqref{eqn:kappa_t} and
\eqref{eqn:Lambda_t} can be rewritten as a single equation in terms of $u_n$, which
is nothing but \eqref{eqn:sdNLS2}. This completes the proof.\quad \qed
\bigskip
%
\subsection{Discrete LIE}\label{subsec:dLIE}
We now formulate the discrete deformation of discrete space curve as a discrete model of LIE
\eqref{eqn:local_induction}. We start from dNLS \eqref{eqn:dNLS}, and identify $u_n^m$ as the
complex curvature of discrete space curve $\gamma_n^m\in\mathbb{R}^3$ where $m$ is the number of
iterations of discrete deformations. We impose the boundary condition as
\begin{equation}\label{eqn:dNLS_bc}
 u_n^m\to 0\ (n\to\pm\infty),\quad \Gamma_n^m\to \Gamma_{\pm\infty}^m\ (n\to\pm\infty).
\end{equation}
Then one of the main statements of this paper is given as follows:
%
\begin{thm}[discrete LIE]\label{thm:discrete_deformation}
For a fixed $m$, let $\gamma_n^m\in\mathbb{R}^3$ be a discrete space curve satisfying
\begin{equation}
 |\gamma_{n+1}^m - \gamma_n^m | = \epsilon,
\end{equation}
and $\Phi_n^m=[T_n^m, N_n^m, B_n^m]\in\mathrm{SO}(3)$ be the discrete Frenet frame defined in
\eqref{eqn:discrete_Frenet_frame} satisfying the discrete Frenet-Serret formula
\begin{equation}\label{eqn:fulldiscrete_Frenet-Serret}
 \Phi_{n+1}^m = \Phi_n^m L_n^m,\quad L_n^m = R_1(-\nu_{n+1}^m)R_3(\kappa_{n+1}^m).
\end{equation}
Let $u_n^m$ be a complex discrete curvature of $\gamma_n^m$. We compute $u_n^{m+1}$ by dNLS
\eqref{eqn:dNLS} under the boundary condition \eqref{eqn:dNLS_bc} and determine $\kappa_n^{m+1}$ and
$\Lambda_n^{m+1}$ by 
$u_n^{m+1}=\frac{2}{\epsilon}\tan\frac{\kappa_n^{m+1}}{2} e^{\sqrt{-1}\Lambda_n^{m+1}}$.  
We define a new curve $\gamma_n^{m+1}\in\mathbb{R}^3$ by
\begin{equation}\label{eqn:discrete_def_Frenet}
 \frac{\gamma_n^{m+1} - \gamma_n^m}{\delta} = \frac{2}{\epsilon^3}(P_n^m T_n^m + Q_n^m N_n^m + R_n^m B_n^m),
\end{equation}
\begin{equation}\label{eqn:discrete_coeffs_Frenet}
 \begin{split}
&P_n^m = \delta\left(-1 + \frac{\Gamma_n^m}{\cos^2\frac{\kappa_n^m}{2}}\right),\\
&Q_n^m = \delta\left[\tan\frac{\kappa_n^m}{2} - \tan\frac{\kappa_{n-1}^{m+1}}{2}
\cos(\Lambda_{n-1}^{m+1}-\Lambda_n^m)\frac{\Gamma_n^m}{\cos^2\frac{\kappa_n^m}{2}}\right],\\
& R_n^m= \epsilon^2\tan\frac{\kappa_n^m}{2} 
- \delta\tan\frac{\kappa_{n-1}^{m+1}}{2}\sin(\Lambda_{n-1}^{m+1}-\Lambda_n^m)\frac{\Gamma_n^m}{\cos^2\frac{\kappa_n^m}{2}}.
 \end{split}
\end{equation}
Suppose that $\Gamma_{\infty}^m$ and $\Gamma_{-\infty}^m$ are either $1$ or
$1+\frac{\epsilon^4}{\delta^2}$.  Then, it follows that
\begin{enumerate}
 \item $|\gamma_{n+1}^{m+1}- \gamma_n^{m+1}|=\epsilon$. Namely, $\gamma_n^{m+1}$ is an
isoperimetric deformation of $\gamma_n^m$.
 \item $u_n^{m+1}$ gives the complex discrete curvature of $\gamma_n^{m+1}$.
\end{enumerate}
\end{thm}
%
Equations \eqref{eqn:discrete_def_Frenet} and \eqref{eqn:discrete_coeffs_Frenet} can
be regarded as a discrete analogue of LIE \eqref{eqn:local_induction}, which will be referred to as
the discrete LIE (dLIE).  We note that dLIE is an implicit scheme in a sense that
$\gamma_n^{m+1}$ is determined by using $\kappa_{n-1}^{m+1}$ and $\Lambda_{n-1}^{m+1}$ which
incorporate the information of $\gamma_{n}^{m+1}$. This is resolved by using dNLS \eqref{eqn:dNLS} to
compute $\kappa_{n-1}^{m+1}$ and $\Lambda_{n-1}^{m+1}$, as described in Theorem \ref{thm:discrete_deformation}. We will explain
the details of how to compute numerically the deformation of curves given by dLIE in Appendix \ref{subsec:numerical}.
We also remark that the deformation
given by dLIE is {\em not} an equidistant deformation
in contrast with the deformation described by dmKdV \cite{IKMO:dmKdV_space_curve}. In fact, one can
show the following proposition:
\begin{prop}\label{prop:distance}
Let $\gamma_n^m\in\mathbb{R}^3$ be the family of discrete space curves given in Theorem
\ref{thm:discrete_deformation}. Then it follows that
\begin{equation}\label{eqn:discrete_displacement_distance}
 \left|\frac{\gamma_n^{m+1}-\gamma_n^m}{\delta}\right|^2 
= \frac{4}{\epsilon^2}\left(-1 + \frac{\Gamma_n^{m}}{\cos^2\frac{\kappa_n^m}{2}}\right).
\end{equation} 
\end{prop}
Equation \eqref{eqn:discrete_displacement_distance} also implies that the solution of dNLS
\eqref{eqn:dNLS} should satisfy the condition $\Gamma_n^{m}\geq \cos^2\frac{\kappa_n^m}{2}$ in order
to be consistent with the curve deformation.

Continuous limit with respect to time can be simply taken as $t=m\delta$ and $\delta\to 0$.  Then
dNLS \eqref{eqn:dNLS} and corresponding deformation equation \eqref{eqn:discrete_def_Frenet} and
\eqref{eqn:discrete_coeffs_Frenet} yields the sdNLS \eqref{eqn:sdNLS2} with $c(t)=-2$ and the
binormal flow \eqref{eqn:discrete_binormal_flow}.

%
The dLIE \eqref{eqn:discrete_def_Frenet} and \eqref{eqn:discrete_coeffs_Frenet} implies the
following deformation of discrete Frenet frame:
\begin{prop}\label{prop:Lax_SO(3)}
Let $\gamma_n^m\in\mathbb{R}^3$ be the family of discrete space curves given in Theorem
\ref{thm:discrete_deformation}, and $\Phi_n^m$ be its discrete Frenet frame. Then
$\Phi_n^m$ satisfies
\begin{equation}\label{eqn:discrete_deformation_Frenet}
\begin{split}
& \Phi_n^{m+1} = \Phi_n^m M^m_n,\\
&M_n^m = \frac{1}{\Gamma^m_{n + 1}}
\begin{bmatrix}
\left|\alpha^m_n\right|^2 - \left|\beta^m_n\right|^2 &
2 \Re \left(\alpha^m_n \beta^m_n{}^\ast\right) &
- 2 \Im \left(\alpha^m_n \beta^m_n{}^\ast{}\right)\\
- 2 \Re \left(\alpha^m_n \beta^m_n\right) &
\Re \left(({\alpha^m_n})^2 - ({\beta^m_n})^2\right) &
- \Im \left(({\alpha^m_n})^2 + ({\beta^m_n})^2\right)\\
- 2 \Im \left(\alpha^m_n \beta^m_n\right) &
\Im \left(({\alpha^m_n})^2 - ({\beta^m_n})^2\right) &
\Re \left(({\alpha^m_n})^2 + ({\beta^m_n})^2\right)
\end{bmatrix}\in\mathrm{SO}(3) ,
\end{split}
\end{equation}
where $\alpha_n^m,\beta_n^m\in\mathbb{C}$ are given by
\begin{equation}\label{eqn:alpha_beta}
\begin{split}
 \alpha_n^m
&=
\sqrt{- 1} \frac{\delta}{\epsilon^2}
\left[ \left(1-\sqrt{-1}\frac{\epsilon^2}{\delta}\right) 
     - \left(1 + \frac{\epsilon^2}{4} u_{n + 1}^m u_n^{m+1}{}^*\right) \Gamma_{n + 1}^m\right] 
e^{\frac{\sqrt{-1}}{2}(\Lambda_n^{m+1} - \Lambda_n^m)},\\
\beta_n^m
&=
\sqrt{- 1} \frac{\delta}{2\epsilon}
\left(u_n^{m+1} - u_{n + 1}^m\right) \Gamma_{n + 1}^m
e^{- \frac{\sqrt{- 1}}{2}(\Lambda_n^m + \Lambda_n^{m+1})},
\end{split}
\end{equation}
respectively. The compatibility condition with the discrete Frenet formula
\eqref{eqn:fulldiscrete_Frenet-Serret} $L_n^{m}M_{n+1}^m=M_{n}^mL_n^{m+1}$ yields dNLS
\eqref{eqn:dNLS}.
\end{prop}
Note that the Frenet-Serret formula \eqref{eqn:fulldiscrete_Frenet-Serret} and the deformation equation
\eqref{eqn:discrete_deformation_Frenet} can be transformed to the $\mathrm{SU}(2)$ version by the
standard correspondence of $\mathrm{SO}(3)$ and $\mathrm{SU}(2)$ (see, for example, \cite{IKMO:dmKdV_space_curve,Rogers-Schief:book}) as
\begin{equation}\label{eqn:Lax_Frenet_SU(2)}
\begin{split}
& \phi_{n+1}^m = \phi_n^m L_n^m,\quad L_n^m=
 \begin{bmatrix}\medskip
\cos \frac{\kappa^m_{n + 1}}{2}\,
e^{- \frac{\sqrt{- 1} }{2}\nu^m_{n + 1}} &
- \sin \frac{\kappa^m_{n + 1}}{2}\,
e^{- \frac{\sqrt{- 1} }{2}\nu^m_{n + 1}}\\
\sin \frac{\kappa^m_{n + 1}}{2}\,
e^{\frac{\sqrt{- 1} }{2}\nu^m_{n + 1}} &
\cos \frac{\kappa^m_{n + 1}}{2}\,
e^{\frac{\sqrt{- 1} }{2}\nu^m_{n + 1}}
\end{bmatrix}\in\text{SU}(2),\\
& \phi_{n}^{m+1} = \phi_n^m M_n^m,\quad M_n^m=
\frac{1}{\sqrt{\Gamma^m_{n + 1}}}
\begin{bmatrix}
\alpha^m_n & \beta^m_n\\
- \beta^m_n{}^\ast & \alpha^m_n{}^\ast
\end{bmatrix}\in\text{SU}(2),
\end{split}
\end{equation}
which is known as the {\em Lax pair} of dNLS \cite{Ablowitz-Ladik:SIAM1976,Ablowitz-Ladik:SIAM1977}. 
%
\subsection{Proof}\label{subsec:proof}
In this section we give the proof of the results in Section \ref{subsec:dLIE}.  The first statement of
Theorem \ref{thm:discrete_deformation} may be verified directly in principle, by computing
$\gamma_{n+1}^{m+1}-\gamma_n^{m+1}$ and its length from \eqref{eqn:discrete_def_Frenet},
\eqref{eqn:discrete_coeffs_Frenet} and the discrete Frenet-Serret formula
\eqref{eqn:fulldiscrete_Frenet-Serret} under the assumption that $u_n^{m+1}$ is determined by dNLS
\eqref{eqn:dNLS}. In fact, $\gamma_{n+1}^{m+1}-\gamma_n^{m+1}$ can be expressed from
\eqref{eqn:discrete_Frenet_frame}, \eqref{eqn:fulldiscrete_Frenet-Serret} and
\eqref{eqn:discrete_def_Frenet} as
\begin{displaymath}
 \gamma_{n+1}^{m+1}-\gamma_n^{m+1} = (\gamma_{n+1}^{m+1} - \gamma_{n+1}^{m}) - ( \gamma_n^{m+1} - \gamma_{n}^{m})
+ (\gamma_{n+1}^m - \gamma_n^m)
= \Phi_n^m
\left\{\ 
\left[\begin{array}{c}\epsilon\\ 0 \\ 0\end{array}\right] 
+ \frac{2\delta}{\epsilon^3}L_n^m\left[\begin{array}{c}P_{n+1}^{m}\\Q_{n+1}^{m}\\R_{n+1}^m \end{array}\right]
-\frac{2\delta}{\epsilon^3}\left[\begin{array}{c}P_{n}^{m}\\Q_{n}^{m}\\R_{n}^m \end{array}\right]\ 
\right\}.
\end{displaymath}
However, each entry has a complicated expression in terms of $\kappa_n^m$, $\nu_n^m$ and
$\Lambda_n^m$ to carry out further calculations. To make it feasible, we change the Frenet frame to
an alternate frame used in 
\cite{Doliwa-Santini:JMP,Hasimoto:JFM1972,Lamb:JMP1977}, which we call the {\em complex
parallel frame} in this paper. Let $F_n^m=[T_n^m,U_n^m,U_n^m{}^\ast]\in\mathrm{U}(3)$ be the complex
parallel frame defined by
\begin{equation}
 U_n^m = \frac{e^{\sqrt{- 1} \Lambda_n^m}}{\sqrt{2}} \left(N_n^m + \sqrt{- 1} B_n^m\right).
\end{equation}
The complex parallel frame $F_n^m$ is related to the discrete Frenet frame $\Phi_n^m$ as
\begin{equation}\label{eqn:rel_Frenet_complex_parallel_discrete}
F_n^m = \Phi_n^m
\begin{bmatrix}
1 & 0 & 0\\
0 & 1 & 1\\
0 & \sqrt{- 1} & - \sqrt{- 1}
\end{bmatrix}
\begin{bmatrix}
1 & 0 & 0\\
0 & \frac{e^{\sqrt{- 1} \Lambda_n^m}}{\sqrt{2}} & 0\\
0 & 0 & \frac{e^{-\sqrt{- 1} \Lambda_n^m}}{\sqrt{2}}
\end{bmatrix}.
\end{equation}
Note that for two real vectors $v,w\in\mathbb{R}^3$ expressed in terms of $F_n^m$ as
\begin{equation}
v = F_n^m \left[\begin{array}{c}p \\ q\\\  q^*\end{array}\right],\quad
w = F_n^m \left[\begin{array}{c}x \\ y\\\  y^*\end{array}\right],
\end{equation}
their scalar product and vector product are given by
\begin{equation}\label{eqn:products_complex_paralell_frame}
 \langle v,w\rangle = px + qy^* + q^*y,\quad
v\times  w
= F_n^m\,\sqrt{-1}\left[
\begin{array}{c}
 q^*y - qy^* \\
 qx - py\\
 -q^*x + py^*
\end{array}\right],
\end{equation}
respectively. In particular, we have
\begin{equation}\label{eqn:norm_complex_paralell_frame}
 |v|^2 = p^2 + 2|q|^2.
\end{equation}
The advantage of complex parallel frame is that the complex curvature $u_n^m$ naturally arises in
this framework; the discrete Frenet-Serret formula \eqref{eqn:fulldiscrete_Frenet-Serret} and the
deformation of the discrete curve are rewritten in terms of $u_n^m$ as
\begin{equation}\label{eqn:discrete_Frenet_complex_parallel}
 F_{n + 1}^m = F_n^m X_n^m,\quad
X_n^m =
\frac{1}{1 + \frac{\epsilon^2}{4} \left|u_{n + 1}^m\right|^2}
\begin{bmatrix}
1 - \frac{\epsilon^2}{4} \left|u_{n + 1}^m\right|^2 &- \frac{\epsilon}{\sqrt{2}}  u_{n + 1}^{m} 
&- \frac{\epsilon}{\sqrt{2}} u_{n + 1}^{m\ \ast}\\[1mm]
\frac{\epsilon}{\sqrt{2}}  u_{n + 1}^{m\ \ast} & 1 & - \frac{\epsilon^2}{4} (u_{n + 1}^{m\ \ast})^2\\[1mm]
\frac{\epsilon}{\sqrt{2}} u_{n + 1}^{m} & - \frac{\epsilon^2}{4} (u_{n + 1}^{m})^2 & 1
\end{bmatrix},
\end{equation}
and 
\begin{equation}\label{eqn:deformation_gamma_complex_parallel}
\gamma_n^{m+1}
= \gamma_n^m
+ \frac{2 \delta^2}{\epsilon^3}
F_n^m
\begin{bmatrix}
- 1 + \left(1 + \frac{\epsilon^2}{4} \left|u_n^m\right|^2\right) \Gamma_n^m\\[2mm]
{\textstyle \frac{\epsilon}{2\sqrt{2}}
\left\{\left({\scriptstyle  -\sqrt{-1}}\frac{\epsilon^2}{\delta} + 1\right)u_n^m{}^\ast 
- u_{n - 1}^{m+1}{}^\ast \left(1 + \frac{\epsilon^2}{4} \left|u_n^m\right|^2\right) \Gamma_n^m\right\}}\\[3mm]
{\textstyle \frac{\epsilon}{2\sqrt{2}}
\left\{\left({\scriptstyle  \sqrt{-1}}\frac{\epsilon^2}{\delta} + 1\right) u_n^m
 - u_{n - 1}^{m+1} \left(1 + \frac{\epsilon^2}{4} \left|u_n^m\right|^2\right) \Gamma_n^m\right\}}
\end{bmatrix},
\end{equation}
respectively. We have
\begin{align}
 \gamma_{n+1}^{m+1} - \gamma_n^{m+1} 
&=  \gamma_{n+1}^{m} - \gamma_n^{m} 
- \frac{\sqrt{2}\delta^2}{\epsilon^2}F_n^m
\left[
\begin{array}{c}
\frac{\sqrt{2}}{4}\epsilon (u_n^{m+1} - u_{n+1}^m)(u_n^{m+1}{}^* - u_{n+1}^{m\, *})\Gamma_{n+1}^m\\[2mm]
-\frac{1}{2}(u_n^{m+1}{}^* - u_{n+1}^{m\, *})\left\{
1+\sqrt{-1}\frac{\epsilon^2}{\delta} - (1+\frac{\epsilon^2}{4} u_{n+1}^{m\, *}\,u_n^{m+1})\Gamma_{n+1}^m
\right\}\\[2mm]
-\frac{1}{2}(u_n^{m+1} - u_{n+1}^{m})\left\{
1-\sqrt{-1}\frac{\epsilon^2}{\delta} - (1+\frac{\epsilon^2}{4} u_{n+1}^{m}\,u_n^{m+1\,*})\Gamma_{n+1}^m
\right\}
\end{array}
\right]\nonumber\\
&= \epsilon F_n^m \left\{\ 
\left[\begin{array}{c}1\\ 0 \\ 0 \end{array}\right] 
- \frac{\sqrt{2}}{\Gamma_{n+1}^m}
\left[
\begin{array}{c}
 \sqrt{2}\big|\,\widehat\beta_n^m\,\big|^2\\[1mm]
\widehat\alpha_n^m{}^*\,\widehat\beta_n^m{}^*\\[1mm]
\widehat\alpha_n^m\,\,\widehat\beta_n^m
\end{array}
\right]\ 
\right\}, \label{eqn:segment_m+1}
\end{align}
where $\widehat\alpha_n^m$ and $\widehat\beta_n^m$ are given by
\begin{equation}\label{eqn:hat_alpha-beta}
\begin{split}
& \widehat\alpha_n^m = \alpha_n^m e^{-\frac{\sqrt{-1}}{2}(\Lambda_n^{m+1} - \Lambda_n^m)}
= \sqrt{- 1} \frac{\delta}{\epsilon^2}
\left[ \left(1-\sqrt{-1}\frac{\epsilon^2}{\delta}\right) 
     - \left(1 + \frac{\epsilon^2}{4} u_{n + 1}^m u_n^{m+1}{}^*\right) \Gamma_{n + 1}^m\right],  \\
&\widehat\beta_n^m
= \beta_n^m e^{\frac{\sqrt{- 1}}{2}(\Lambda_n^m + \Lambda_n^{m+1})}
=\sqrt{- 1} \frac{\delta}{2\epsilon}
\left(u_n^{m+1} - u_{n + 1}^m\right) \Gamma_{n + 1}^m,
\end{split}
\end{equation} 
respectively. Then we obtain
\begin{equation}\label{eqn:iso1}
 |\gamma_{n+1}^{m+1} - \gamma_n^{m+1}|^2 
= \epsilon^2\left\{
1+\frac{4\big|\,\widehat\beta_n^m\,\big|^2}{\Gamma_{n+1}^{m\ 2}}
\left(\left|\,\widehat\alpha_n^m\,\right|^2 + \big|\,\widehat\beta_n^m\,\big|^2 - \Gamma_{n+1}^m\right)
\right\}.
\end{equation}
In order to evaluate $\left|\,\widehat\alpha_n^m\,\right|^2 + \big|\,\widehat\beta_n^m\,\big|^2$, 
we introduce $\psi_n^m$ by 
\begin{equation}\label{eqn:gauge_Frenet_parallel_su(2)}
 \psi_n^m=\phi_n^m R_1(-\Lambda_n^m),\quad R_1(\theta) 
= \left[
\begin{array}{cc}
 e^{\frac{\sqrt{-1}}{2}\theta} & 0 \\  0 & e^{-\frac{\sqrt{-1}}{2}\theta}
\end{array}
\right],
\end{equation}
and consider a Lax pair of dNLS
\begin{equation}\label{eqn:Lax_dNLS_2}
\begin{split}
& \psi_{n+1}^m = \psi_n^m \widehat{L}_n^m,\quad \widehat{L}_n^m=
\frac{1}{\sqrt{1+\frac{\epsilon^2}{4}|u_{n+1}^m|^2}} 
\begin{bmatrix}\medskip
1 & - \frac{\epsilon}{2} u_{n+1}^m\\
\frac{\epsilon}{2} u_{n+1}^{m\ *} & 1
\end{bmatrix},\\
& \psi_{n}^{m+1} = \psi_n^m \widehat{M}_n^m,\quad \widehat{M}_n^m=
\frac{1}{\sqrt{\Gamma^m_{n + 1}}}
\begin{bmatrix}\smallskip
\widehat\alpha^m_n & \widehat\beta^m_n\\
- \widehat\beta^m_n{}^\ast & \widehat\alpha^m_n{}^\ast
\end{bmatrix}.
\end{split}
\end{equation}
Note that the gauge transformation \eqref{eqn:gauge_Frenet_parallel_su(2)} preserves the
compatibility conditions of associated linear problems; $\widehat L_n^{m}\widehat M_{n+1}^m =
\widehat M_{n}^m\widehat L_n^{m+1}$ is equivalent to $L_n^{m}M_{n+1}^m = M_{n}^m L_n^{m+1}$. Indeed,
one can verify directly that dNLS \eqref{eqn:dNLS} is obtained from those compatibility
conditions. Taking the determinant of both sides of $\widehat L_n^{m}\widehat M_{n+1}^m = \widehat
M_{n}^m\widehat L_n^{m+1}$, we see from $\det \widehat L_n^m=1$ that $ \det \widehat M_{n+1}^m =
\det\widehat M_n^m$, which implies
\begin{equation}
\frac{1}{\Gamma_{n+1}^m}
\left(\left|\,\widehat\alpha_n^m\,\right|^2 + \left|\,\widehat\beta_n^m\,\right|^2\right)
=C_m,
\end{equation}
for some $C_m$ depending only on $m$. Substituting \eqref{eqn:hat_alpha-beta} and taking the limit
$n\to\pm\infty$, we see from the boundary condition \eqref{eqn:dNLS_bc} that $C_m=1$, namely
\begin{equation}\label{eqn:alpha_beta_squaredsum}
\left|\,\widehat\alpha_n^m\,\right|^2 + \bigl|\,\widehat\beta_n^m\,\bigr|^2 = \Gamma_{n+1}^m,
\end{equation}
and thus \eqref{eqn:iso1} gives  $|\gamma_{n+1}^{m+1} - \gamma_n^{m+1} | = \epsilon$.  This completes the 
proof of the first statement of Theorem \ref{thm:discrete_deformation}.
Consequently \eqref{eqn:segment_m+1} yields
\begin{equation}\label{tangent^{m+1}}
T_n^{m+1}
=
\frac{\gamma_{n+1}^{m+1} - \gamma_n^{m+1}}{\epsilon}
= F_n^m
\frac{1}{\Gamma_{n+1}^m}
\left[
\begin{array}{c}
\left|\,\widehat\alpha_n^m\,\right|^2
-\big|\,\widehat\beta_n^m\,\big|^2\\[2mm]
-\sqrt{2}\,\widehat\alpha_n^m{}^*\,\widehat\beta_n^m{}^*\\[2mm]
-\sqrt{2}\,\widehat\alpha_n^m\,\widehat\beta_n^m
\end{array}
\right].
\end{equation}

\par\bigskip

We next proceed to the proof of the second statement. Our goal is to show
\begin{equation}\label{eqn:goal}
 \langle T_{n-1}^{m+1}, T_{n}^{m+1}\rangle = \cos\kappa_n^{m+1},\quad
  \langle B_{n-1}^{m+1}, B_{n}^{m+1}\rangle = \cos\nu_n^{m+1},\quad
  \langle N_{n-1}^{m+1}, B_{n}^{m+1}\rangle = \sin\nu_n^{m+1},
\end{equation}
where $\kappa_n^{m+1}$ and $\nu_{n}^{m+1}$ are determined from dNLS \eqref{eqn:dNLS},
$T_{n-1}^{m+1}$, $B_{n-1}^{m+1}$ and $B_n^{m+1}$ are constructed from the Frenet frame $\Phi_n^m$ by
using the deformation equation \eqref{eqn:discrete_def_Frenet} and
\eqref{eqn:discrete_coeffs_Frenet}.  Note that from the compatibility condition $\widehat
L_n^{m}\widehat M_{n+1}^m = \widehat M_{n}^m\widehat L_n^{m+1}$ of \eqref{eqn:Lax_dNLS_2} we have
the following equality:
\begin{equation}\label{eqn:alpha_beta_contiguity}
 \left[\begin{array}{c} \widehat{\alpha}_{n-1}^m  \\[2mm]
\widehat{\beta}_{n-1}^m\end{array}\right]
= \frac{1}{1+\frac{\epsilon^2}{4}|u_n^m|^2} 
\left[\begin{array}{cc}
1 & -\frac{\epsilon}{2} u_n^{m+1\ \ast}\\[2mm]
\frac{\epsilon}{2} u_{n}^{m+1} & 1
\end{array}\right]
\left[
\begin{array}{c}
 \widehat{\alpha}_n^m + \frac{\epsilon}{2} u_n^m \widehat{\beta}_n^{m\ \ast}\\[2mm]
 \widehat{\beta}_n^m - \frac{\epsilon}{2} u_n^m \widehat{\alpha}_n^{m\ \ast}
\end{array}
\right],
\end{equation}
which will be frequently used in the calculations below.  Shifting $n \to n - 1$ in
\eqref{tangent^{m+1}}, the discrete Frenet-Serret formula
\eqref{eqn:discrete_Frenet_complex_parallel} and \eqref{eqn:alpha_beta_contiguity} yields
\begin{align}
T^{m+1}_{n-1}
&=
F^m_n\frac{1}{1+\frac{\epsilon^2}{4}\left|u^m_n\right|^2}
\begin{bmatrix}
1 - \frac{\epsilon^2}{4} \left|u_n^m\right|^2 &
\frac{\epsilon}{\sqrt{2}} u_n^m &
\frac{\epsilon}{\sqrt{2}} u_n^{m\, \ast}\\[2mm]
- \frac{\epsilon}{\sqrt{2}} u_n^{m\, \ast} & 1 & - \frac{\epsilon^2}{4} (u_n^{m\,\ast})^2\\[2mm]
- \frac{\epsilon}{\sqrt{2}} u_n^m & - \frac{\epsilon^2}{4} (u_n^m)^2 & 1
\end{bmatrix}
\frac{1}{\Gamma^m_n}
\begin{bmatrix}
\left|\,\widehat\alpha_{n-1}^m\,\right|^2
- \big|\,\widehat\beta_{n-1}^m\,\big|^2\\[2mm]
- \sqrt{2}\,\widehat\alpha_{n-1}^{m\, *}\,\widehat\beta_{n-1}^{m\, *}\\[2mm]
- \sqrt{2}\,\widehat\alpha_{n-1}^{m}\,\widehat\beta_{n-1}^{m}
\end{bmatrix} \nonumber\\
&=
\frac{1-\frac{\epsilon^2}{4}\left|u_n^{m+1}\right|^2}{1+\frac{\epsilon^2}{4}\big|u_n^{m+1}\big|^2} T^{m+1}_n
- F_n^m \frac{1}{\Big(1+\frac{\epsilon^2}{4} \left|u_n^m\right|^2\Big)\Gamma_n^m}
\begin{bmatrix}
x_n^m\\[2mm]
\sqrt{2} y_n^{m\, *}\\[2mm]
\sqrt{2} y_n^m
\end{bmatrix}, \label{eqn:T(n-1,m+1)}
\end{align}
where
\begin{equation}
x^m_n =  \epsilon \left(u_n^{m+1}\, \widehat{\alpha}_n^m\,\widehat{\beta}_n^{m\,*}
+ u_n^{m+1\, *}\, \widehat{\alpha}_n^{m\, *}
\widehat{\beta}_n^m\right),\quad 
y_n^m =
\frac{\epsilon}{2} \left(u_n^{m+1}\, {\widehat{\alpha}_n^m}{}^2- u^{m+1\,\ast}_n\, {\widehat{\beta}_n^m}{}^2\right).
\end{equation}
Then we compute $ \langle T_{n-1}^{m+1}, T_{n}^{m+1}\rangle $ by using
\eqref{eqn:products_complex_paralell_frame}, \eqref{tangent^{m+1}} and \eqref{eqn:T(n-1,m+1)} as
\begin{align*}
 \langle T_{n-1}^{m+1}, T_{n}^{m+1}\rangle 
&= \frac{1-\frac{\epsilon^2}{4}\left|u_n^{m+1}\right|^2}{1+\frac{\epsilon^2}{4}\left|u_n^{m+1}\right|^2}
 - \frac{\left(\left|\widehat\alpha_n^m\right|^2 - \big|\widehat\beta_n^m\big|^2\right) x_n^m
- 2y_n^m{}^*\,\widehat\alpha_n^m\,\widehat\beta_n^m 
- 2y_n^m\,\widehat\alpha_n^m{}^*\,\widehat\beta_n^m{}^*}
{\left(1+\frac{\epsilon^2}{4} \left|u_n^m\right|^2\right)\Gamma_n^m\Gamma_{n+1}^m}
\\
&= \frac{1-\frac{\epsilon^2}{4}\left|u_n^{m+1}\right|^2}{1+\frac{\epsilon^2}{4}\left|u_n^{m+1}\right|^2} = \cos \kappa_{n}^{m+1},
\end{align*}
which is the first equation of \eqref{eqn:goal}. In order to show the second equation of
\eqref{eqn:goal}, first we need to compute $B_{n}^{m+1}$. Noticing
\eqref{eqn:products_complex_paralell_frame} and \eqref{eqn:alpha_beta_squaredsum}, we have from
\eqref{tangent^{m+1}}, \eqref{eqn:T(n-1,m+1)}
\begin{align*}
T_{n - 1}^{m+1} \times T_n^{m+1}
&=
\frac{1}{\big(1 + \frac{\epsilon^2}{4} |u_n^m|^2\big)
\Gamma_n^m \Gamma_{n + 1}^m}
F_n^m
\sqrt{- 1}
\begin{bmatrix}
2 y_n^m\, \widehat{\alpha}^m_n{}^\ast\, \widehat{\beta}_n^m{}^\ast
- 2 y_n^m{}^\ast\, \widehat{\alpha}_n^m\, \widehat{\beta}_n^m\\[2mm]
- \sqrt{2} y_n^m{}^* \left(\left|\widehat{\alpha}_n^m\right|^2
- \big|\widehat{\beta}_n^m\big|^2\right)
- \sqrt{2} x_n^m\, \widehat{\alpha}_n^m{}^*\,\widehat{\beta}_n^m{}^*\\[2mm]
\sqrt{2} y_n^m \left(\left|\widehat{\alpha}_n^m\right|^2
- \big|\widehat{\beta}_n^m\big|^2\right)
+ \sqrt{2} x_n^m\, \widehat{\alpha}_n^m\, \widehat{\beta}_n^m
\end{bmatrix}\\
&=
\frac{ \epsilon}{\sqrt{2}\big(1 + \frac{\epsilon^2}{4} |u_n^m|^2\big) \Gamma_{n}^m}
F_n^m
\sqrt{- 1}
\begin{bmatrix}
\sqrt{2} \left(u_n^{m+1}
\widehat{\alpha}_n^m\, \widehat{\beta}_n^m{}^*
- u_n^{m+1}{}^*
\widehat{\alpha}_n^m{}^*\, \widehat{\beta}_n^m\right)\\[2mm]
- u_n^{m+1}{}^* (\widehat{\alpha}_n^m{}^*{})^2
- u_n^{m+1} (\widehat{\beta}_n^m{}^*{})^2\\[2mm]
u_n^{m+1} (\widehat{\alpha}_n^m{})^2
+ u_n^{m+1\ \ast} (\widehat{\beta}_n^m{})^2
\end{bmatrix}.
\end{align*}
Then we obtain by using \eqref{eqn:dNLS}
\begin{displaymath}
 \big|T_{n - 1}^{m+1} \times T_n^{m+1}\big| = \frac{\epsilon |u_n^{m+1}|}{1 + \frac{\epsilon^2}{4} \big|u_n^{m+1}\big|^2},
\end{displaymath}
which gives
\begin{equation}\label{eqn:B(n,m+1)}
 B_n^{m+1} = 
F_n^m 
\frac{\sqrt{- 1}}{\sqrt{2} \Gamma_{n+1}^m}
\begin{bmatrix}
\sqrt{2} \left(e^{\sqrt{-1}\Lambda_n^{m+1}}\widehat{\alpha}_n^m\, \widehat{\beta}_n^m{}^*
- e^{-\sqrt{-1}\Lambda_n^{m+1}}
\widehat{\alpha}_n^m{}^*\, \widehat{\beta}_n^m\right)\\[2mm]
- e^{-\sqrt{-1}\Lambda_n^{m+1}} (\widehat{\alpha}_n^m{}^*{})^2
- e^{\sqrt{-1}\Lambda_n^{m+1}} (\widehat{\beta}_n^m{}^*{})^2\\[2mm]
e^{\sqrt{-1}\Lambda_n^{m+1}} (\widehat{\alpha}_n^m{})^2
+ e^{-\sqrt{-1}\Lambda_n^{m+1}} (\widehat{\beta}_n^m{})^2
\end{bmatrix}.
\end{equation}
Computing $ B_{n-1}^{m+1}$ by using \eqref{eqn:discrete_Frenet_complex_parallel} and
\eqref{eqn:alpha_beta_contiguity}, we have after long but straightforward calculations
\begin{align}
 B_{n-1}^{m+1} & =
 F_n^m (X_{n-1}^m)^{-1}
\frac{\sqrt{- 1}}{\sqrt{2} \Gamma_n^m}
\begin{bmatrix}
\sqrt{2} \left(e^{\sqrt{-1}\Lambda_{n-1}^{m+1}}\widehat{\alpha}_{n-1}^m\, \widehat{\beta}_{n-1}^{m\,*}
- e^{-\sqrt{-1}\Lambda_{n-1}^{m+1}}
\widehat{\alpha}_{n-1}^{m\,*}\, \widehat{\beta}_{n-1}^m\right)\\[2mm]
- e^{-\sqrt{-1}\Lambda_{n-1}^{m+1}} (\widehat{\alpha}_{n-1}^{m\,*}{})^2
- e^{\sqrt{-1}\Lambda_{n-1}^{m+1}} (\widehat{\beta}_{n-1}^{m\,*}{})^2\\[2mm]
e^{\sqrt{-1}\Lambda_{n-1}^{m+1}} (\widehat{\alpha}_{n-1}^m{})^2
+ e^{-\sqrt{-1}\Lambda_{n-1}^{m+1}} (\widehat{\beta}_{n-1}^m{})^2
 \end{bmatrix}\nonumber\\
& =  F_n^m 
\frac{\sqrt{- 1}}{\big(1+\frac{\epsilon^2}{4} |u_{n}^m|^2\big)\sqrt{2} \Gamma_n^m}
\begin{bmatrix}
\sqrt{2} \left(
e^{\sqrt{-1}\Lambda_{n-1}^{m+1}}C_{n}^m\, D_{n}^{m\,*}
- e^{-\sqrt{-1}\Lambda_{n-1}^{m+1}}C_{n}^{m\,*}\, D_{n}^m
\right)\\[2mm]
- e^{\sqrt{-1}\Lambda_{n-1}^{m+1}} (D_{n}^{m\,*}{})^2
- e^{-\sqrt{-1}\Lambda_{n-1}^{m+1}} (C_{n}^{m\,*}{})^2\\[2mm]
e^{\sqrt{-1}\Lambda_{n-1}^{m+1}} (C_{n}^m{})^2
 + e^{-\sqrt{-1}\Lambda_{n-1}^{m+1}} (D_{n}^m{})^2
\end{bmatrix},
\end{align}
where
\begin{equation}
C_n^m = \widehat{\alpha}_n^m - \frac{\epsilon}{2} u_{n}^{m+1\,*}\widehat{\beta}_n^m,\quad
D_n^m = \widehat{\beta}_n^m + \frac{\epsilon}{2} u_n^{m+1}\widehat{\alpha}_n^m.
\end{equation}
Then we obtain $\langle B_{n-1}^{m+1}, B_{n}^{m+1}\rangle$ by using
\eqref{eqn:products_complex_paralell_frame} and \eqref{eqn:alpha_beta_squaredsum} as
\begin{equation}
 \langle  B_{n-1}^{m+1},  B_{n}^{m+1}\rangle 
= \frac{e^{\sqrt{-1}(\Lambda_n^{m+1}-\Lambda_{n-1}^{m+1} )} + e^{-\sqrt{-1}(\Lambda_n^{m+1}-\Lambda_{n-1}^{m+1} )}}{2}
= \frac{e^{-\sqrt{-1}\nu_{n}^{m+1}} + e^{\sqrt{-1}\nu_n^{m+1}}}{2} =\cos\nu_{n}^{m+1},
\end{equation}
which proves the second equation of \eqref{eqn:goal}. The third equation can be proved in a similar manner starting from
\begin{equation}\label{eqn:N(n,m+1)}
 N_n^{m+1}
= B_n^{m+1} \times T_n^{m+1}
=
F_n^m
\frac{1}{\sqrt{2}\, \Gamma_{n + 1}^m}
\begin{bmatrix}
\sqrt{2} \left(e^{\sqrt{-1}\Lambda_n^{m+1}}\widehat\alpha_n^m\, \widehat\beta_n^m{}^\ast 
+ e^{-\sqrt{-1}\Lambda_n^{m+1}}\widehat\alpha_n^m{}^\ast\,\widehat\beta_n^m\right)\\[2mm]
e^{-\sqrt{-1}\Lambda_n^{m+1}}(\widehat\alpha_n^m{}^\ast)^2 
- e^{\sqrt{-1}\Lambda_n^{m+1}}(\widehat\beta_n^m{}^\ast)^2\\[2mm]
e^{\sqrt{-1}\Lambda_n^{m+1}}(\widehat\alpha_n^m)^2 - e^{-\sqrt{-1}\Lambda_n^{m+1}}(\widehat\beta_n^m)^2
\end{bmatrix}.
\end{equation}
This proves the second statement of Theorem \ref{thm:discrete_deformation}.

Now let us next prove Proposition \ref{prop:Lax_SO(3)}. 
From $T_n^{m+1}$ \eqref{tangent^{m+1}}, $B_n^{m+1}$ \eqref{eqn:B(n,m+1)}, $N_n^{m+1}$ \eqref{eqn:N(n,m+1)}, 
one can write $F_n^{m+1}=[T_n^{m+1}, U_n^{m+1}, U_n^{m+1\,*}]$ as
\begin{equation}
F_n^{m+1}
= F_n^{m} Y_n^{m},\quad
= \frac{1}{\Gamma_{n + 1}^m}
\begin{bmatrix}
\left|\widehat{\alpha}_n^{m}\right|^2 - \left|\widehat{\beta}_n^{m}\right|^2 &
\sqrt{2} \widehat{\alpha}_n^{m}{}^\ast\, \widehat{\beta}_n^{m} &
\sqrt{2} \widehat{\alpha}_n^{m}\, \widehat{\beta}_n^{m}{}^\ast\\[2mm]
- \sqrt{2} \widehat{\alpha}_n^{m}{}^\ast\, \widehat{\beta}_n^{m}{}^\ast &
(\widehat{\alpha}_n^{m}{}^\ast{})^2 & - (\widehat{\beta}_n^{m}{}^\ast{})^2\\[2mm]
- \sqrt{2} \widehat{\alpha}_n^{m}\, \widehat{\beta}_n^{m} & - (\widehat{\beta}_n^{m}{})^2 & (\widehat{\alpha}_n^{m}{})^2
\end{bmatrix}.
\end{equation}
Noticing from \eqref{eqn:hat_alpha-beta} that 
\begin{equation}
\widehat{\alpha}_n^m \widehat{\beta}^{m}_n{}^*
= e^{-\sqrt{-1}\Lambda_n^{m+1}}\alpha_n^m \beta_n^m{}^*,\quad 
(\widehat{\alpha}_n^m{})^2
= (\alpha_n^m)^2e^{-\sqrt{-1}(\Lambda_n^{m+1}-\Lambda_n^m)},\quad 
(\widehat{\beta}_n^m{})^2
= (\beta_n^m)^2e^{\sqrt{-1}(\Lambda_n^{m+1} + \Lambda_n^m)} ,
\end{equation}
and using \eqref{eqn:rel_Frenet_complex_parallel_discrete}, 
we obtain the deformation of the Frenet frame as
\begin{equation}
\Phi_n^{m+1}
= \Phi_n^{m} M_n^{m},
\end{equation}
\begin{align}
M_n^{m} &=
\frac{1}{\Gamma_{n + 1}^m}
\begin{bmatrix}
\left|\alpha_n^{m}\right|^2 - \left|\beta_n^{m}\right|^2 &
\alpha_n^{m} \beta_n^{m}{}^\ast + \alpha_n^{m}{}^\ast \beta_n^{m} &
\sqrt{- 1}
\left(\alpha_n^{m} \beta_n^{m}{}^\ast - \alpha_n^{m}{}^\ast \beta_n^{m}\right)\\[2mm]
- \alpha_n^{m} \beta_n^{m} - \alpha_n^{m}{}^\ast \beta_n^{m}{}^\ast &
\frac{(\alpha_n^{m})^2 - (\beta_n^{m})^2
+ (\alpha_n^{m}{}^\ast)^2 - (\beta_n^{m}{}^\ast)^2}{2} &
\frac{\sqrt{- 1}\left((\alpha_n^{m})^2 + (\beta_n^{m})^2
- (\alpha_n^{m}{}^\ast)^2 - (\beta_n^{m}{}^\ast)^2\right)}{2}\\[2mm]
\sqrt{- 1}
\left(\alpha_n^{m} \beta_n^{m} - \alpha_n^{m}{}^\ast \beta_n^{m}{}^\ast\right) &
- 
\frac{\sqrt{- 1}\left((\alpha_n^{m})^2 - (\beta_n^{m})^2
- (\alpha_n^{m}{}^\ast)^2 + (\beta_n^{m}{}^\ast)^2\right)}{2} &
\frac{(\alpha_n^{m})^2 + (\beta_n^{m})^2
+ (\alpha_n^{m}{}^\ast)^2 + (\beta_n^{m}{}^\ast)^2}{2}
\end{bmatrix}\nonumber\\
&=
\frac{1}{\Gamma_{n + 1}^m}
\begin{bmatrix}
\left|\alpha_n^{m}\right|^2 - \left|\beta_n^{m}\right|^2 & 2\Re (\alpha_n^{m} \beta_n^{m}{}^\ast)& -2\Im (\alpha_n^{m} \beta_n^{m\,*})\\[2mm]
-2\Re(\alpha_n^{m} \beta_n^{m}) &
\Re ((\alpha_n^{m})^2 - (\beta_n^{m})^2)
 &-\Im ((\alpha_n^{m})^2 + (\beta_n^{m})^2)\\[2mm]
-2\Im (\alpha_n^{m} \beta_n^{m}) &
\Im ((\alpha_n^{m})^2 - (\beta_n^{m})^2)) &
\Re ((\alpha_n^{m})^2 + (\beta_n^{m})^2)
\end{bmatrix},
\end{align}
which proves Proposition \ref{prop:Lax_SO(3)}. 

Finally, Proposition \ref{prop:distance} can be verified by direct computation by using
\eqref{eqn:norm_complex_paralell_frame}, \eqref{eqn:deformation_gamma_complex_parallel} and
\eqref{eqn:alpha_beta_squaredsum}. Therefore we have proved all the statements in Section
\ref{subsec:dLIE}.
%
\section{Explicit formulas}\label{sec:explicit_formulas}

It is well-known that the $N$-soliton solution of NLS \eqref{eqn:NLS} and sdNLS \eqref{eqn:sdNLS}
can be expressed in terms of double Wronskians or double Casorati determinants which are the
(reduced) $\tau$ functions of the two-component KP (or Toda) hierarchy
\cite{Ablowitz-Prinari-Trubach:NLS_book,Ohta:RIMS1989,Sadakane:sdNLS}. However, dNLS
\eqref{eqn:dNLS} is not studied well compared to NLS \eqref{eqn:NLS} and sdNLS
\eqref{eqn:sdNLS}. Hirota and Ohta presented \eqref{eqn:dNLS} in \cite{Hirota-Ohta:PSJ_91Spring} and
constructed soliton solutions by the perturbational technique. In \cite{Tsujimoto:App_book},
solutions in terms of double Casorati determinants have been given without complex structure.
In this section, we aim at constructing the regular soliton type solutions to those
equations in terms of the double Wronski/Casorati determinants.

In order to reconstruct the space curves for given complex curvature, we still need
nontrivial steps; we have to solve the system of linear partial differential/difference equations
satisfied by the Frenet frame to obtain the tangent vector, from which the position vector the curve
is constructed by integration. However, the Sym-Tafel formula \cite{Sym:Sym-Tafel} enables us to
reconstruct the position vector of the curve from the Frenet frame by using the differentiation with
respect to the spectral parameter instead of integration. This observation may suggest the
possibility of constructing explicit formula for the position vector of the curve in terms of the
$\tau$ functions without integration. Actually such explicit formula was formulated for the plane
curves \cite{IKMO:KJM}, although the relationship to the Sym-Tafel formula has not been established
yet.

Based on those backgrounds, in this section we first establish a parametrization
of both smooth and discrete space curves in terms of $\tau$ functions of two-component KP hierarchy
with suitable reduction.  Introducing deformation parameters appropriately, we construct explicit
formulas of smooth/discrete space curves deformed by NLS, sdNLS and dNLS, together with
the regular soliton type solutions to those equations.

\subsection{Explicit formulas for space curves}\label{subset:explicit_formulas_for_space_curves}
%
The following Proposition can be verified by direct calculation:
\begin{prop}\label{prop:Frame_tau}
For arbitrary parameters $f, g, v\in \mathbb{C}$ satisfying $ff^*+gg^*\neq 0$ and $|v|=\frac{1}{2}$, we
define $T,N,B\in\mathbb{R}^3$ as
\begin{align}
&T=\frac{1}{ff^* + gg^*}
\begin{bmatrix}
f^{*}g+fg^{*} \\[2mm]
(f^{*}g-fg^{*})/\sqrt{-1} \\[2mm]
ff^{*}-gg^{*}
\end{bmatrix} ,\label{eqn:T_tau}\\
&N=\frac{1}{ff^* + gg^*}
\begin{bmatrix}
\Big\{(f^{*})^2-(g^{*})^2\Big\}\,v
+\Big\{(f)^2-(g)^2\Big\}\,v^{*}
\\[1mm]
\left[\Big\{(f^{*})^2+(g^{*})^2\Big\}v
-\Big\{(f)^2+(g)^2\Big\}v^{*}\right]/\sqrt{-1}
\\[1mm]
-2(f^{*}g^{*}v+fgv^{*})
\end{bmatrix},\label{eqn:N_tau}\\
&B=\frac{1}{ff^* + gg^*}
\begin{bmatrix}
\left[-\Big\{(f^{*})^2-(g^{*})^2\Big\}\,v
+\Big\{(f)^2-(g)^2\Big\}\,v^{*}\right]/\sqrt{-1}\\[1mm]
\Big\{(f^{*})^2+(g^{*})^2\Big\}v
+\Big\{(f)^2+(g)^2\Big\}v^{*}\\[1mm]
2(f^{*}g^{*}v-fgv^{*})/\sqrt{-1}
\end{bmatrix}\label{eqn:B_tau}.
\end{align}   
Then $\Phi=[T,N,B]\in\mbox{{\rm SO}}(3)$.
\end{prop}
%
\noindent Then it is possible to regard $\Phi$ as the Frenet frame of space curves by
introducing appropriate dependence on arc-length parameter of the curves.
\begin{prop}\label{prop:Frame_continuous_tau}
Let $f, g,G\in\mathbb{C}$, $F\in\mathbb{R}_{>0}$ 
be the $\tau$ functions satisfying the following bilinear equations:
\begin{align}
& F^2 = ff^* + gg^*,\label{eqn:bl1}\\
& D_x\, f\cdot F = -g^* G,\label{eqn:bl2}\\
& D_x\, g\cdot F = f^* G.\label{eqn:bl3}
\end{align}
Under the identification
\begin{equation}\label{eqn:v_continuous}
v=\frac{1}{2}\,\frac{G}{|G|},
\end{equation}
in equations \eqref{eqn:T_tau}--\eqref{eqn:B_tau}, it follows that 
\begin{equation}\label{eqn:Frame_continuous_tau_2}
N = \frac{T'}{|T'|},\quad B = T\times N,
\end{equation}
so that $\Phi$ satisfies the Frenet-Serret formula 
\eqref{eqn:Frenet-Serret}.
\end{prop}
%
\begin{rem}
Due to \eqref{eqn:bl1}--\eqref{eqn:v_continuous}, the binormal vector $B$ admits an alternative
expression as
\begin{equation}\label{eqn:B_alternative}
B=\frac{1}{2|G|F}
\begin{bmatrix}\medskip
(D_x\,(g^*\cdot f - g\cdot f^*))/\sqrt{-1}\\
D_x\,(g^*\cdot f + g\cdot f^*) \\[1mm]
(D_x\, (f^*\cdot f - g^*\cdot g))/\sqrt{-1}
\end{bmatrix},
\end{equation}
which will be used later.
\end{rem}
%
Then we have an explicit formula in terms of the $\tau$ functions for the position vector of the
curve $\gamma$ as follows:
\begin{prop}\label{prop:gamma_continuous_tau}
Let $H\in\mathbb{C}$ be the $\tau$ function satisfying the following bilinear equation
\begin{equation}\label{eqn:bl5}
 D_x\, H\cdot F = f^{*}g.
\end{equation}
We also introduce an auxiliary independent variable $z$ by
\begin{equation}\label{eqn:bl6}
\frac{1}{2}D_zD_x\, F \cdot F = gg^{*}.
\end{equation}
Defining $\gamma$ as
\begin{equation}\label{eqn:gamma_by_tau}
\gamma = 
\begin{bmatrix}\medskip
\dfrac{H +H^{*}}{F} \\
\medskip
\dfrac{1}{\sqrt{-1}}\,\dfrac{H-H^{*}}{F} \\[1mm]
x - 2\dfrac{\partial \log F}{\partial z}
\end{bmatrix},
\end{equation}
it follows that $\gamma'=T$.
\end{prop}
%
The curvature $\kappa$, the torsion $\lambda$, and the complex curvature 
$u=\kappa e^{\sqrt{-1}\int\lambda~dx}$ also admit the explicit formulas in terms of the $\tau$ functions.
\begin{prop}\label{prop:kappa_formula_continuous}
It follows that
\begin{equation}\label{eqn:kappa_formula_continuous}
 \kappa =2\frac{|G|}{F} ,\quad 
\lambda = \frac{1}{2\sqrt{-1}}\frac{\partial }{\partial x} \left(\log\frac{G}{G^*}\right),\quad 
u=2\frac{G}{F} .
\end{equation} 
\end{prop}
%
It is possible to construct the similar formulas for discrete space curves as follows.
\begin{prop}\label{prop:Frame_discrete_tau}
 Let $f_n, g_n,G_n\in\mathbb{C}$, $F_n\in\mathbb{R}_{>0}$ 
be the $\tau$ functions satisfying the following bilinear equations:
\begin{align}
&F_{n+1}F_n=f_nf_n^{*}+g_ng_n^{*},\label{eqn:dbl1}\\
&f_nF_n-f_{n-1}F_{n+1}=-g_n^{*}G_n,\label{eqn:dbl2}\\
&g_nF_n-g_{n-1}F_{n+1}=f_n^{*}G_n.\label{eqn:dbl3}
\end{align}
Under the identification
\begin{equation}
 f=f_n,\quad g=g_n,\quad v=\frac{1}{2}\,\frac{G_n}{|G_n|},\quad T=T_n,\quad N=N_n,\quad B=B_n,
\end{equation}
in equations \eqref{eqn:T_tau}--\eqref{eqn:B_tau}, it follows that 
\begin{equation}\label{eqn:discrete_B_and_N}
B_n = \frac{T_{n-1}\times T_n}{|T_{n-1}\times T_n|},\quad
N_n = B_n\times T_n,
\end{equation}
so that $\Phi_n$ satisfies the discrete Frenet-Serret formula 
\eqref{eqn:discrete_Frenet-Serret}--\eqref{eqn:kappa_nu}.
\end{prop}
%
\begin{rem}\label{rem:discrete_altenative_B}
By using \eqref{eqn:dbl2} and \eqref{eqn:dbl3} and their complex conjugate,
the binormal vector $B_n$ admits an alternative expression as
\begin{equation}\label{eqn:discrete_alternative_B}
B_n=\frac{1}{2|G_n|F_n}
\begin{bmatrix}
(g_n^{*}f_{n-1}-f_ng_{n-1}^{*}
-g_nf_{n-1}^{*}+f_n^{*}g_{n-1})/\sqrt{-1}
\\[1mm]
g_n^{*}f_{n-1}-f_ng_{n-1}^{*}
+g_nf_{n-1}^{*}-f_n^{*}g_{n-1}
\\[1mm]
(f_n^{*}f_{n-1} - f_nf_{n-1}^{*} - g_n^{*}g_{n-1} + g_ng_{n-1}^{*})/\sqrt{-1}
\end{bmatrix}.
\end{equation}
\end{rem}
%
Then we have an explicit formula in terms of the $\tau$ functions for the position vector of the
discrete curve $\gamma_n$ as follows:
\begin{prop}\label{prop:gamma_discrete_tau}
Let $H_n\in\mathbb{C}$ be the $\tau$ function satisfying the following bilinear equation
\begin{equation}\label{eqn:dbl7}
H_{n+1}F_n-H_nF_{n+1}=f_n^{*}g_n.
\end{equation}
We also introduce an auxiliary independent variable $z\in\mathbb{R}$ by
\begin{equation}\label{eqn:dbl8}
D_z\,F_{n+1}\cdot F_n = g_n g_n^{*}.
\end{equation}
Defining $\gamma_n$ as
\begin{equation}
\gamma_n=\epsilon
\begin{bmatrix}\medskip
\dfrac{H_n+H_n^{*}}{F_n}
\\
\medskip
\dfrac{1}{\sqrt{-1}}\,\dfrac{H_n-H_n^{*}}{F_n}
\\
n + \rho - 2\dfrac{\partial \log F_n}{\partial z}
\end{bmatrix},
\end{equation}
where $\rho$ is an arbitrary constant in $n$, it follows that
$(\gamma_{n+1}-\gamma_n)/\epsilon=T_n$.
\end{prop}
%
\begin{prop}\label{prop:kappa_formula_discrete}
It follows that
\begin{equation}\label{eqn:kappa_formula_discrete}
\kappa_n = 2\arctan\frac{|G_n|}{F_n} ,\quad 
\nu_n = \frac{1}{2\sqrt{-1}} \log\frac{G_n^*G_{n-1}}{G_nG_{n-1}^*},\quad 
u_n=\frac{2}{\epsilon}\frac{G_n}{F_n} .
\end{equation} 
\end{prop}
%
\subsection{Proof of explicit formulas}\label{proof_of_explicit_formulas}
\paragraph{\it Proof of Proposition \ref{prop:Frame_continuous_tau}} \hfill\\
We first replace the denominators of \eqref{eqn:T_tau}--\eqref{eqn:B_tau} by $F^2$
by using \eqref{eqn:bl1} and compute $T'$ by using the bilinear equations
\eqref{eqn:bl2}, \eqref{eqn:bl3} and their complex conjugate. For instance, the first entry
is computed as:
\begin{align*}
& \left(\frac{f^{*}g+fg^{*}}{F^2}\right)' 
= \left(\frac{D_x\, f^*\cdot F}{F^2}\right)\frac{g}{F} + \frac{f^*}{F}\left(\frac{D_x\, g\cdot F}{F^2} \right)
+ \left(\frac{D_x\, f\cdot F}{F^2}\right)\frac{g^*}{F} + \frac{f}{F}\left(\frac{D_x\, g^*\cdot F}{F^2}\right)\\
=& -\left(\frac{gG^*}{F^2}\right)\frac{g}{F} + \frac{f^*}{F}\left(\frac{f^*G}{F^2} \right)
- \left(\frac{g^*G}{F^2}\right)\frac{g^*}{F} + \frac{f}{F}\left(\frac{fG^*}{F^2}\right).
\end{align*}
Similarly, the second and third entries can be computed and we obtain $|T'|$ as
\begin{equation}\label{eqn:T'_continous}
 |T'| = 2\frac{|G|}{F},
\end{equation}
which proves the first equation of \eqref{eqn:Frame_continuous_tau_2}. The second equation of 
\eqref{eqn:Frame_continuous_tau_2} can be justified by the fact that $\Phi\in{\rm SO}(3)$. 
\qed
%
\paragraph{\it Proof of Proposition \ref{prop:gamma_continuous_tau}} \hfill\\
The derivative of the first and second entries of $\gamma$ can be computed by using \eqref{eqn:bl5}
and its complex conjugate.  The third entry is computed by using \eqref{eqn:bl6}. Equation
\eqref{eqn:bl1} is used to rewrite the denominators.\qed
%
\paragraph{\it Proof of Proposition \ref{prop:kappa_formula_continuous}} \hfill\\
The first equation in \eqref{eqn:kappa_formula_continuous} follows from \eqref{eqn:T'_continous}
since $\kappa = |T'|$. Adding \eqref{eqn:bl2} multiplied by $f^*$, \eqref{eqn:bl3} multiplied by
$g^*$ and \eqref{eqn:bl1} differentiated by $x$ and multiplied by $F/2$, we get
\begin{equation}\label{eqn:bl7}
 D_x\,(f^*\cdot f + g^*\cdot g)=0.
\end{equation}
Denoting $B={}^t[B_1,B_2,B_3]$ and $N={}^t[N_1,N_2,N_3]$ we have
\begin{align*}
& \lambda = -\langle B',N\rangle = -\frac{1}{2}\sum_{i=1}^3 D_x\,B_i\cdot N_i\\
& = \frac{1}{\sqrt{-1}F^4} D_x\,\Big(\Big\{(f^{*})^2-(g^{*})^2\Big\}v\cdot
\Big\{(f)^2-(g)^2\Big\}v^{*}
+ \Big\{(f^{*})^2+(g^{*})^2\Big\}v\cdot
\Big\{(f)^2+(g)^2\Big\}v^{*}
+4f^{*}g^{*}v\cdot fgv^*\Big)\\
&= 
\frac{vv^*}{\sqrt{-1}F^4} D_x\,\Big(\Big\{(f^{*})^2-(g^{*})^2\Big\}\cdot
\Big\{(f)^2-(g)^2\Big\}
+ \Big\{(f^{*})^2+(g^{*})^2\Big\}\cdot
\Big\{(f)^2+(g)^2\Big\}
+4f^{*}g^{*}\cdot fg\Big)\\
&+\frac{1}{\sqrt{-1}F^4}\Big(\Big\{(f^{*})^2-(g^{*})^2\Big\}\Big\{(f)^2-(g)^2\Big\}
+ \Big\{(f^{*})^2+(g^{*})^2\Big\}\Big\{(f)^2+(g)^2\Big\}+4f^{*}g^{*}fg\Big)\, (D_x\,v\cdot\,v^{*}),
\end{align*}
where the first term of the most right hand side vanishes due to \eqref{eqn:bl7}. Thus we have
\begin{displaymath}
 \lambda = \frac{2(ff^* + gg^*)^2}{\sqrt{-1}F^4}\, (D_x\,v\cdot\,v^{*})
= \frac{2vv^*}{\sqrt{-1}}\, \left(\log\frac{v}{v^*}\right)',
\end{displaymath}
where we used \eqref{eqn:bl1}. By using \eqref{eqn:v_continuous} we obtain
the second equation of \eqref{eqn:kappa_formula_continuous}
\begin{equation}
 \lambda = \frac{1}{2\sqrt{-1}}\left(\log\frac{G}{G^*}\right)'.
\end{equation}
The third equation
follows automatically from the first and second equations. \qed
%
\paragraph{\it Proof of Proposition \ref{prop:Frame_discrete_tau}} \hfill\\
Adding \eqref{eqn:dbl2} multiplied by $f_n^*$ and \eqref{eqn:dbl3} multiplied by $g_n^*$, 
we get
\begin{equation}
 F_nF_n=f_n^{*}f_{n-1}+g_n^{*}g_{n-1}. \label{eqn:dbl4}
\end{equation}
Similarly, subtracting \eqref{eqn:dbl2} multiplied by $g_{n-1}$ from 
\eqref{eqn:dbl3} multiplied by $f_{n-1}$, we get
\begin{equation}
 g_nf_{n-1} - f_ng_{n-1} = G_nF_n,\label{eqn:dbl5}
\end{equation}
where we used \eqref{eqn:dbl4}.

We first replace the denominators of \eqref{eqn:T_tau}--\eqref{eqn:B_tau} by $F_{n+1}F_n$
by using \eqref{eqn:dbl1} and compute $T_{n-1}\times T_n$. For example, the first entry turns out to be
\begin{align*}
& \frac{1}{\sqrt{-1}F_{n+1}F_n^2F_{n-1}}
\left[
\left(
  (g_n^*f_{n-1} + f_n^*g_{n-1})(f_nf_{n-1}^* + g_ng_{n-1}^*) 
- (g_nf_{n-1}^* + f_ng_{n-1}^*)(f_n^*f_{n-1} + g_n^*g_{n-1})
\right)
\right]\\
=&  \frac{1}{\sqrt{-1}F_{n+1}F_{n-1}}
  (g_n^*f_{n-1} + f_n^*g_{n-1}
- g_nf_{n-1}^* - f_ng_{n-1}^*),
\end{align*}
where we used \eqref{eqn:dbl4} and its complex conjugate. Similarly, the second and third entries
are computed, and we obtain
\begin{equation}
 |T_{n-1}\times T_{n}|^2 = \frac{4}{(F_{n+1}F_{n-1})^2}\left(g_nf_{n-1} - f_ng_{n-1}\right)
\left(g_n^*f_{n-1}^* - f_n^*g_{n-1}^*\right)
 = \frac{4G_nG_n^*F_n^2}{(F_{n+1}F_{n-1})^2},
\end{equation}
where we have used \eqref{eqn:dbl5}. This proves that $(T_{n-1}\times T_{n})/|T_{n-1}\times T_{n}|$
coincides with the expression of $B_n$ \eqref{eqn:discrete_alternative_B} in Remark
\ref{rem:discrete_altenative_B}, thus the first equation of \eqref{eqn:discrete_B_and_N} is
justified. The second equation of \eqref{eqn:discrete_B_and_N} is just a consequence of
$\Phi_n=[T_n,N_n,B_n]\in{\rm SO(3)}$.
\qed
%
\paragraph{\it Proof of Proposition \ref{prop:gamma_discrete_tau}} \hfill\\
The first and second entries of $\gamma_{n+1}-\gamma_n$ can be computed by using \eqref{eqn:dbl7}
and its complex conjugate.  The third entry is computed by using \eqref{eqn:dbl8}. Equation
\eqref{eqn:dbl1} is used to rewrite the denominators.\qed
%
\paragraph{\it Proof of Proposition \ref{prop:kappa_formula_discrete}} \hfill\\
From \eqref{eqn:dbl1}, \eqref{eqn:dbl4} and \eqref{eqn:dbl5}, we have
\begin{equation}
 F_{n+1}F_n^2F_{n-1} - F_n^4 = (g_nf_{n-1} - f_ng_{n-1})(g_n^*f_{n-1}^* - f_n^*g_{n-1}^*)
= G_nG_n^*F_n^2.
\end{equation}
Thus we get
\begin{equation}
F_{n+1}F_{n-1}-F_nF_n=G_nG_n^{*}.\label{eqn:dbl6}
\end{equation}

We prove the first equation of \eqref{eqn:kappa_formula_discrete} by showing \eqref{eqn:kappa_nu}.
We have 
\begin{displaymath}
\begin{split}
 \langle T_{n-1},T_n\rangle =& \frac{1}{F_{n+1}F_n^2F_{n-1}}\\[1mm]
&\times \left\{
(f_nf_n^*+g_ng_n^*)(f_{n-1}f_{n-1}^*+g_{n-1}g_{n-1}^*)
-2(g_nf_{n-1} - f_ng_{n-1})
(g_n^*f_{n-1}^* - f_n^*g_{n-1}^*)
\right\}, 
\end{split}
\end{displaymath}
which is rewritten by using \eqref{eqn:dbl1}, \eqref{eqn:dbl5} and \eqref{eqn:dbl6} as
\begin{equation}
  \langle T_{n-1},T_n\rangle = 1-2 \frac{G_nG_n^*}{F_{n+1}F_{n-1}}
=\frac{1- \frac{G_nG_n^*}{(F_n)^2}}{1+ \frac{G_nG_n^*}{(F_n)^2}} = \cos\kappa_n,
\end{equation}
so that
\begin{equation}
 \frac{|G_n|^2}{(F_n)^2} = \tan^2\frac{\kappa_n}{2},
\end{equation}
which is the first equation of \eqref{eqn:kappa_formula_discrete}. Similarly, we get
\begin{align*}
 \langle B_n,B_{n-1}\rangle = \frac{G_n^*G_{n-1} + G_nG_{n-1}^*}{2|G_nG_{n-1}|},\quad
 \langle B_n,N_{n-1}\rangle = \frac{G_n^*G_{n-1} - G_nG_{n-1}^*}{2\sqrt{-1}|G_nG_{n-1}|},
\end{align*}
which proves the second equation of \eqref{eqn:kappa_formula_discrete}. The third equation
automatically follows from \eqref{eqn:u_discrete}.\qed
%
\subsection{Parametrization by $\tau$ functions of two-component KP hierarchy}\label{subset:parametrization}
The bilinear equations \eqref{eqn:bl1} -- \eqref{eqn:bl3}, \eqref{eqn:bl5} and \eqref{eqn:bl6} are
simultaneously solved by the $\tau$ functions of two-component KP hierarchy. Also, equations
\eqref{eqn:dbl1}--\eqref{eqn:dbl3}, \eqref{eqn:dbl7} and \eqref{eqn:dbl8} are solved by those of
discrete two-component KP hierarchy.
%
\begin{prop}\label{prop:continuous_tau_formula}
We define a $2N\times 2N$ determinant $\tau(\nu,r)$ by
\begin{equation}\label{eqn:continuous_tau}
 \tau(\nu,r)=\left|\begin{matrix}
\phi_1^{(0)}(r) &\phi_1^{(1)}(r)
 &\cdots &\phi_1^{(N+\nu-1)}(r) &\vbl{5}
 &-1 &-(-p_1) &\cdots &-(-p_1)^{N-\nu-1}
\\
\phi_2^{(0)}(r) &\phi_2^{(1)}(r)
 &\cdots &\phi_2^{(N+\nu-1)}(r) &\vbl{5}
 &-1 &-(-p_2) &\cdots &-(-p_2)^{N-\nu-1}
\\
\vdots &\vdots &&\vdots &\vbl{6} &\vdots &\vdots &&\vdots
\\
\phi_N^{(0)}(r) &\phi_N^{(1)}(r)
 &\cdots &\phi_N^{(N+\nu-1)}(r) &\vbl{5}
 &-1 &-(-p_N) &\cdots &-(-p_N)^{N-\nu-1}
\\[2mm]
\multispan{9}\hblfil\
\\
1 &(-p_1^*) &\cdots &(-p_1^*)^{N+\nu-1} &\vbl{5}
 &\bar\phi_1^{(0)}(r) &\bar\phi_1^{(1)}(r)
 &\cdots &\bar\phi_1^{(N-\nu-1)}(r)
\\
1 &(-p_2^*) &\cdots &(-p_2^*)^{N+\nu-1} &\vbl{5}
 &\bar\phi_2^{(0)}(r) &\bar\phi_2^{(1)}(r)
 &\cdots &\bar\phi_2^{(N-\nu-1)}(r)
\\
\vdots &\vdots &&\vdots &\vbl{6} &\vdots &\vdots &&\vdots
\\
1 &(-p_N^*) &\cdots &(-p_N^*)^{N+\nu-1} &\vbl{5}
 &\bar\phi_N^{(0)}(r) &\bar\phi_N^{(1)}(r)
 &\cdots &\bar\phi_N^{(N-\nu-1)}(r)
\end{matrix}\right|,
\end{equation}
where
\begin{equation}\label{eqn:continuous_tau_entries}
\phi_i^{(n)}(r) = p_i^n(-p_i)^{-r}e^{\zeta_i},\quad 
\bar\phi_i^{(n)}(r) = (p_i^*)^{n+r} e^{\zeta_i^*},\quad
\zeta_i= \frac{1}{p_i}z + p_i x +\zeta_{i0},
\end{equation}
and $p_i$, $\zeta_{i0}$ are constants. Then the following $\tau$
functions satisfy the bilinear equations \eqref{eqn:bl1} --
\eqref{eqn:bl3}, \eqref{eqn:bl5} and \eqref{eqn:bl6}:
\begin{equation}\label{eqn:continuous_tau_identification}
\begin{array}{lll}
G = \tau(1,0)& g = -\tau(1,1) & H = \tau(1,2) \\[2mm]
f = \tau(0,-1) &F = \tau(0,0)  &f^* = \tau(0,1)   \\[2mm]
H^* = \tau(-1,-2) & g^* = \tau(-1,-1) & G^* = \tau(-1,0).
\end{array}
\end{equation}
\end{prop}
%
\begin{prop}\label{prop:discrete_tau_formula}
 We define a $2N\times 2N$ determinant $\tau_n(\nu,r)$ by
\begin{equation}\label{eqn:discrete_tau}
 \tau_n(\nu,r)=\left|\begin{matrix}
\phi_1^{(n)}(r) &\phi_1^{(n+1)}(r)
 &\cdots &\phi_1^{(n+N+\nu-1)}(r) &\vbl{5}
 &-1 &-\frac{1}{p_1} &\cdots &-(\frac{1}{p_1})^{N-\nu-1}
\\
\phi_2^{(n)}(r) &\phi_2^{(n+1)}(r)
 &\cdots &\phi_2^{(n+N+\nu-1)}(r) &\vbl{5}
 &-1 &-\frac{1}{p_2} &\cdots &-(\frac{1}{p_2})^{N-\nu-1}
\\
\vdots &\vdots &&\vdots &\vbl{6} &\vdots &\vdots &&\vdots
\\
\phi_N^{(n)}(r) &\phi_N^{(n+1)}(r)
 &\cdots &\phi_N^{(n+N+\nu-1)}(r) &\vbl{5}
 &-1 &-\frac{1}{p_N} &\cdots &-(\frac{1}{p_N})^{N-\nu-1}
\\[2mm]
\multispan{9}\hblfil\
\\
1 &\frac{1}{p_1^*} &\cdots &(\frac{1}{p_1^*})^{N+\nu-1} &\vbl{5}
 &\bar\phi_1^{(n)}(r) &\bar\phi_1^{(n+1)}(r)
 &\cdots &\bar\phi_1^{(n+N-\nu-1)}(r)
\\
1 &\frac{1}{p_2^*} &\cdots &(\frac{1}{p_2^*})^{N+\nu-1} &\vbl{5}
 &\bar\phi_2^{(n)}(r) &\bar\phi_2^{(n+1)}(r)
 &\cdots &\bar\phi_2^{(n+N-\nu-1)}(r)
\\
\vdots &\vdots &&\vdots &\vbl{6} &\vdots &\vdots &&\vdots
\\
1 &\frac{1}{p_N^*} &\cdots &(\frac{1}{p_N^*})^{N+\nu-1} &\vbl{5}
 &\bar\phi_N^{(n)}(r) &\bar\phi_N^{(n+1)}(r)
 &\cdots &\bar\phi_N^{(n+N-\nu-1)}(r)
\end{matrix}\right|,
\end{equation}
where
\begin{equation}\label{eqn:discrete_tau_entries}
\phi_i^{(n)}(r)=p_i^n\,(1-p_i)^{-r}\,e^{\zeta_i},\quad  
\bar\phi_i^{(n)}(r)=(p_i^*)^n\,\bigg(1-\frac{1}{p_i^*}\bigg)^r\,e^{\zeta_i^*},\quad
\zeta_i=\frac{p_i+1}{p_i-1}\frac{z}{2}+\zeta_{i0},
\end{equation}
and $p_i$, $\zeta_{i0}$ are constants. Then the following $\tau$
functions satisfy the discrete bilinear equations
\eqref{eqn:dbl1}--\eqref{eqn:dbl3}, \eqref{eqn:dbl7} and
\eqref{eqn:dbl8}:
\begin{equation}\label{eqn:discrete_tau_identification}
\begin{array}{lll}
G_n = \tau_n(1,0)& g_n = -\tau_{n+1}(1,1) & H_n = \tau_{n+1}(1,2) \\[2mm]
f_n = \tau_n(0,-1) &F_n = \tau_{n}(0,0)  &f_n^* = \tau_{n+1}(0,1) \\[2mm]
 H^*_n = \tau_{n-1}(-1,-2) & g^*_n = \tau_n(-1,-1) &G_n^* = \tau_n(-1,0).
\end{array}
\end{equation}
\end{prop}
The proofs of Propositions \ref{prop:continuous_tau_formula} and
\ref{prop:discrete_tau_formula} will be given in Appendix.
%
\subsection{Deformations of curves}\label{subsec:formula_deformation}
It is possible to describe the deformations of curves by introducing appropriate time evolutions
in the $\tau$ functions through the explicit formulas.
%
\begin{prop}\label{prop:time_evolution_continuous}
In \eqref{eqn:continuous_tau} and \eqref{eqn:continuous_tau_entries}, we introduce the dependence 
on time $t$ by
\begin{equation}\label{eqn:phi_continuous}
\zeta_i= \frac{1}{p_i}z+ p_i x +\sqrt{-1}p_i^2 t+ \zeta_{i0}.
\end{equation}
We define the $\tau$ functions as \eqref{eqn:continuous_tau_identification}.
Then 
\begin{equation}
 u=2\frac{G}{F},
\end{equation}
satisfy NLS \eqref{eqn:NLS}. Moreover, the Frenet frame $\Phi$ in
Proposition \ref{prop:Frame_tau} and $\gamma$ in Proposition \ref{prop:gamma_continuous_tau}
satisfy the deformation equation \eqref{eqn:continuous_binormal_flow}.
\end{prop}
%
\begin{prop}\label{prop:time_evolution_semi-discrete}
In \eqref{eqn:discrete_tau} and \eqref{eqn:discrete_tau_entries}, we introduce the time
 $t$ dependence by
\begin{equation}\label{eqn:phi_semi-discrete}
\begin{split}
&\zeta_i=\frac{p_i+1}{p_i-1}\frac{z}{2}+ \frac{\sqrt{-1}}{\epsilon^2}\Big(p_i + \frac{1}{p_i}\Big) t + \zeta_{i0}.
\end{split}
\end{equation}
We define the $\tau$ functions as
\begin{equation}\label{eqn:sdNLS_tau_identification}
\begin{array}{lll}
G_n = A\tau_n(1,0)& g_n = -A\tau_{n+1}(1,1) &H_n = A\tau_{n+1}(1,2) \\[2mm]
f_n = \tau_n(0,-1)&F_n = \tau_{n}(0,0)  &f_n^* = \tau_{n+1}(0,1)  \\[2mm]
H_n^* =  A^{-1}\tau_{n-1}(-1,-2) & g_n^* =  A^{-1}\tau_n(-1,-1) &G_n^* = A^{-1}\tau_n(-1,0),
\end{array}
\end{equation}
where $A=e^{-\frac{2\sqrt{-1}}{\epsilon^2}t}$ . Then
\begin{equation}
 u_n=\frac{2}{\epsilon}\frac{G_n}{F_n},
\end{equation}
satisfy sdNLS \eqref{eqn:sdNLS}. Moreover, the Frenet frame $\Phi_n$ in
Proposition \ref{prop:Frame_tau} and $\gamma_n$ in Proposition \ref{prop:gamma_discrete_tau} with
$\rho=0$ satisfy the deformation equation \eqref{eqn:discrete_binormal_flow}.
\end{prop}
%
\begin{prop}\label{prop:time_evolution_discrete}
In \eqref{eqn:discrete_tau} and \eqref{eqn:discrete_tau_entries}, we introduce the discrete time
 $m$ dependence by replacing $\zeta_i$ with
\begin{equation}\label{eqn:phi_discrete}
\begin{split}
&\zeta_i=\frac{p_i+1}{p_i-1}\frac{z}{2} + m \log \left(\frac{1-ap_i}{1-\frac{a^*}{p_i}}\right)+\zeta_{i0},
\end{split}
\end{equation}
where
\begin{equation}\label{eqn:def_a}
 a = \frac{\Gamma_\infty}{\sqrt{-1}\frac{\epsilon^2}{\delta}+1}.
\end{equation}
We suppose that all the variables depend also on $m$, and write as $\tau_n^m(\nu,r)$.
We define the $\tau$ functions as
\begin{equation}\label{eqn:dNLS_tau_identification}
\begin{array}{lll}
 G_n^m = A^m\tau_n^m(1,0)& g_n^m = -A^m B^m \tau_{n+1}^m(1,1) &H_n^m = A^m B^{2m}\tau_{n+1}^m(1,2) \\[2mm]
f_n^m = B^{-m}\tau_n^m(0,-1) &F_n^m = \tau_{n}^m(0,0)  &f_n^{m\,*} = B^m \tau_{n+1}^m(0,1) \\[2mm]
H^{m\,*}_n =  A^{-m} B^{-2m}\tau_{n-1}^m(-1,-2) & g^{m\,*}_n =  A^{-m} B^{-m}\tau_n^m(-1,-1) & G_n^{m\,*} = A^{-m}\tau_n^m(-1,0),
\end{array}
\end{equation}
where
\begin{equation}
 A= \frac{\sqrt{-1}\frac{\epsilon^2}{\delta} + 1}{\sqrt{-1}\frac{\epsilon^2}{\delta}-1},\quad
B= \left(\frac{\sqrt{-1}\frac{\epsilon^2}{\delta} - 1 + \Gamma_{\infty}}{\sqrt{-1}\frac{\epsilon^2}{\delta}+1-\Gamma_{\infty}}\right)^{1/2},
\end{equation}
and $\Gamma_\infty$ is either $1$ or $1+\epsilon^4/\delta^2$. Then 
\begin{equation}\label{eqn:variable_transformation_dNLS}
 u_n^m=\frac{2}{\epsilon}
\frac{G_n^m}{F_n^m},\quad
\Gamma_n^m=\Gamma_{\infty}
\frac{F_n^mF_{n-1}^{m+1}}{F_{n-1}^mF_n^{m+1}}
\end{equation}
satisfy dNLS \eqref{eqn:dNLS}. Moreover, the Frenet frame $\Phi_n^m$ in
Proposition \ref{prop:Frame_tau} and $\gamma_n^m$ in Proposition \ref{prop:gamma_discrete_tau}
with $\rho=\rho_m 
= \frac{2\delta^2}{\epsilon^4}(\Gamma_\infty - 1)m$ satisfy the deformation equation
\eqref{eqn:discrete_def_Frenet} and \eqref{eqn:discrete_coeffs_Frenet}.
\end{prop}

It is noted that in the above Proposition, only the solutions for
$\Gamma_{\infty}^m=\Gamma_{-\infty}^m$ are given and those with
the different constant boundary condition, for instance,
$\Gamma_{\infty}^m=1$ and $\Gamma_{-\infty}^m=1+\frac{\epsilon^4}{\delta^2}$,
are not yet known.
Proofs of Proposition \ref{prop:time_evolution_continuous}, \ref{prop:time_evolution_semi-discrete} and
\ref{prop:time_evolution_discrete} will be given in Appendix.
%
\subsection{Regularity of solutions}\label{subset:regularity}
The formulas given in Section \ref{subsec:formula_deformation} give rise to regular solutions of
NLS \eqref{eqn:NLS}, sdNLS \eqref{eqn:sdNLS} and dNLS \eqref{eqn:dNLS} so that the corresponding
curve deformations are also regular. In this section, we establish the regularity of solutions
which is guaranteed by the strict positivity of the $\tau$ functions $F$, $F_n$ and $F_n^m$
in Propositions \ref{prop:time_evolution_continuous}, \ref{prop:time_evolution_semi-discrete} and
\ref{prop:time_evolution_discrete}, respectively.
%
\begin{lemma}\label{lem:detM}
 Define a $2N\times 2N$ matrix $M$ as
\begin{equation}
 M=\begin{bmatrix}
a_1 &p_1a_1 &\cdots &p_1^{N-1}a_1 &\vbl{5}
 &-1 &-q_1 &\cdots &-q_1^{N-1}
\\
a_2 &p_2a_2 &\cdots &p_2^{N-1}a_2 &\vbl{5}
 &-1 &-q_2 &\cdots &-q_2^{N-1}
\\
\vdots &\vdots &&\vdots &\vbl{6} &\vdots &\vdots &&\vdots
\\
a_N &p_Na_N &\cdots &p_N^{N-1}a_N &\vbl{5}
 &-1 &-q_N &\cdots &-q_N^{N-1}
\\[2mm]
\multispan{9}\hblfil\
\\
1 &q_1^* &\cdots &(q_1^*)^{N-1} &\vbl{5}
 &a_1^* &p_1^*a_1^* &\cdots &(p_1^*)^{N-1}a_1^*
\\
1 &q_2^* &\cdots &(q_2^*)^{N-1} &\vbl{5}
 &a_2^* &p_2^*a_2^* &\cdots &(p_2^*)^{N-1}a_2^*
\\
\vdots &\vdots &&\vdots &\vbl{6} &\vdots &\vdots &&\vdots
\\
1 &q_N^* &\cdots &(q_N^*)^{N-1} &\vbl{5}
 &a_N^* &p_N^*a_N^* &\cdots &(p_N^*)^{N-1}a_N^*
\end{bmatrix},
\end{equation}
where $p_i$, $q_i$, $a_i$ ($i=1,\ldots,N$) are arbitrary complex parameters. Then we have
\begin{equation}\label{eqn:detM}
\det M =\left|\begin{matrix}
\displaystyle\frac{\mu_i}{p_i-q_j^*} &\vbl{14} &-\delta_{ij}
\\
\multispan{3}\hblfil\
\\
\delta_{ij} &\vbl{14} &\displaystyle\frac{\mu_i^*}{p_i^*-q_j}
\end{matrix}\right|
\prod_{1\le i<j\le N}(q_i-q_j)(q_i^*-q_j^*),
\end{equation}
where
\begin{equation}
 \mu_i = a_i\frac{\displaystyle\prod_{k=1}^N(p_i-q_k^*)}
{\displaystyle\prod_{\genfrac{}{}{0pt}{2}{k=1}{k\ne i}}^N(q_i-q_k)}.
\end{equation}
\end{lemma}
{\em Proof.} \quad We introduce $2N\times 2N$ matrices $P_k$ and $Q_k$ as
$$
P_k=\begin{bmatrix}
J_k^* &O
\\
O &J_k
\end{bmatrix},
\qquad
Q_k=\begin{bmatrix}
K_k^* &O
\\
O &K_k
\end{bmatrix},
$$
for $1\le k\le N-1$, where $O$ is zero matrix of size $N$, $J_k$ and $K_k$ are $N\times N$ matrices defined by
$$
J_k=\begin{bmatrix}
\quad I_{k-1}\quad &\vbl{10}
\\
\multispan{3}\hblfil\
\\
&\vbl{21} &\begin{matrix}
1 &-q_k
\\
&1 &-q_k
\\
&&1 &\ddots
\\
&&&\ddots &-q_k
\\
&&&&1
\end{matrix}
\end{bmatrix},
\quad
K_k=\begin{bmatrix}
{\scriptstyle q_1-q_{k+1}} &&&&\vbl{4}
\\
&{\scriptstyle q_2-q_{k+1}} &&&\vbl{4}
\\
&&\ddots &&\vbl{5}
\\
&&&{\scriptstyle q_k-q_{k+1}} &\vbl{4}
\\
\multispan{6}\hblfil\
\\
\begin{matrix} 1 \\ 0 \\ \vdots \\ 0 \end{matrix}
&\begin{matrix} 1 \\ 0 \\ \vdots \\ 0 \end{matrix}
&\begin{matrix} \cdots \\ \cdots \\ \\ \cdots \end{matrix}
&\begin{matrix} 1 \\ 0 \\ \vdots \\ 0 \end{matrix} &\vbl{16}
&\quad I_{N-k} \quad
\end{bmatrix},
$$
respectively. Here, $I_k$ is unit matrix of size $k$.
Then we have
$$
MP_1P_2\cdots P_{N-1}=\begin{bmatrix}
\displaystyle a_i\prod_{k=1}^{j-1}(p_i-q_k^*) &\vbl{12}
&\displaystyle -\prod_{k=1}^{j-1}(q_i-q_k)
\\
\multispan{3}\hblfil\
\\
\displaystyle\prod_{k=1}^{j-1}(q_i^*-q_k^*) &\vbl{12}
&\displaystyle a_i^*\prod_{k=1}^{j-1}(p_i^*-q_k)
\end{bmatrix},
$$
and
\begin{eqnarray*}
&&MP_1P_2\cdots P_{N-1}Q_1Q_2\cdots Q_{N-1}=\begin{bmatrix}
\displaystyle
a_i\prod_{\genfrac{}{}{0pt}{2}{k=1}{k\ne j}}^N(p_i-q_k^*)
&\vbl{14}
&\displaystyle
-\delta_{ij}\prod_{\genfrac{}{}{0pt}{2}{k=1}{k\ne i}}^N(q_i-q_k)
\\
\multispan{3}\hblfil\
\\
\displaystyle
\delta_{ij}\prod_{\genfrac{}{}{0pt}{2}{k=1}{k\ne i}}^N(q_i^*-q_k^*)
&\vbl{14}
&\displaystyle
a_i^*\prod_{\genfrac{}{}{0pt}{2}{k=1}{k\ne j}}^N(p_i^*-q_k)
\end{bmatrix}
\\
&&=\begin{bmatrix}
\displaystyle\frac{\mu_i}{p_i-q_j^*}
\prod_{\genfrac{}{}{0pt}{2}{k=1}{k\ne i}}^N(q_i-q_k)
&\vbl{14}
&\displaystyle -\delta_{ij}
\prod_{\genfrac{}{}{0pt}{2}{k=1}{k\ne i}}^N(q_i-q_k)
\\
\multispan{3}\hblfil\
\\
\displaystyle\delta_{ij}
\prod_{\genfrac{}{}{0pt}{2}{k=1}{k\ne i}}^N(q_i^*-q_k^*)
&\vbl{14}
&\displaystyle\frac{\mu_i^*}{p_i^*-q_j}
\prod_{\genfrac{}{}{0pt}{2}{k=1}{k\ne i}}^N(q_i^*-q_k^*)
\end{bmatrix}.
\end{eqnarray*}
Thus we get \eqref{eqn:detM}.\qed

We note that $\det M$ is nonnegative in Lemma \ref{lem:detM} because
$M$ has the form of $M=\left[\begin{array}{cc}A & -B \\ B^* & A^*\end{array}\right]$ 
with $N\times N$ matrices $A$ and $B$. In the following Propositions we give two cases 
of $\det M$ being strictly positive.
%
\begin{prop}\label{prop:regularity_continuous}
We assume $p_i\ne p_j$ for $1\le i\ne j\le N$
and $\Re p_i>0$ for $1\le i\le N$.
Then $F$ in Proposition \ref{prop:time_evolution_continuous} is strictly positive.
\end{prop}
\noindent{\em Proof.} By taking $ q_i = -p_i $ and $a_i=\phi_i^{(0)}(0)$ in Lemma \ref{lem:detM},
where $\phi_i^{(n)}(r)$ is given in \eqref{eqn:continuous_tau_entries}, $\det M$ coincides with
$F$. Then \eqref{eqn:detM} implies $$ \det M =\left|\begin{matrix}
\displaystyle\frac{\mu_i\mu_j^*}{p_i+p_j^*} &\vbl{14} &-\delta_{ij}
\\
\multispan{3}\hblfil\
\\
\delta_{ij} &\vbl{14} &\displaystyle\frac{1}{p_i^* + p_j}
\end{matrix}\right|
\prod_{1\le i<j\le N}
(p_i-p_j)(p_i^* - p_j^*).
$$
When $\Re p_i>0$, we can show that both of two $N\times N$ Hermite matrices
$$
\begin{bmatrix}
\\
\ \displaystyle\frac{\mu_i\mu_j^*}{p_i+p_j^*}\
\\
\\
\end{bmatrix}_{1\le i,j\le N},
\qquad
\begin{bmatrix}
\\
\ \displaystyle\frac{1}{p_i^* + p_j}\
\\
\\
\end{bmatrix}_{1\le i,j\le N}
$$
are positive definite. Actually, for arbitrary non-zero vector $[v_1,v_2,\ldots, v_N]$ we have
\begin{align*}
& [v_1,v_2,\ldots, v_N]
\begin{bmatrix}
\\
\ \displaystyle\frac{\mu_i\mu_j^*}{p_i+p_j^*}\
\\
\\
\end{bmatrix}
\left[\begin{array}{c} v_1^*\\v_2^*\\\vdots\\ v_N^*    \end{array}\right]
=\sum_{i,j=1}^N\frac{\mu_i\mu_j^*}{p_i + p_j^*}v_iv_j^*
= \int_{-\infty}^0 \sum_{i,j=1}^N\mu_i\mu_j^*  v_iv_j^* e^{(p_i+p_j^*)y}~dy\\
&=\int_{-\infty}^0 \left|\sum_{i=1}^N \mu_i  v_i e^{p_i y}\right|^2~dy \geq 0.
\end{align*}
Therefore $\det M$ is strictly positive, i.e., $\det M>0$. This completes the proof.\qed
%
\begin{prop}\label{prop:regularity_discrete}
We assume $p_i\ne p_j$ for $1\le i\ne j\le N$ and $|p_i|>1$ for $1\le i\le N$.  Then $F_n$ in
Proposition \ref{prop:time_evolution_semi-discrete} and $F_n^m$ in Proposition
\ref{prop:time_evolution_discrete} are strictly positive.
\end{prop}

\noindent{\em Proof.} \quad By taking $ q_i = 1/p_i $ and $a_i=\phi_i^{(n)}(0)$ in Lemma
\ref{lem:detM}, where $\phi_i^{(n)}(r)$ is given in \eqref{eqn:discrete_tau_entries} with
\eqref{eqn:phi_semi-discrete} or \eqref{eqn:phi_discrete}, then $\det M$ coincides with $F_n$ or
$F_n^m$. Then we have from \eqref{eqn:detM} $$ \det M =\left|\begin{matrix}
\displaystyle\frac{p_i\mu_ip_j^*\mu_j^*}{p_ip_j^*-1} &\vbl{14} &-\delta_{ij}
\\
\multispan{3}\hblfil\
\\
\delta_{ij} &\vbl{14} &\displaystyle\frac{1}{p_i^*p_j-1}
\end{matrix}\right|
\prod_{1\le i<j\le N}
\Big(\frac{1}{p_i}-\frac{1}{p_j}\Big)\Big(\frac{1}{p_i^*}-\frac{1}{p_j^*}\Big).
$$
When $|p_i|>1$, we can show that both of two $N\times N$ Hermite matrices
$$
\begin{bmatrix}
\\
\ \displaystyle\frac{p_i\mu_ip_j^*\mu_j^*}{p_ip_j^*-1}\
\\
\\
\end{bmatrix}_{1\le i,j\le N},
\qquad
\begin{bmatrix}
\\
\ \displaystyle\frac{1}{p_i^*p_j-1}\
\\
\\
\end{bmatrix}_{1\le i,j\le N}
$$
are positive definite. Actually, for arbitrary non-zero vector $[v_1,v_2,\ldots, v_N]$ we have
by noting $|p_i|>1$
\begin{align*}
& [v_1,v_2,\ldots, v_N]
\begin{bmatrix}
\\
\ \displaystyle\frac{p_i\mu_ip_j^*\mu_j^*}{p_ip_j^*-1}\
\\
\\
\end{bmatrix}
\left[\begin{array}{c} v_1^*\\v_2^*\\\vdots\\ v_N^*    \end{array}\right]
=\sum_{i,j=1}^N\frac{\mu_i\mu_j^*}{1-\frac{1}{p_ip_j^*}}v_iv_j^*
=\sum_{i,j=1}^N \mu_iv_i\mu_j^*v_j^* \sum_{n=0}^\infty \left(\frac{1}{p_ip_j^*}\right)^n\\
&= \sum_{n=0}^\infty \left|\sum_{i=1}^N \mu_iv_i  \frac{1}{p_i^n}\right|^2\geq 0.
\end{align*}
Therefore $\det M$ is strictly positive, i.e., $\det M>0$. This completes the proof.\qed
\section{Concluding Remarks}\label{sec:concluding_remarks}
In this paper, we have constructed a discrete analogue of local induction equation describing the deformation of
discrete space curves whose complex curvature is governed by the dNLS equation.
We have shown some numerical simulations for both open and closed curves. We have constructed explicit formulas
of space curves and related quantities in terms of the $\tau$ functions of the two-component KP hierarchy. By introducing
appropriate time dependence, we have constructed regular soliton type solutions to the NLS, sdNLS and dNLS equations
and explicit formulas for the deformation of space curves associated with those solutions.

The relationship to ddIHM \cite{Hoffmann:dNLS_HM,Hoffmann:dNLS} is not yet clear in this paper.
Actually, it can be shown that ddIHM is equivalent to the isotropic version of the discrete
Landau-Lifschitz equation presented by Nijhoff et al \cite{Nijhoff:discrete_LL}. Then the complex
curvature of the discrete curves deformed by ddIHM is governed by a variant of dNLS type equation
which is closely related to the dNLS \eqref{eqn:dNLS}. Since the Landau-Lifschitz equation belongs
to the BKP hierarchy \cite{DJKM:LL}, it is expected that we have an alternate explicit formulas of
space curves in terms of the Pfaffians, namely, $\tau$ functions of the BKP hierarchy.  These
results will be reported in the forthcoming paper.

\section*{Acknowledgements}
This work has been supported by JSPS KAKENHI Grant Numbers JP19K03461, JP16H03941, JP16K13763, JP24340029,
JP26610029, JP15K04909, JP15K04862.  It was also supported by 2016 IMI Joint Use Research Program
Short-term Joint Research ``Construction of directable discrete models of physical phenomena''. The
authors would like to thank Prof. Konrad Polthier for encouragement and valuable suggestions.
\appendix
\section{Numerical computations}\label{subsec:numerical}
\begin{figure}[ht]
 \begin{center}
\includegraphics[bb=0 0 1071 3001,scale=0.075]{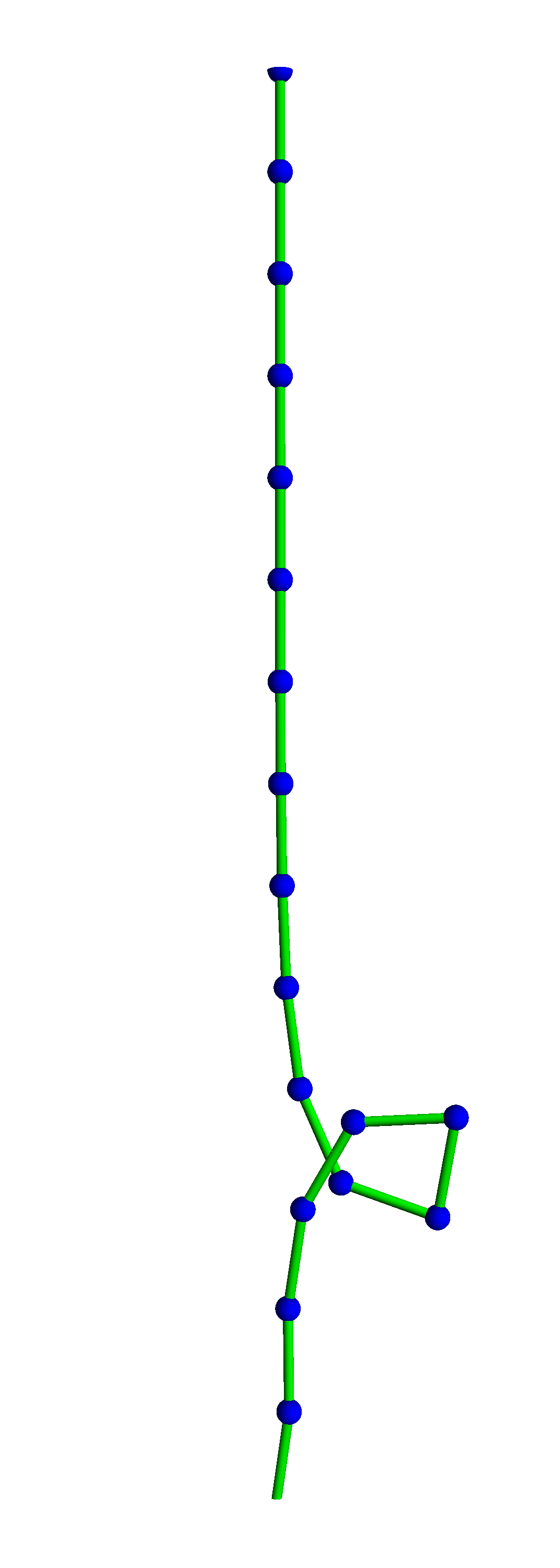}
\includegraphics[bb=0 0 1071 3001,scale=0.075]{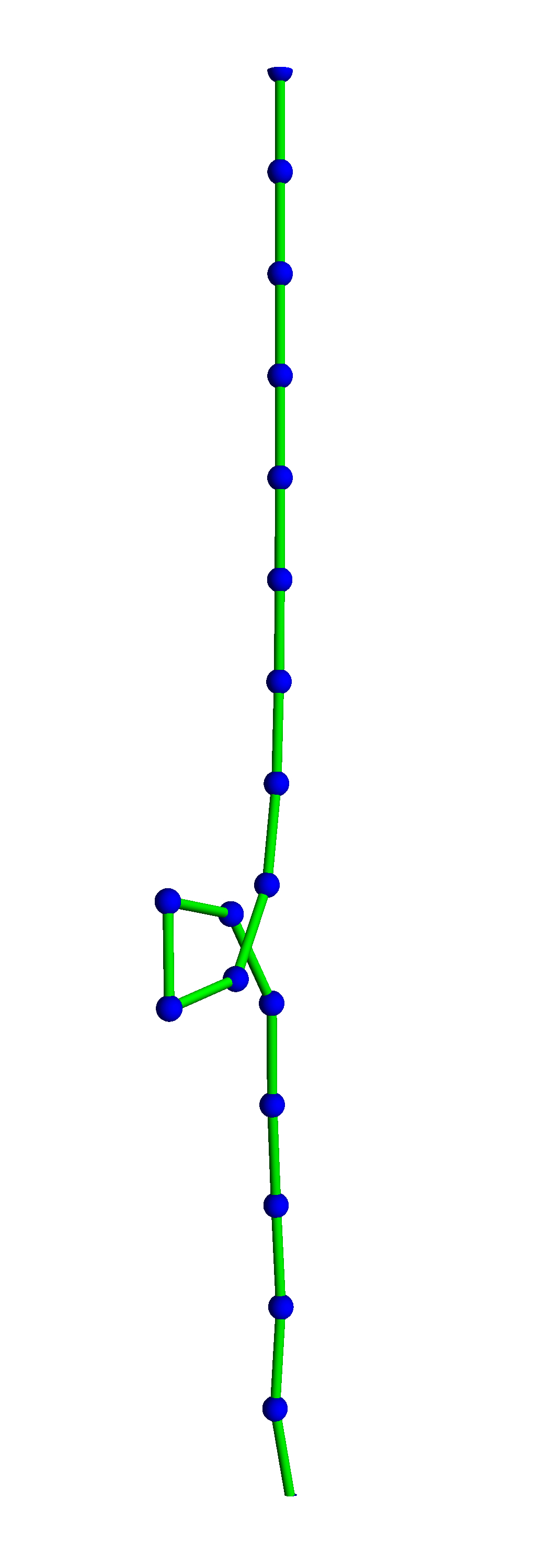}
\includegraphics[bb=0 0 1071 3001,scale=0.075]{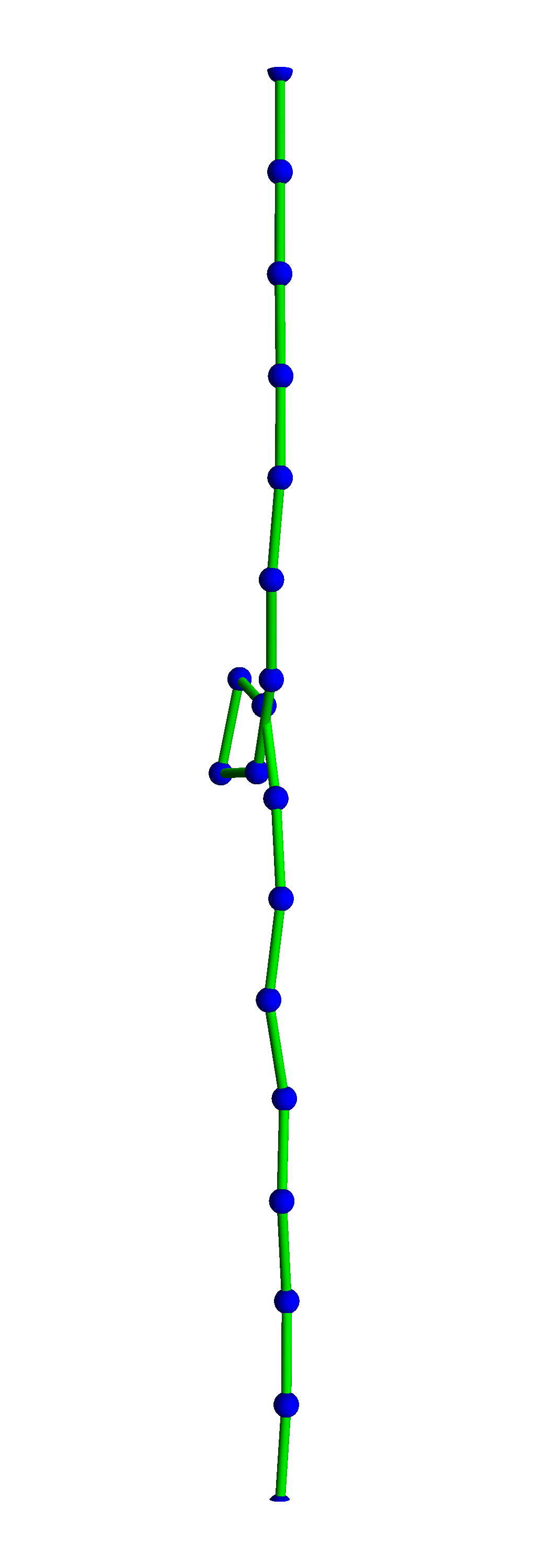}
\includegraphics[bb=0 0 1071 3001,scale=0.075]{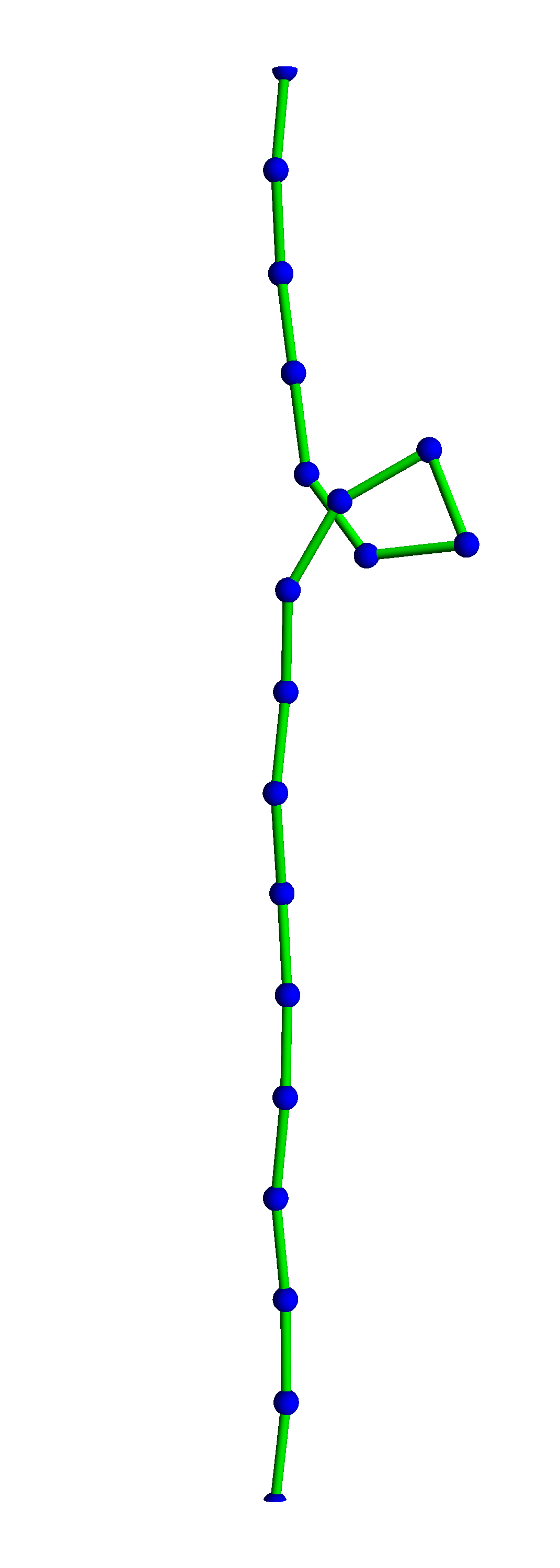}
\includegraphics[bb=0 0 1071 3001,scale=0.075]{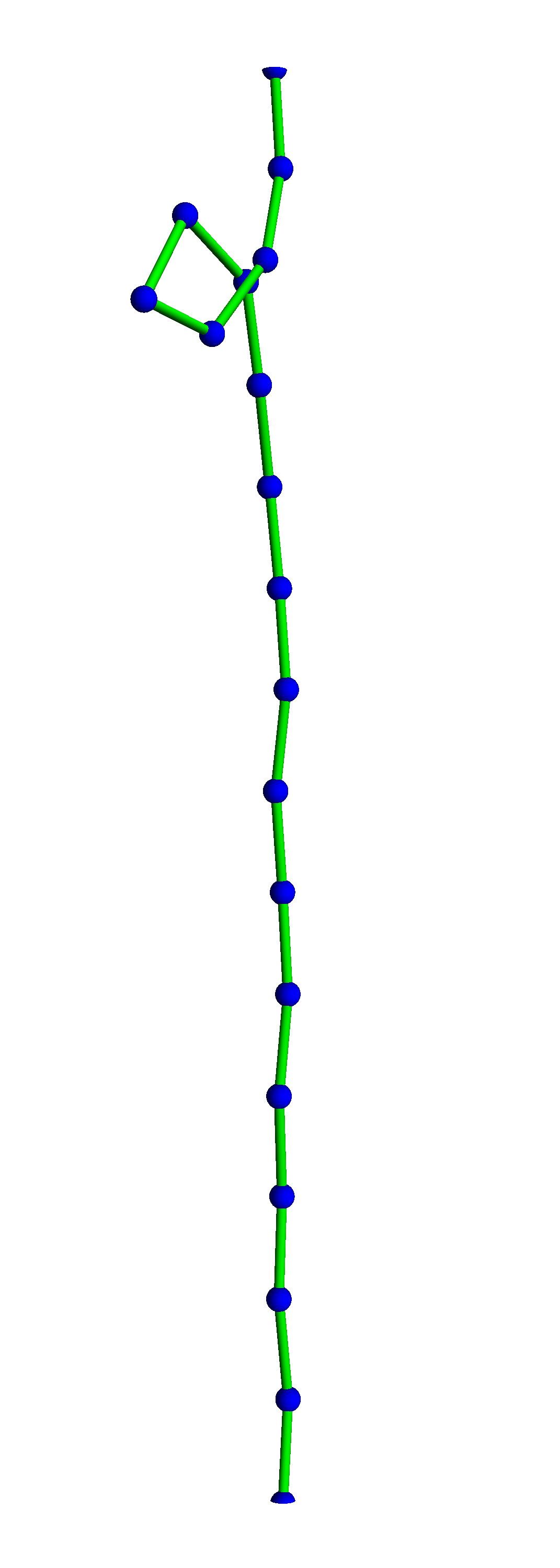}
\end{center}
\caption{Numerical simulation of dLIE. }\label{fig:dLIE_simulation1}
\end{figure}

Let us describe how to compute numerically the deformation of curves given by dLIE from a given
initial curve. We first consider the case of sufficiently long curve with $n_0\leq n\leq n_1$ for
some fixed $n_0$, $n_1$ under the vanishing boundary condition, where the calculation is carried out
from small $n$ to large $n$.  Then, for a given $m$, an algorithm to compute the deformation
$\gamma_n^{m+1}$ from a given initial curve $\gamma_n^m$ is described as follows:
\begin{enumerate}
 \item Give an initial curve $\gamma_n^m$ for $n_0\leq n\leq n_1$ for some fixed $n_0$ and
       $n_1$. At boundaries, give $B_{n_0}^m$, $u_{n_0}^m$, $u_{n_0-1}^{m+1}$, $\Gamma_{n_0}^m$ and
       $T_{n_1}^m$. $\Gamma_{n_0}^m$ may be chosen close to either $1$ or $1+\epsilon^4/\delta^2$.
\end{enumerate}
At the left edge ($n=n_0$):
\begin{enumerate}
\setcounter{enumi}{1}
 \item Compute $\Phi_{n_0}^m=[T_{n_0}^m, N_{n_0}^m,B_{n_0}^m]$ from $\gamma_{n_0}^m$,
       $\gamma_{n_0+1}^m$ and $B_{n_0}^m$ by using \eqref{eqn:discrete_Frenet_frame}.
 \item Compute $\gamma_{n_0}^{m+1}$ from $u_{n_0}^m$, $u_{n_0-1}^{m+1}$, $\Gamma_{n_0}^m$ and $\Phi_{n_0}^m$ by
       using \eqref{eqn:discrete_def_Frenet} and \eqref{eqn:discrete_coeffs_Frenet}.
\end{enumerate}
Repeat from $n=n_0+1$ to $n_1-1$:
\begin{enumerate}
\setcounter{enumi}{3}
\item Compute $\Phi_n^m=[T_{n}^m, N_{n}^m,B_{n}^m]$ from $\gamma_{n+1}^m$, $\gamma_{n}^m$, $\Phi_{n-1}^m$ by using \eqref{eqn:discrete_Frenet_frame}. 
\item Compute $u_n^m$ from $\Phi_n^m$ and $\Phi_{n-1}^m$.\\
More precisely, compute $\kappa_n^m$ and $\nu_n^m$ by using \eqref{eqn:kappa_nu}. 
Then compute $u_n^m$ by \eqref{eqn:u_discrete}.
\item Compute $u_{n-1}^{m+1}$ and $\Gamma_{n}^m$ from $u_{n-1}^m$, $u_{n-2}^{m+1}$, $u_n^m$ and $\Gamma_{n-1}^m$ by using dNLS \eqref{eqn:dNLS}.
\item Compute $\gamma_{n}^{m+1}$ from $u_n^m$, $u_{n-1}^{m+1}$, $\Gamma_n^m$ and $\Phi_n^m$ by using
      \eqref{eqn:discrete_def_Frenet} and \eqref{eqn:discrete_coeffs_Frenet}.
\end{enumerate}
At the right edge ($n=n_1$):
\begin{enumerate}
\setcounter{enumi}{7}
\item Compute $\Phi_{n_1}^m=[T_{n_1}^m, N_{n_1}^m,B_{n_1}^m]$ from $T_{n_1}^m$ and
      $\Phi_{n_1-1}^m$ by using \eqref{eqn:discrete_Frenet_frame}.
\item Compute $u_{n_1}^m$ from $\Phi_{n_1}^m$ and $\Phi_{n_1-1}^m$.
\item Compute $u_{n_1-1}^{m+1}$ and $\Gamma_{n_1}^m$ from $u_{n_1-1}^m$, $u_{n_1-2}^{m+1}$, 
$u_{n_1}^m$ and $\Gamma_{n_1-1}^m$ by using dNLS \eqref{eqn:dNLS}.
\item Compute $\gamma_{n_1}^{m+1}$ from $u_{n_1}^m$, $u_{n_1-1}^{m+1}$, $\Gamma_{n_1}^m$ and $\Phi_{n_1}^m$ by using
      \eqref{eqn:discrete_def_Frenet} and \eqref{eqn:discrete_coeffs_Frenet}.
\end{enumerate}
Figure \ref{fig:dLIE_simulation1} illustrates a result of numerical computation according to this algorithm.
%

\begin{figure}[ht]
 \begin{center}
\includegraphics[bb=0 0 877 1500,scale=0.1]{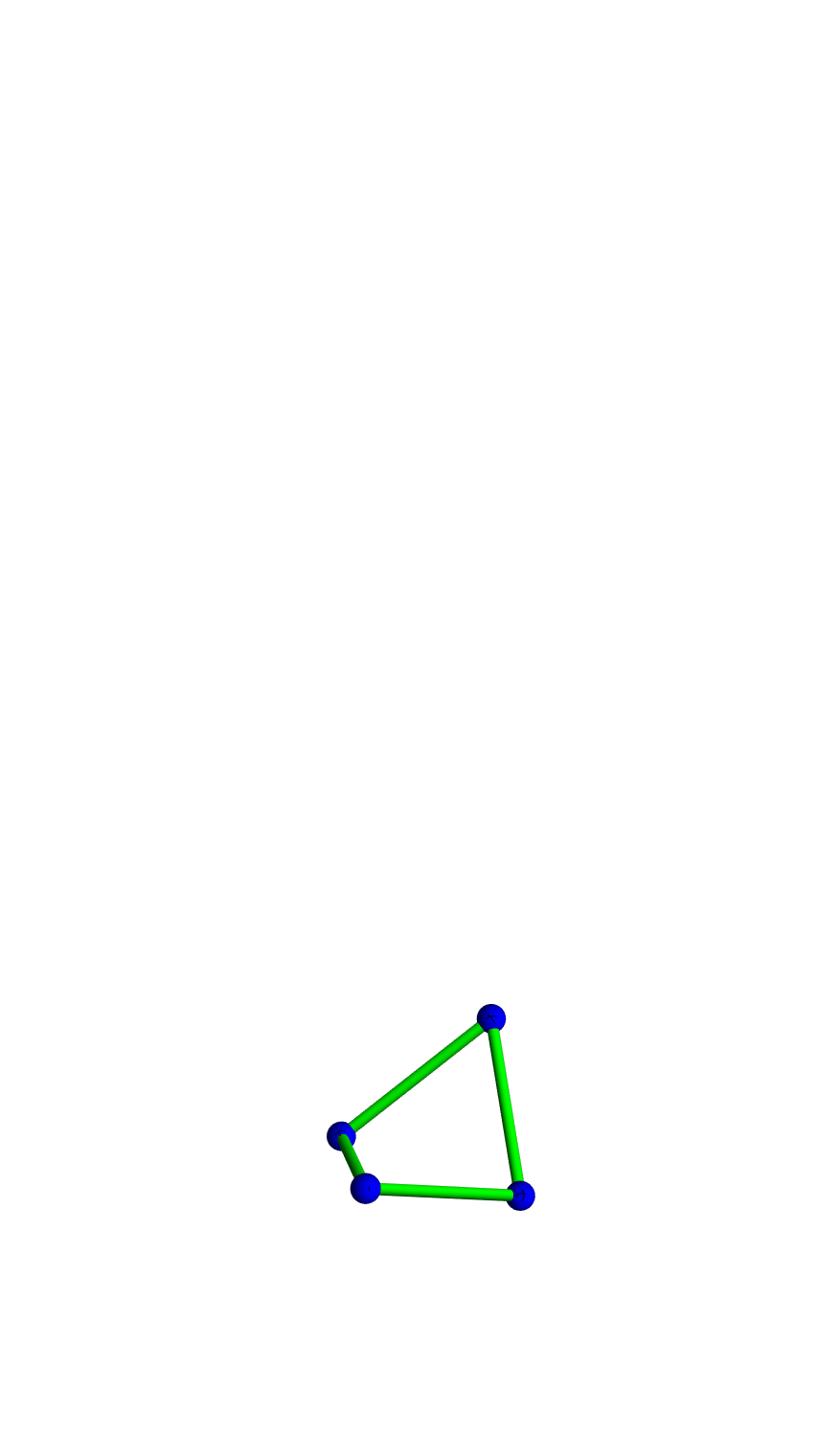}
\includegraphics[bb=0 0 877 1500,scale=0.1]{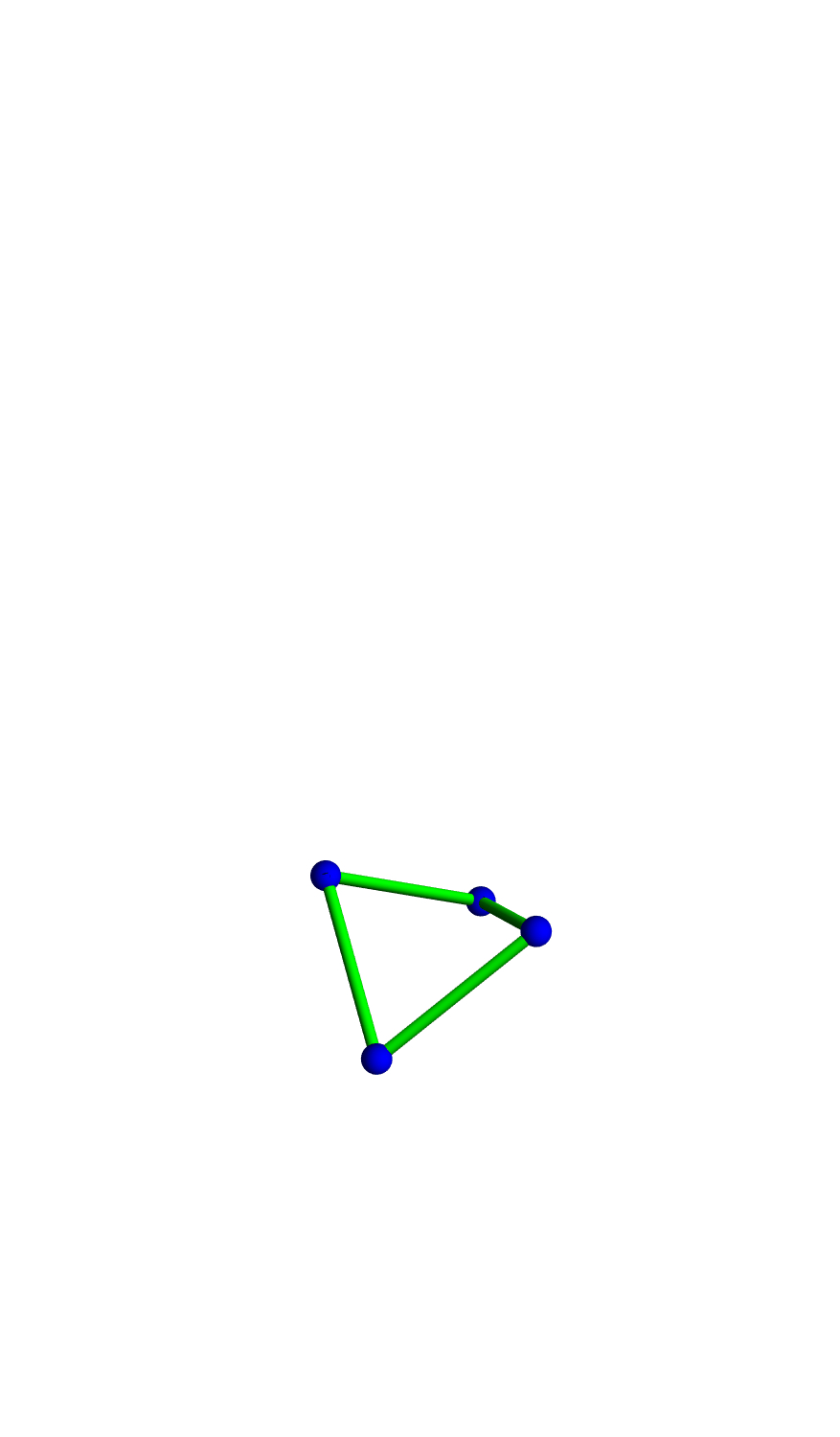}
\includegraphics[bb=0 0 877 1500,scale=0.1]{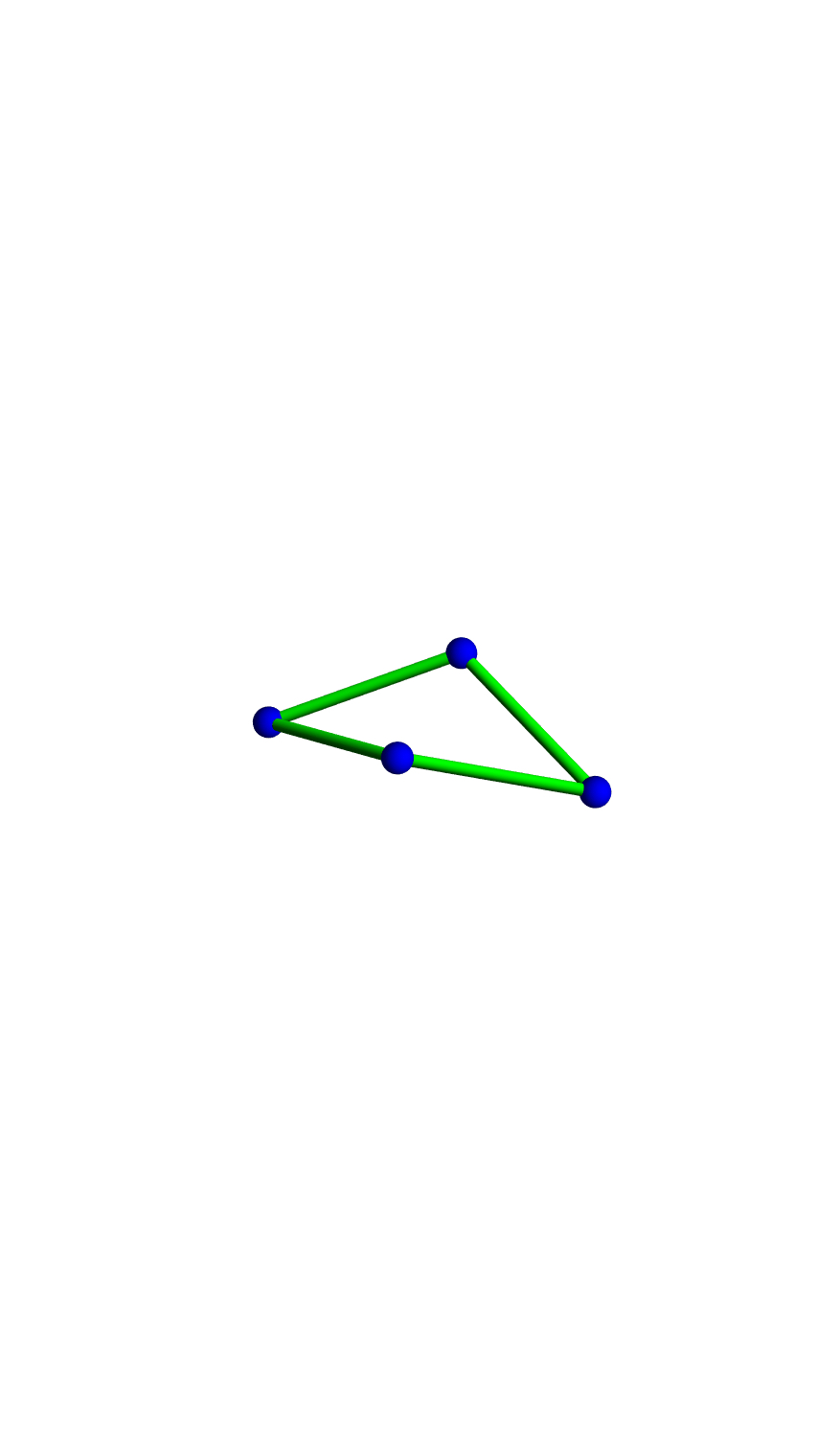}
\includegraphics[bb=0 0 877 1500,scale=0.1]{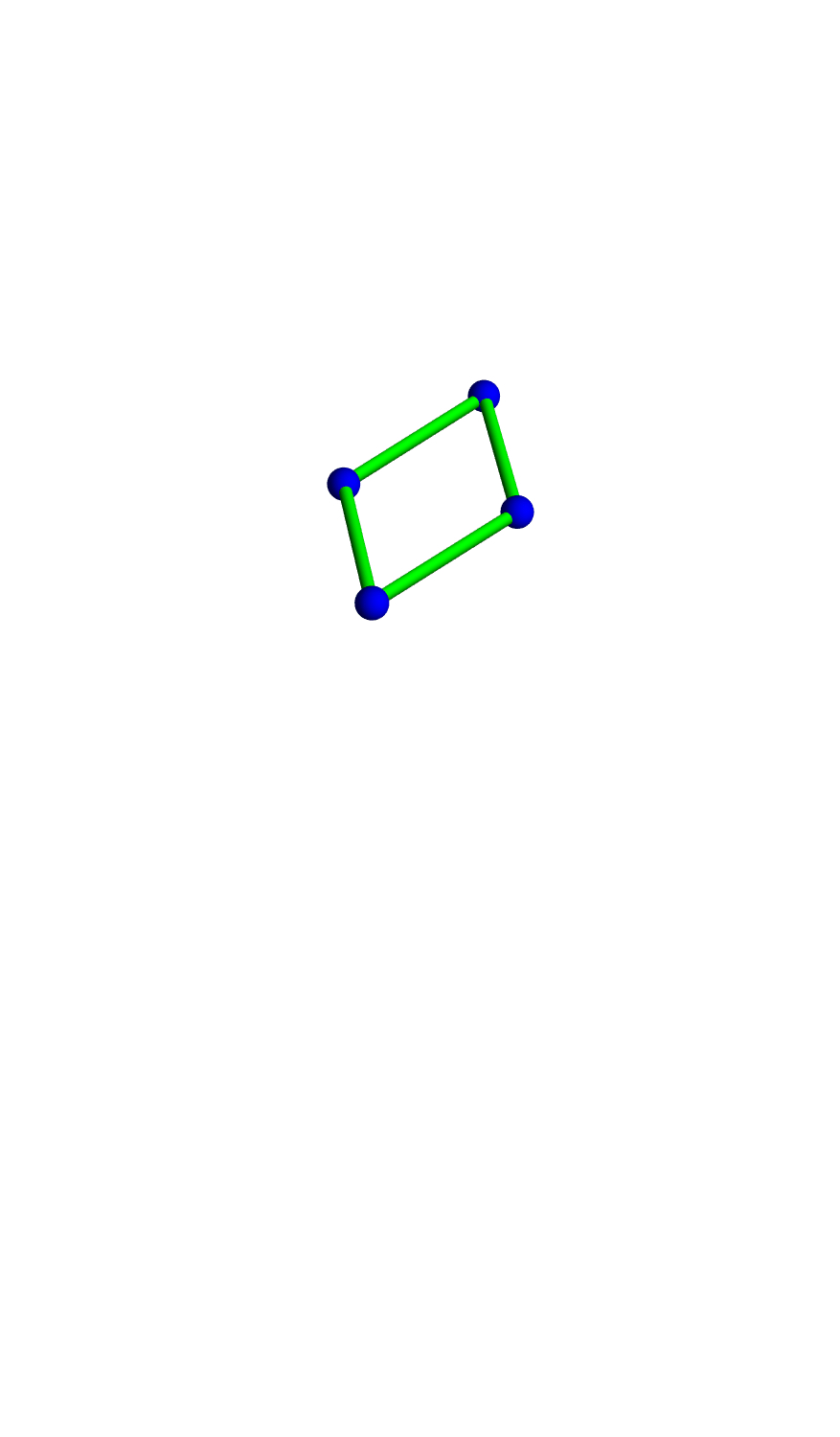}
\includegraphics[bb=0 0 877 1500,scale=0.1]{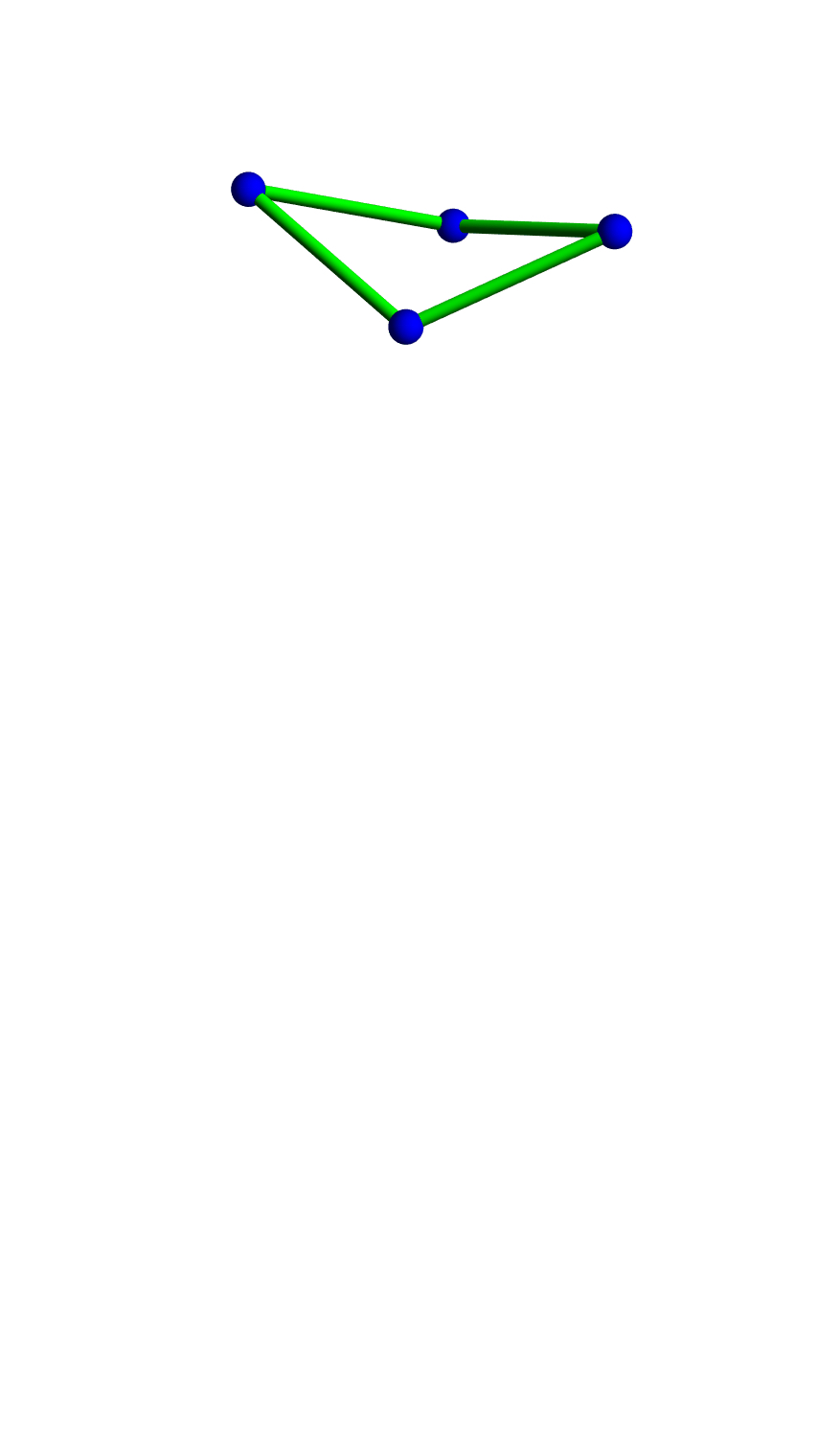}
\end{center}
\caption{Numerical simulation of dLIE for a closed curve of period $l = 4$. }\label{fig:dLIE_simulation2}
\end{figure}

\begin{figure}[ht]
 \begin{center}
\includegraphics[bb=0 0 877 1500,scale=0.1]{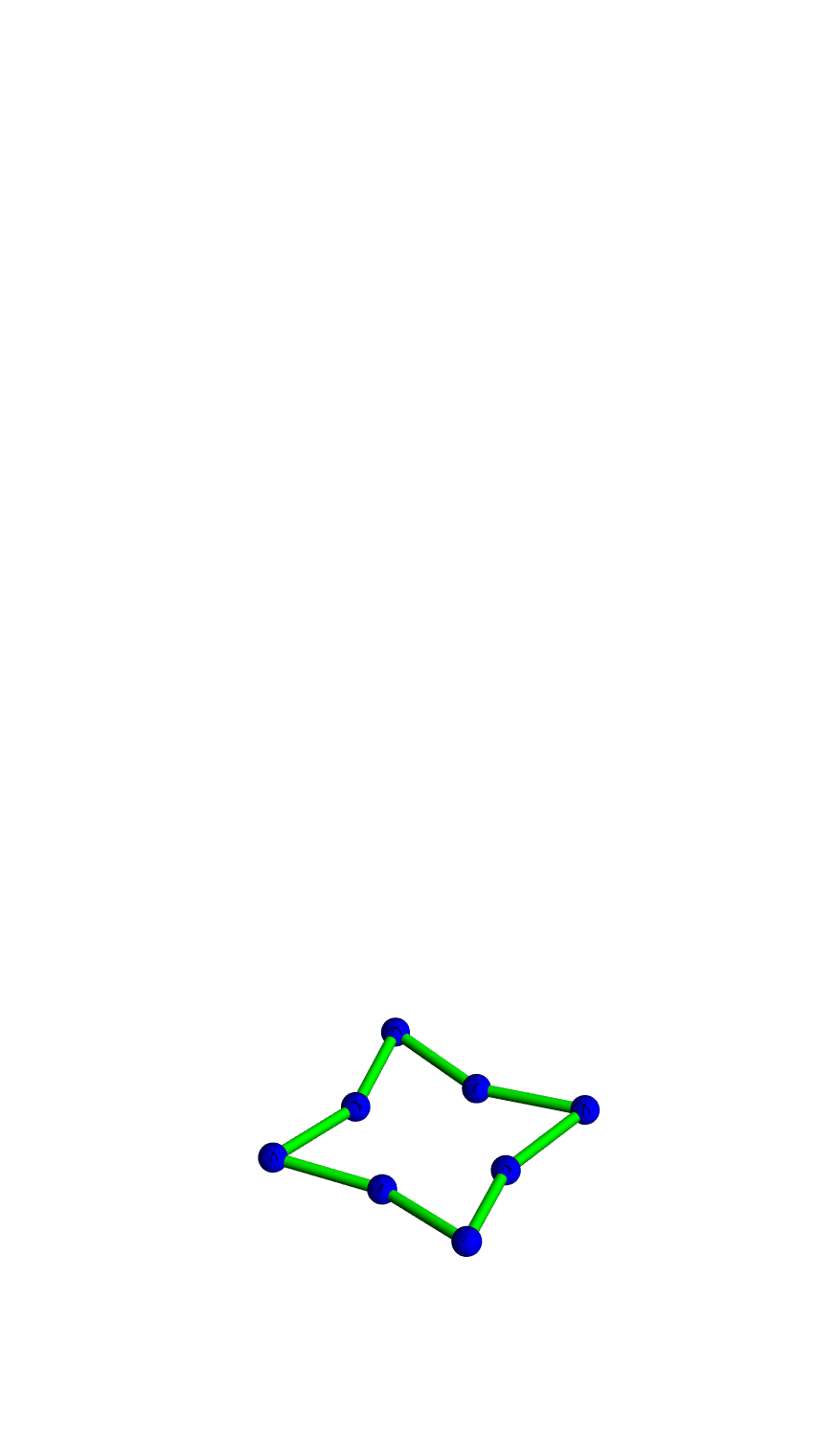}
\includegraphics[bb=0 0 877 1500,scale=0.1]{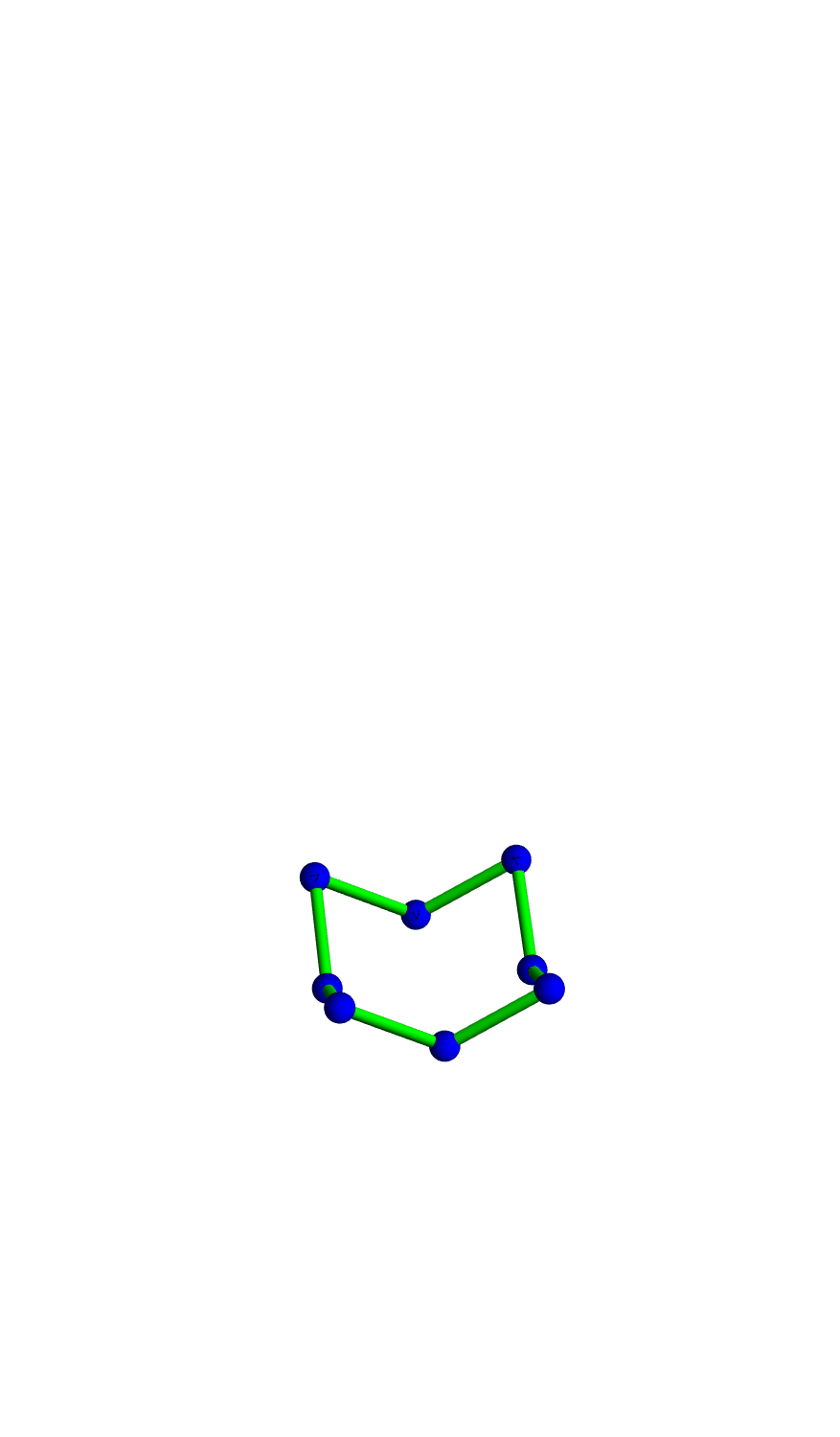}
\includegraphics[bb=0 0 877 1500,scale=0.1]{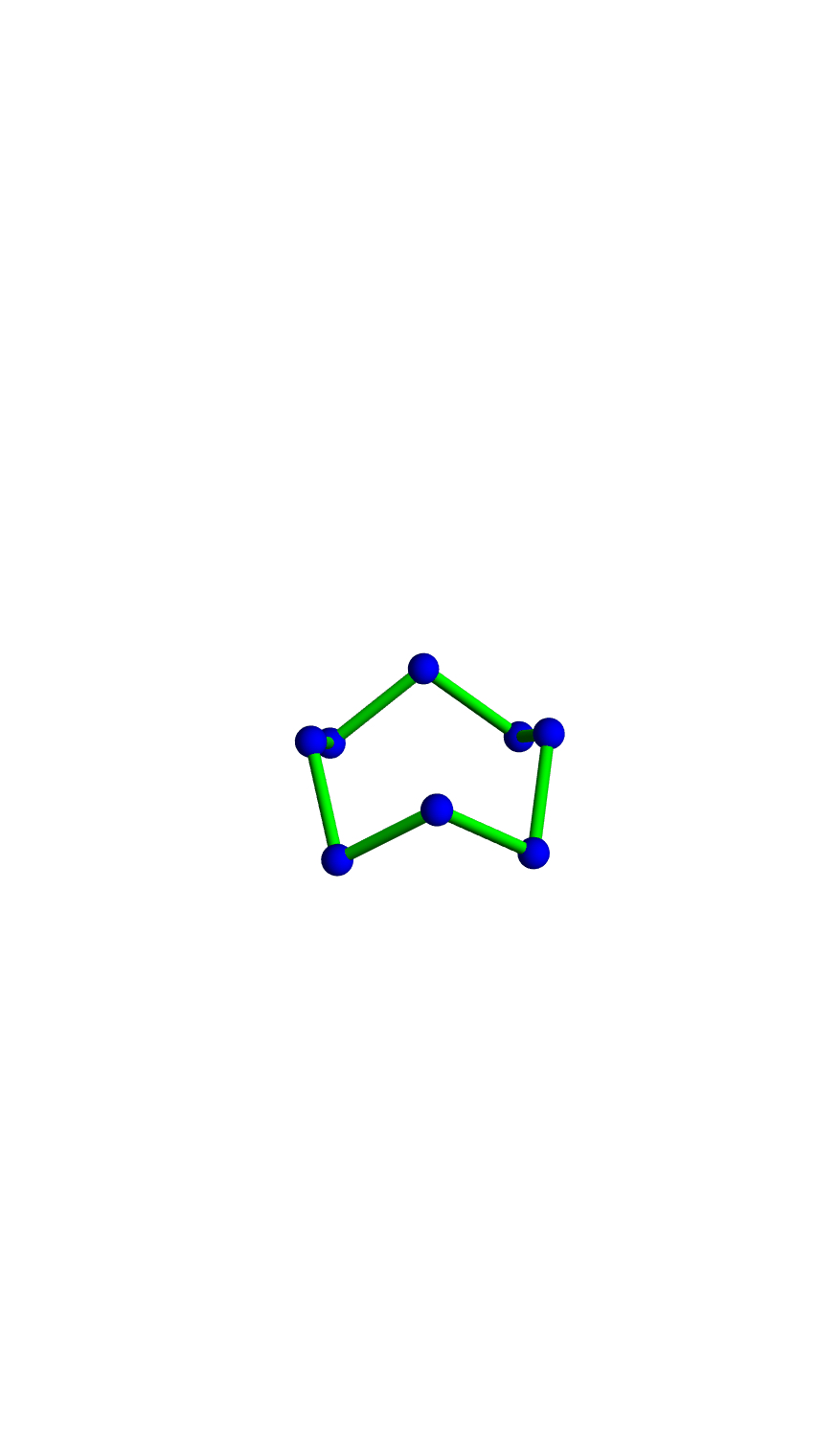}
\includegraphics[bb=0 0 877 1500,scale=0.1]{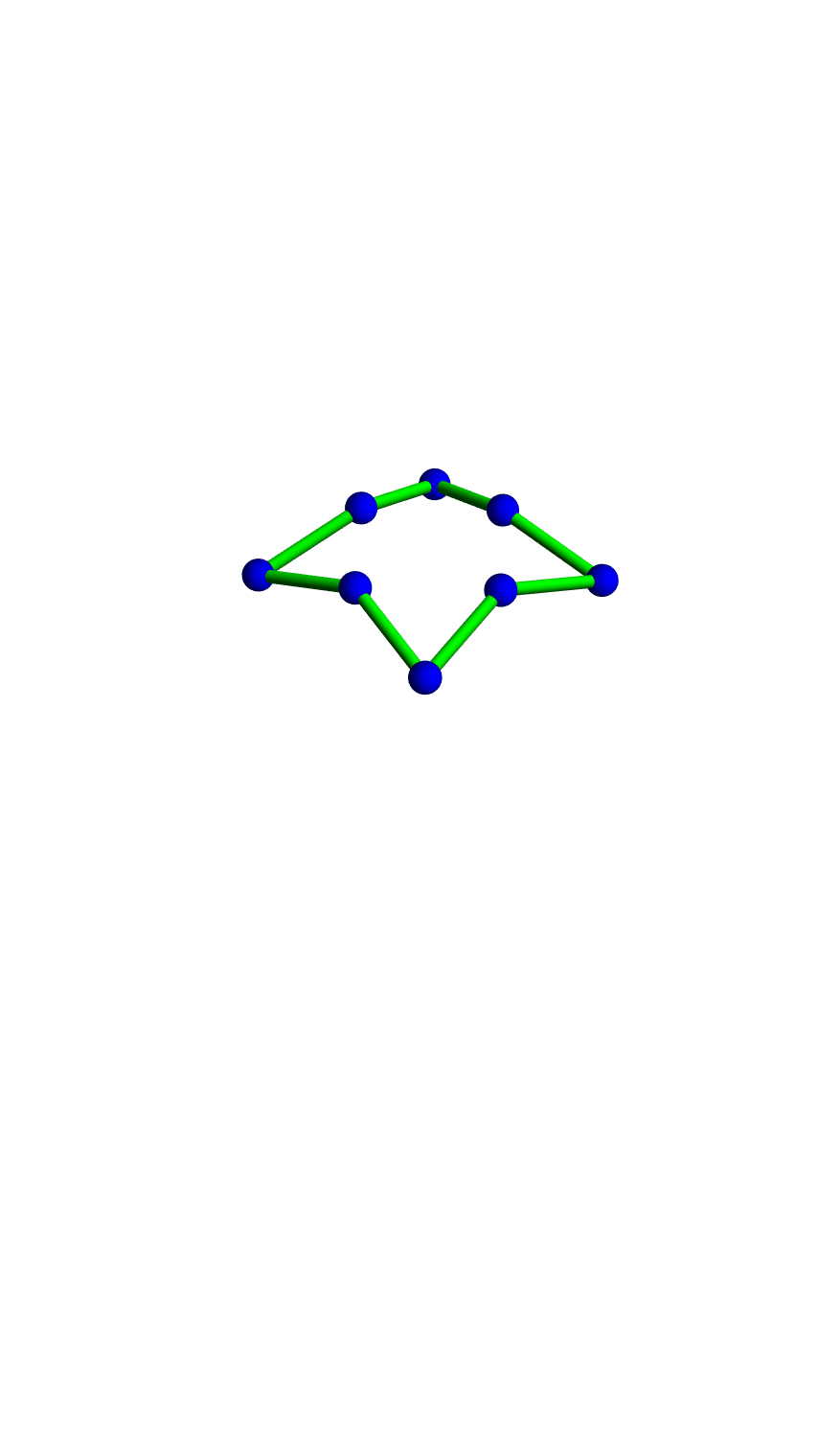}
\includegraphics[bb=0 0 877 1500,scale=0.1]{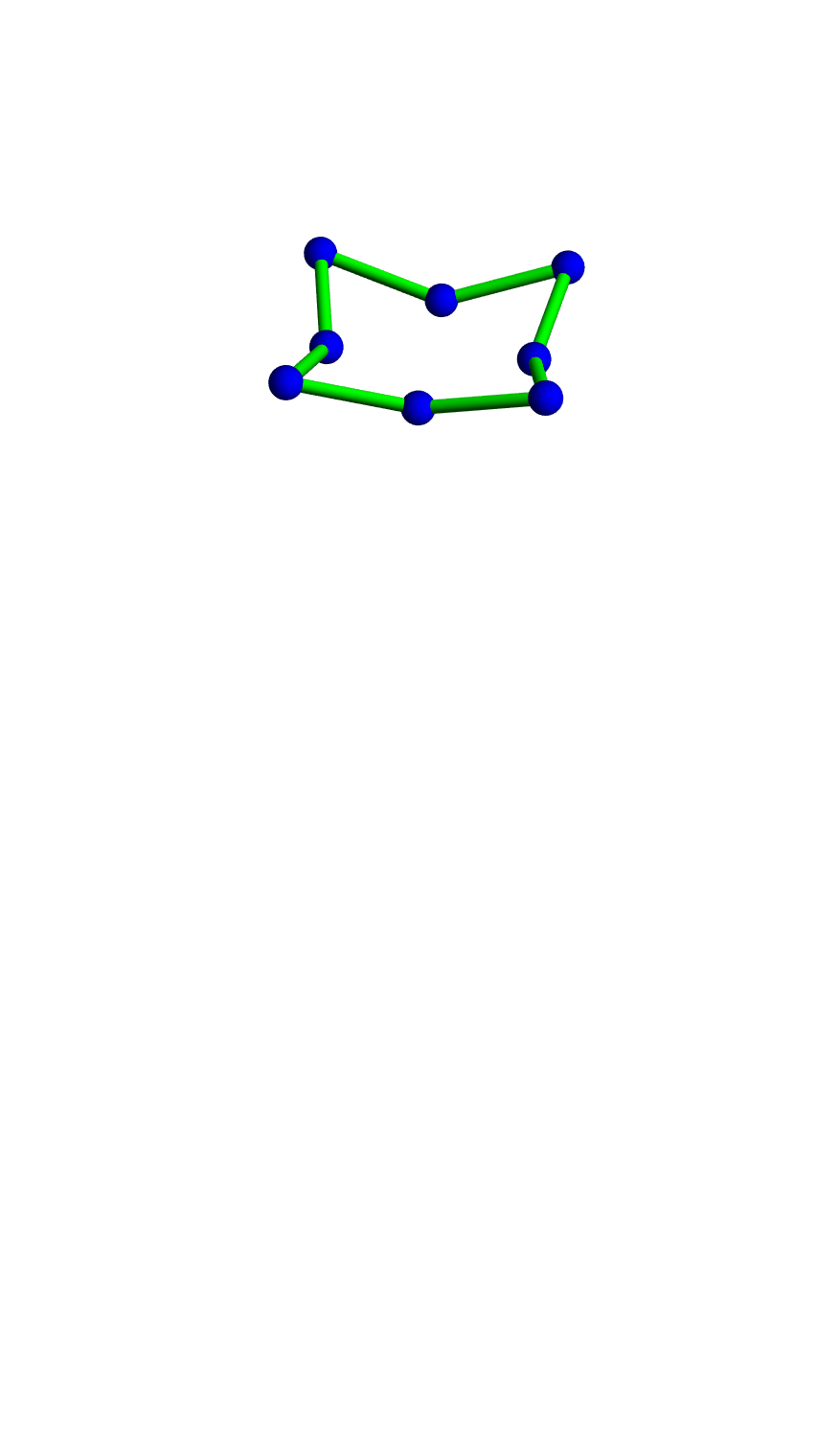}
\end{center}
\caption{Numerical simulation of dLIE for a closed curve of period $l = 8$. }\label{fig:dLIE_simulation3}
\end{figure}

We next describe how to compute the deformation of a closed curve $\gamma_n^m$ of period $l$ in $n$,
namely $\gamma_{n+l}^m = \gamma_n^m$.
We note that it is possible to simulate such a periodic case by using dLIE
\eqref{eqn:discrete_def_Frenet}--\eqref{eqn:discrete_coeffs_Frenet},
although we derived it under the vanishing boundary condition
\eqref{eqn:dNLS_bc}.
Since the computation requires a solution to dNLS which has
the same period as $\gamma_n^m$, we fix a positive number $c$ that plays a role of tolerance in
constructing $l$-periodic numerical solution $(u^{m+1}_n, \Gamma^m_{n+1})$ to dNLS.  An algorithm to
compute $\gamma_n^{m+1}$ is described as follows:
\begin{enumerate}
 \item Give an initial closed curve $\gamma_n^m$ and
positive numbers $\delta$, $c$.
 \item Compute $\Phi_n^m=[T_{n}^m, N_{n}^m,B_{n}^m]$ from $\gamma_{n+1}^m$, $\gamma_{n}^m$ and
       $\gamma_{n-1}^m$ by using \eqref{eqn:discrete_Frenet_frame}.
 \item Compute $u_n^m$ from $\Phi_n^m$ and $\Phi_{n-1}^m$ by using \eqref{eqn:u_discrete} with $\Lambda_0^m = 0$.
 \item Give a pair of numbers $(u_1^{m+1}, \Gamma_2^m)
\in \mathbb{C} \times \mathbb{R}_{>0}$
which satisfies \eqref{eqn:alpha_beta_squaredsum} with $n=1$, that is,
 \begin{equation}
 \begin{split}
 &\frac{\delta^2}{\epsilon^4} \left[ \left(1-\sqrt{-1}\frac{\epsilon^2}{\delta}\right) - \left(1 + \frac{\epsilon^2}{4} u_2^m u_1^{m+1}{}^*\right) \Gamma_2^m\right] \left[ \left(1+\sqrt{-1}\frac{\epsilon^2}{\delta}\right) - \left(1 + \frac{\epsilon^2}{4} u_2^m{}^* u_1^{m+1}\right) \Gamma_2^m\right] \\
 &+ \frac{\delta^2}{4\epsilon^2} \left(u_1^{m+1} - u_2^m\right) \left(u_1^{m+1}{}^* - u_2^m{}^*\right) (\Gamma_2^m)^2 = \Gamma_2^m.
 \end{split}
 \end{equation}
This procedure incorporates the isoperimetricity of the deformation.
 \item Compute an $l$-periodic solution $(u_n^{m+1}, \Gamma_{n+1}^m)$ to dNLS \eqref{eqn:dNLS} as follows:
For $2 \leq n \leq l+1$,
compute $(u_n^{m+1}, \Gamma_{n+1}^m)$ by
\begin{equation*}
\begin{split}
&u_{n}^{m+1} = \frac{1}{{\textstyle {\scriptstyle \sqrt{-1}}\frac{\epsilon^2}{\delta} - 1}} 
\left[{\textstyle \left({\scriptstyle\sqrt{-1}}\frac{\epsilon^2}{\delta} + 1\right)}u_n^m 
- (u_{n+1}^m+u_{n-1}^{m+1}) \Big({\textstyle 1+\frac{\epsilon^2}{4}|u_n^m|^2}\Big)\Gamma_n^m\right], \\
&\Gamma_{n+1}^m = \frac{1+\frac{\epsilon^2}{4}|u_n^m|^2}{1+\frac{\epsilon^2}{4}|u_n^{m+1}|^2} \Gamma_n^m
\end{split}
\end{equation*}
with the initial value $(u_1^{m+1}, \Gamma_2^m)$.
If $\big|u_1^{m+1} - u_{l+1}^{m+1}\big| > c$, we recompute $(u_n^{m+1}, \Gamma_{n+1}^m)$ for $2 \leq n \leq l+1$ 
with the revised initial value $(u_1^{m+1}, \Gamma_2^m) = (u_{l+1}^{m+1},\Gamma_{l+2}^m)$.
\item Compute $\gamma_{n}^{m+1}$ from $u_n^m$, $u_{n-1}^{m+1}$, $\Gamma_n^m$ and $\Phi_n^m$ by 
using \eqref{eqn:discrete_def_Frenet} and \eqref{eqn:discrete_coeffs_Frenet}.
\end{enumerate}
Figure \ref{fig:dLIE_simulation2} and Figure \ref{fig:dLIE_simulation3} illustrate the result of
numerical computations according to this algorithm.
%
\section{Proofs of Proposition \ref{prop:continuous_tau_formula} and \ref{prop:discrete_tau_formula}}\label{proofs_of_parametrization}
%
We give a proof of Proposition \ref{prop:continuous_tau_formula}.
Firstly we observe that $\tau(\nu,r)$ defined by
\eqref{eqn:continuous_tau} and \eqref{eqn:continuous_tau_entries} satisfies
\begin{equation}\label{eqn:appendix_1_1}
\tau(\nu,r)^*=(-1)^{r\nu}\tau(-\nu,-r),
\end{equation}
which verifies that $\tau(0,0)$ is real, $\tau(0,-1)^*=\tau(0,1)$, $\tau(1,1)^*=-\tau(-1,-1)$,
$\tau(1,0)^*=\tau(-1,0)$ and $\tau(1,2)^*=\tau(-1,-2)$ in \eqref{eqn:continuous_tau_identification}.
Let us define a $2N\times 2N$ determinant $\sigma(\nu,r,s)$ by
\begin{equation}\label{eqn:appendix_1_2}
 \sigma(\nu,r,s)=\left|\begin{matrix}
\phi_1^{(0)}(r) &\phi_1^{(1)}(r)
 &\cdots &\phi_1^{(N+\nu-1)}(r) &\vbl{5}
 &\psi_1^{(0)}(s) &\psi_1^{(1)}(s) &\cdots &\psi_1^{(N-\nu-1)}(s)
\\
\phi_2^{(0)}(r) &\phi_2^{(1)}(r)
 &\cdots &\phi_2^{(N+\nu-1)}(r) &\vbl{5}
 &\psi_2^{(0)}(s) &\psi_2^{(1)}(s) &\cdots &\psi_2^{(N-\nu-1)}(s)
\\
\vdots &\vdots &&\vdots &\vbl{6} &\vdots &\vdots &&\vdots
\\
\phi_{2N}^{(0)}(r) &\phi_{2N}^{(1)}(r)
 &\cdots &\phi_{2N}^{(N+\nu-1)}(r) &\vbl{5}
 &\psi_{2N}^{(0)}(s) &\psi_{2N}^{(1)}(s) &\cdots &\psi_{2N}^{(N-\nu-1)}(s)
\end{matrix}\right|,
\end{equation}
where
\begin{equation}\label{eqn:appendix_1_3}
\phi_i^{(n)}(r) = p_i^{n-r}e^{\zeta_i},\quad 
\psi_i^{(n)}(s) = (-p_i)^{n-s}e^{\eta_i},\quad
\zeta_i = \frac{1}{p_i}z + p_i x +\zeta_{i0},
\end{equation}
and $\eta_i$ are constants. It is easy to see that $\sigma(\nu,r,s)$ satisfies
\begin{equation}\label{eqn:appendix_1_4}
\sigma(\nu,r-1,s-1)=\sigma(\nu,r,s)\, (-1)^{N-\nu}\prod_{i=1}^{2N}p_i.
\end{equation}

Now the following bilinear equations can be proved straightforwardly
by the Laplace expansion technique,
\begin{align}
& \sigma(\nu,r+1,s+1)\,\sigma(\nu,r,s)
- \sigma(\nu,r+1,s)\,\sigma(\nu,r,s+1)\nonumber\\
&\hskip40pt = \sigma(\nu+1,r+1,s)\,\sigma(\nu-1,r,s+1),
\label{eqn:appendix_1_5}\\[1mm]
& D_x\, \sigma(\nu,r,s+1)\cdot\sigma(\nu,r,s)
= - \sigma(\nu+1,r,s)\,\sigma(\nu-1,r,s+1),
\label{eqn:appendix_1_6}\\[1mm]
& D_x\, \sigma(\nu+1,r+1,s)\cdot\sigma(\nu,r,s)
= \sigma(\nu+1,r,s)\,\sigma(\nu,r+1,s),
\label{eqn:appendix_1_7}\\[1mm]
& D_x\, \sigma(\nu+1,r+1,s-1)\cdot\sigma(\nu,r,s)
= \sigma(\nu+1,r,s-1)\,\sigma(\nu,r+1,s),
\label{eqn:appendix_1_8}\\[1mm]
& \Big(\frac{1}{2}D_zD_x - 1\Big)\, \sigma(\nu,r,s)\cdot\sigma(\nu,r,s)
= -\sigma(\nu,r+1,s)\,\sigma(\nu,r-1,s).
\label{eqn:appendix_1_9}
\end{align}
For notational simplicity, let us denote columns of determinant as
\begin{equation}
\sigma(\nu,r,s)=\left|
\ -r\ \ 1-r\ \ \cdots\ N+\nu-1-r\ ;
\ -s\ \ 1-s\ \ \cdots\ N-\nu-1-s\ \right|,
\end{equation}
where ``$n-r$'' in the left block means the column vector
$^t(\phi_1^{(n)}(r), \phi_2^{(n)}(r), \cdots, \phi_{2N}^{(n)}(r))$
and ``$n-s$'' in the right block means
$^t(\psi_1^{(n)}(s), \psi_2^{(n)}(s), \cdots, \psi_{2N}^{(n)}(s))$.
In this notation, we have the expressions
\begin{align}
& \frac{\partial}{\partial x}\sigma(\nu,r,s)=\left|
{\scriptstyle \ -r\ \ \cdots\ N+\nu-2-r\ \ N+\nu-r\ ;
\ -s\ \ 1-s\ \ \cdots\ N-\nu-1-s\ }\right|, \\
&  \frac{\partial}{\partial z}\sigma(\nu,r,s)=\left|
{\scriptstyle \ -1-r\ \ 1-r\ \ \cdots\ N+\nu-1-r\ ;
\ -s\ \ 1-s\ \ \cdots\ N-\nu-1-s\ }\right|, \\
& \bigg(\frac{\partial^2}{\partial z\partial x}-1\bigg)\,\sigma(\nu,r,s)
=\left|{\scriptstyle \ -1-r\ \ 1-r\ \ \cdots\ N+\nu-2-r\ \ N+\nu-r\ ;\ -s\ \ 1-s\ \ \cdots\ N-\nu-1-s\ }\right|.
\end{align}
For example \eqref{eqn:appendix_1_6} is proved by applying
the Laplace expansion to the $4N\times 4N$ vanishing determinant
in the left-hand side of the following identity,
\begin{equation}
\left|\begin{matrix}
{\scriptstyle -r\ \cdots\ N+\nu-2-r} &\vbl{5}  &{\scriptstyle N+\nu-r}
& ; 
& {\scriptstyle -1-s} &\vbl{5} &{\scriptstyle -s\ \cdots\ N-\nu-2-s }&\vbl{5}
&\hbox{ \O} &\vbl{5} &{\scriptstyle N-\nu-1-s}
\\[2mm]
\multispan{11}\hblfil\
\\
\hbox{\O} &\vbl{5} &{\scriptstyle N+\nu-r}
& ; 
& {\scriptstyle -1-s} &\vbl{5} &\hbox{ \O} &\vbl{5}
&{\scriptstyle -r\ \cdots\ N+\nu-1-r\ } ;\ {\scriptstyle -s\ \cdots\ N-\nu-2-s }&\vbl{5} &{\scriptstyle N-\nu-1-s}
\end{matrix}\right|=0.
\end{equation}
In each term of the bilinear equation \eqref{eqn:appendix_1_6},
two determinants $\sigma$'s have common columns
\begin{equation}
-r, \cdots, N+\nu-2-r\ ;\ -s, \cdots, N-\nu-2-s,
\end{equation}
and distinct columns
\begin{equation}
N+\nu-1-r,\ N+\nu-r\ ;\  -1-s,\  N-\nu-1-s
\end{equation}
which are dealt to each $\sigma$.
The other bilinear equations are proved in the same way with the
different choice of common and distinct columns as shown below.
\begin{center}
\begin{tabular}{|c|c|c|}
\hline
Equation & common columns & distinct columns \\
\hline
\eqref{eqn:appendix_1_5}
& {\footnotesize $-r, \cdots, N+\nu-2-r\ ;\ -s, \cdots, N-\nu-2-s$}
& {\footnotesize $-1-r, N+\nu-1-r\ ;\ -1-s, N-\nu-1-s$}\\
\hline
\eqref{eqn:appendix_1_7}
& {\footnotesize $-r, \cdots, N+\nu-2-r\ ;\ -s, \cdots, N-\nu-2-s$}
& {\footnotesize $-1-r, N+\nu-1-r, N+\nu-r\ ;\ N-\nu-1-s$}\\
\hline
\eqref{eqn:appendix_1_8} 
& {\footnotesize $-r, \cdots, N+\nu-2-r\ ;\ 1-s, \cdots, N-\nu-1-s$}
& {\footnotesize $-1-r, N+\nu-1-r, N+\nu-r\ ;\ -s$}\\
\hline
\eqref{eqn:appendix_1_9}
& {\footnotesize $1-r, \cdots, N+\nu-2-r\ ;\ -s, \cdots, N-\nu-1-s$}
& {\footnotesize $-1-r, -r, N+\nu-1-r, N+\nu-r\ ;$}\\
\hline
\end{tabular} 
\end{center}
By taking
\begin{equation}\label{eqn:appendix_reduction_continuous}
p_{N+i}=-p_i^*,\quad
\zeta_{N+i,0}=-\zeta_{i0}^*,\quad
e^{\eta_i}=-1,\quad
e^{\eta_{N+i}}=1,\quad
\end{equation}
for $1\le i\le N$, $\sigma$ defined by
\eqref{eqn:appendix_1_2} and \eqref{eqn:appendix_1_3}
reduces to $\tau$ in \eqref{eqn:continuous_tau} up to gauge, i.e.,
\begin{equation}\label{eqn:appendix_reduction_continuous_tau}
\tau(\nu,r)=\sigma(\nu,r,0)\, (-1)^{r\nu}\prod_{i=1}^N
(-p_i^*)^r\,e^{\zeta_i^*}.
\end{equation}
Thus with the help of reduction condition \eqref{eqn:appendix_1_4}, the bilinear equations
\eqref{eqn:bl1}--\eqref{eqn:bl3} and \eqref{eqn:bl5} are derived from
\eqref{eqn:appendix_1_5}--\eqref{eqn:appendix_1_8}, respectively, and \eqref{eqn:bl6} is derived
from \eqref{eqn:appendix_1_5} and \eqref{eqn:appendix_1_9}. \qed

Proposition \ref{prop:discrete_tau_formula} can be proved in the same manner as above.  We observe
that $\tau_n(\nu,r)$ defined by \eqref{eqn:discrete_tau} and \eqref{eqn:discrete_tau_entries}
satisfies
\begin{equation}\label{eqn:appendix_2_1}
\tau_n(\nu,r)^*=(-1)^{r\nu}\tau_{n-r}(-\nu,-r),
\end{equation}
which verifies that $\tau_n(0,0)$ is real, $\tau_n(0,-1)^*=\tau_{n+1}(0,1)$,
$\tau_{n+1}(1,1)^*=-\tau_n(-1,-1)$, $\tau_n(1,0)^*=\tau_n(-1,0)$ and
$\tau_{n+1}(1,2)^*=\tau_{n-1}(-1,-2)$ in \eqref{eqn:discrete_tau_identification}.  We define a
$2N\times 2N$ determinant $\sigma_{nn'}(\nu,r,s)$ by
\begin{equation}\label{eqn:appendix_2_2}
\hskip-15pt 
\sigma_{nn'}(\nu,r,s)=\left|\begin{matrix}
\phi_1^{(n)}(r) &\phi_1^{(n+1)}(r)
 &\cdots &\phi_1^{(n+N+\nu-1)}(r) &\vbl{5}
 &\psi_1^{(n')}(s) &\psi_1^{(n'+1)}(s) &\cdots &\psi_1^{(n'+N-\nu-1)}(s)
\\
\phi_2^{(n)}(r) &\phi_2^{(n+1)}(r)
 &\cdots &\phi_2^{(n+N+\nu-1)}(r) &\vbl{5}
 &\psi_2^{(n')}(s) &\psi_2^{(n'+1)}(s) &\cdots &\psi_2^{(n'+N-\nu-1)}(s)
\\
\vdots &\vdots &&\vdots &\vbl{6} &\vdots &\vdots &&\vdots
\\
\phi_{2N}^{(n)}(r) &\phi_{2N}^{(n+1)}(r)
 &\cdots &\phi_{2N}^{(n+N+\nu-1)}(r) &\vbl{5}
 &\psi_{2N}^{(n')}(s) &\psi_{2N}^{(n'+1)}(s) &\cdots &\psi_{2N}^{(n'+N-\nu-1)}(s)
\end{matrix}\right|,
\end{equation}
where
\begin{equation}\label{eqn:appendix_2_3}
\phi_i^{(n)}(r)=p_i^n\,(1-p_i)^{-r}\,e^{\zeta_i},\quad  
\psi_i^{(n')}(s)=\bigg(\frac{1}{p_i}\bigg)^{n'}\,(1-p_i)^{-s}\,e^{\eta_i},\quad
\zeta_i=\frac{p_i+1}{p_i-1}\frac{z}{2}+\zeta_{i0},
\end{equation}
and $\eta_i$ are constants. Then $\sigma_{nn'}(\nu,r,s)$ satisfies
\begin{equation}\label{eqn:appendix_2_4}
\sigma_{n+1,n'-1}(\nu,r,s)=\sigma_{nn'}(\nu,r,s)\prod_{i=1}^{2N}p_i,\quad
\sigma_{nn'}(\nu,r-1,s-1)=\sigma_{nn'}(\nu,r,s)\prod_{i=1}^{2N}(1-p_i).
\end{equation}

The following bilinear equations can be proved by the same technique as Proposition
\ref{prop:continuous_tau_formula}:
\begin{align}
& \sigma_{n+1,n'}(\nu,r+1,s)\,\sigma_{nn'}(\nu,r,s+1)
- \sigma_{n+1,n'}(\nu,r+1,s+1)\,\sigma_{nn'}(\nu,r,s) \nonumber\\
& = \sigma_{n,n'+1}(\nu+1,r+1,s)\,\sigma_{n+1,n'-1}(\nu-1,r,s+1),
\label{eqn:appendix_2_5}\\[1mm]
& \sigma_{n+1,n'+1}(\nu,r+1,s)\,\sigma_{nn'}(\nu,r,s)
- \sigma_{n+1,n'}(\nu,r+1,s)\,\sigma_{n,n'+1}(\nu,r,s) \nonumber\\
& = \sigma_{n,n'+1}(\nu+1,r+1,s)\,\sigma_{n+1,n'}(\nu-1,r,s),
\label{eqn:appendix_2_6}\\[1mm]
& \sigma_{nn'}(\nu+1,r+1,s)\,\sigma_{n+1,n'}(\nu,r,s)
- \sigma_{n+1,n'}(\nu+1,r+1,s)\,\sigma_{nn'}(\nu,r,s) \nonumber\\
& = \sigma_{nn'}(\nu+1,r,s)\,\sigma_{n+1,n'}(\nu,r+1,s),
\label{eqn:appendix_2_7}\\[1mm]
& \sigma_{nn'}(\nu+1,r+1,s)\,\sigma_{nn'}(\nu,r,s+1)
- \sigma_{nn'}(\nu+1,r+1,s+1)\,\sigma_{nn'}(\nu,r,s) \nonumber\\
& = \sigma_{n,n'+1}(\nu+1,r+1,s)\,\sigma_{n,n'-1}(\nu,r,s+1),
\label{eqn:appendix_2_8}\\[1mm]
& D_z\, \sigma_{n,n'+1}(\nu,r,s)\cdot\sigma_{nn'}(\nu,r,s)
= -\sigma_{n,n'+1}(\nu+1,r+1,s)\,\sigma_{nn'}(\nu-1,r-1,s).
\label{eqn:appendix_2_9}
\end{align}
We denote columns of determinant as
\begin{equation}\label{eqn:sd_FN}
\sigma_{nn'}(\nu,r,s)=\left|
\ \col{n}{r}\ \ \col{n+1}{r}\ \ \cdots\ \col{n+N+\nu-1}{r}\ ;
\ \col{n'}{s}\ \ \col{n'+1}{s}\ \ \cdots\ \col{n'+N-\nu-1}{s}\ \right|,
\end{equation}
where $\col{n}{r}$ in the left block means the column vector
$^t(\phi_1^{(n)}(r), \phi_2^{(n)}(r), \cdots, \phi_{2N}^{(n)}(r))$
and $\col{n'}{s}$ in the right block means
$^t(\psi_1^{(n')}(s), \psi_2^{(n')}(s), \cdots, \psi_{2N}^{(n')}(s))$.
In this notation, we have the expressions
\begin{align}
& \sigma_{nn'}(\nu,r+1,s)=\left|
\ \col{n}{r}\ \ \cdots\ \col{n+N+\nu-2}{r}\ \ \col{n+N+\nu-1}{r+1}\ ;
\ \col{n'}{s}\ \ \col{n'+1}{s}\ \ \cdots\ \col{n'+N-\nu-1}{s}\ \right| \\
&\phantom{\sigma_{nn'}(\nu,r+1,s)}= \left|
\ \col{n}{r}\ \ \cdots\ \col{n+N+\nu-2}{r}\ \ \col{n+N+\nu-2}{r+1}\ ;
\ \col{n'}{s}\ \ \col{n'+1}{s}\ \ \cdots\ \col{n'+N-\nu-1}{s}\ \right|, \\
& \sigma_{nn'}(\nu,r,s+1)=\left|
\ \col{n}{r}\ \ \col{n+1}{r}\ \ \cdots\ \col{n+N+\nu-1}{r}\ ;
\ \col{n'}{s+1}\ \ \col{n'+1}{s}\ \ \cdots\ \col{n'+N-\nu-1}{s}\ \right| \\
&\phantom{\sigma_{nn'}(\nu,r,s+1)}=\left|
\ \col{n}{r}\ \ \col{n+1}{r}\ \ \cdots\ \col{n+N+\nu-1}{r}\ ;
\ \col{n'+1}{s+1}\ \ \col{n'+1}{s}\ \ \cdots\ \col{n'+N-\nu-1}{s}\ \right|, \\
& \sigma_{nn'}(\nu,r+1,s+1)=\left|
\ \col{n}{r}\ \ \cdots\ \col{n+N+\nu-2}{r}\ \ \col{n+N+\nu-1}{r+1}\ ;
\ \col{n'}{s+1}\ \ \col{n'+1}{s}\ \ \cdots\ \col{n'+N-\nu-1}{s}\ \right|.
\end{align}
For example \eqref{eqn:appendix_2_5} is proved by applying
the Laplace expansion to the $4N\times 4N$ vanishing determinant
in the left-hand side of the following identity,
\begin{equation}
\left|\begin{matrix}
\col{n+1}{r} \cdots \col{n+N+\nu-1}{r} &\vbl{5}  &\col{n+N+\nu}{r+1}
& ; 
& \col{n'}{s}&\vbl{5} &\col{n'+1}{s} \cdots \col{n'+N-\nu-1}{s} &\vbl{5} &\col{n}{r} &\vbl{5} & \hbox{ \O}
\\[2mm]
\multispan{11}\hblfil\
\\
\hbox{\O} &\vbl{5} &\col{n+N+\nu}{r+1}
& ; 
& \col{n'}{s} 
&\vbl{5} &\hbox{ \O} &\vbl{5}
&\col{n}{r} &\vbl{5} &\col{n+1}{r} \cdots \col{n+N+\nu-1}{r}\ ;\ 
\col{n'}{s+1} \col{n'+1}{s} \cdots \col{n'+N-\nu-1}{s}
\end{matrix}\right|=0.
\end{equation}
The common and distinct columns of each bilinear equation are shown below.
\begin{center}
\begin{tabular}{|c|c|c|}
\hline
Equation & common columns & distinct columns \\
\hline
\eqref{eqn:appendix_2_5} & $\col{n+1}{r}, \cdots, \col{n+N+\nu-1}{r}\ ;\ \col{n'+1}{s}, \cdots, \col{n'+N-\nu-1}{s}$ &
$\col{n}{r},\col{n+N+\nu}{r+1}\ ;\  \col{n'}{s+1},\  \col{n'}{s}$\\
\hline
\eqref{eqn:appendix_2_6} & 
$\col{n+1}{r}, \cdots, \col{n+N+\nu-1}{r}\ ;\  \col{n'+1}{s}, \cdots, \col{n'+N-\nu-1}{s}$ & 
$\col{n}{r}, \col{n+N+\nu}{r+1}\ ;\ \col{n'}{s}, \col{n'+N-\nu}{s}$\\
\hline
\eqref{eqn:appendix_2_7} & 
$\col{n+1}{r}, \cdots, \col{n+N+\nu-1}{r}\ ;\ \col{n'}{s}, \cdots, \col{n'+N-\nu-2}{s}$
& $\col{n}{r}, \col{n+N+\nu}{r}, \col{n+N+\nu}{r+1}\ ;\ \col{n'+N-\nu-1}{s}$\\
\hline
\eqref{eqn:appendix_2_8} 
& $\col{n}{r}, \cdots, \col{n+N+\nu-1}{r}\ ;\ \col{n'+1}{s}, \cdots, \col{n'+N-\nu-2}{s}$
& $\col{n+N+\nu}{r+1}\ ; \col{n'}{s+1}, \col{n'}{s}, \col{n'+N-\nu-1}{s}$\\
\hline
\eqref{eqn:appendix_2_9}
& $\col{n}{r-N-\nu+1}, \cdots, \col{n}{r-1}; \col{n'+1}{s}, \cdots, \col{n'+N-\nu-1}{s}$
& $\col{n}{r}, \col{n}{r+1}; \col{n'}{s}, \col{n'+N-\nu}{s}$\\
\hline
\end{tabular} 
\end{center}
Note that we use the following formulas for proving \eqref{eqn:appendix_2_9}:
\begin{align}
& \sigma_{nn'}(\nu,r,s)=\left|
\ \col{n}{r-N-\nu+1}\ \ \cdots\ \col{n}{r-1}\ \ \col{n}{r}\ ;
\ \col{n'}{s}\ \ \col{n'+1}{s}\ \ \cdots\ \col{n'+N-\nu-1}{s}\ \right|, \\
& \bigg(-\frac{\partial}{\partial z}+\frac{N+\nu}{2}\bigg)\,\sigma_{nn'}(\nu,r,s)=\left|
\ \col{n}{r-N-\nu+1}\ \ \cdots\ \col{n}{r-1}\ \ \col{n}{r+1}\ ;
\ \col{n'}{s}\ \ \col{n'+1}{s}\ \ \cdots\ \col{n'+N-\nu-1}{s}\ \right|. 
\end{align}
By taking
\begin{equation}\label{eqn:sd_specialization}
p_{N+i}=\frac{1}{p_i^*},\quad
\zeta_{N+i,0}=-\zeta_{i0}^*,\quad
e^{\eta_i}=-1,\quad
e^{\eta_{N+i}}=1,\quad
\end{equation}
for $1\le i\le N$, $\sigma_{nn'}$ defined by
\eqref{eqn:appendix_2_2} and \eqref{eqn:appendix_2_3}
reduces to $\tau_n$ in \eqref{eqn:discrete_tau} up to gauge, i.e.,
\begin{equation}\label{eqn:appendix_reduction_semidiscrete_tau}
\tau_n(\nu,r)=\sigma_{n0}(\nu,r,0)\prod_{i=1}^N
(p_i^*)^n\,\bigg(1-\frac{1}{p_i^*}\bigg)^r\,e^{\zeta_i^*}.
\end{equation}
Thus with the help of reduction condition \eqref{eqn:appendix_2_4},
the discrete bilinear equations
\eqref{eqn:dbl1}--\eqref{eqn:dbl3}, \eqref{eqn:dbl7} and
\eqref{eqn:dbl8} are derived from
\eqref{eqn:appendix_2_5}--\eqref{eqn:appendix_2_9}, respectively.\qed
%
\section{Proofs of Proposition \ref{prop:time_evolution_continuous}, 
\ref{prop:time_evolution_semi-discrete} and \ref{prop:time_evolution_discrete}}\label{proofs_of_formula_deformation}
%
We first prove Proposition \ref{prop:time_evolution_continuous}.
Under the $t$ dependence introduced in \eqref{eqn:phi_continuous}, it can be shown that the $\tau$ functions
satisfy the following bilinear equations:
\begin{align}
& \sqrt{-1} D_t\,H\cdot F = D_x\,f^*\cdot g, \label{eqn:appendix_bl1}\\
& \frac{\sqrt{-1}}{2}D_zD_t\,F\cdot F = D_x\,f\cdot f^*,\label{eqn:appendix_bl2}\\
& \Big(\sqrt{-1}D_t+D_x^2\Big)\,G\cdot F = 0,\label{eqn:appendix_bl3}\\
& \frac{1}{2}D_x^2\, F\cdot F = GG^*. \label{eqn:appendix_bl4}
\end{align}
Equations \eqref{eqn:NLS} and \eqref{eqn:continuous_binormal_flow} are verified directly 
from \eqref{eqn:appendix_bl1} -- \eqref{eqn:appendix_bl4} by using 
\eqref{eqn:B_alternative}, \eqref{eqn:gamma_by_tau} and \eqref{eqn:kappa_formula_continuous}.
For example, dividing \eqref{eqn:appendix_bl1} by $\sqrt{-1} F^2$ we have
\begin{equation}\label{eqn:appendix_n1}
\frac{\partial}{\partial t} \left(\frac{H}{F}\right) 
= \kappa\,\frac{1}{2|G|F}\,\frac{1}{\sqrt{-1}}D_x\,f^*\cdot g.
\end{equation}
Taking the real part of \eqref{eqn:appendix_n1}, we obtain
\begin{equation}
 \frac{\partial}{\partial t}\left(\frac{H+H^*}{F}\right) = \kappa\,\frac{1}{2|G|F}
\frac{1}{\sqrt{-1}}D_x\,(g^*\cdot f-g\cdot f^*),
\end{equation}
which gives the first entry of \eqref{eqn:continuous_binormal_flow} by using
\eqref{eqn:B_alternative} and \eqref{eqn:gamma_by_tau}. The second entry of 
\eqref{eqn:continuous_binormal_flow} is verified in the same way. The 
third entries is obtained by dividing \eqref{eqn:appendix_bl2} by $F^2$
and using \eqref{eqn:bl7}. Note that \eqref{eqn:bl7} can be derived from
\eqref{eqn:bl1}--\eqref{eqn:bl3}. NLS \eqref{eqn:NLS} follows from
\eqref{eqn:appendix_bl3} and \eqref{eqn:appendix_bl4}.

We consider the determinant $\sigma$ given in \eqref{eqn:appendix_1_2} with $\zeta_i$ in 
\eqref{eqn:phi_continuous}. We also introduce $y$ dependence in $\eta_i$ as
\begin{equation}
 \eta_i = -p_i y.
\end{equation}
Then we see that $\sigma$ satisfies
\begin{equation}\label{eqn:appendix_reduction_y}
\Big(\frac{\partial}{\partial x} - \frac{\partial}{\partial y}\Big)\,\sigma(\nu,r,s) = 
\left(\sum_{i=1}^{2N}p_i\right)\,\sigma(\nu, r, s).
\end{equation}
The bilinear equations \eqref{eqn:appendix_bl1}--\eqref{eqn:appendix_bl4} are derived from the
following bilinear equations for $\sigma$ through the conditions
\eqref{eqn:appendix_reduction_continuous}, \eqref{eqn:appendix_reduction_continuous_tau} and the
identification given in \eqref{eqn:continuous_tau_identification}:
\begin{align}
& \sqrt{-1} D_t\, \sigma(\nu+1,r+1,s-1)\cdot\sigma(\nu,r,s)
= D_x\,\sigma(\nu,r+1,s)\cdot\sigma(\nu+1,r,s-1),
\label{eqn:appendix_bl6}\\
& \frac{\sqrt{-1}}{2}D_zD_t\, \sigma(\nu,r,s)\cdot\sigma(\nu,r,s)
= D_x\,\sigma(\nu,r-1,s)\cdot\sigma(\nu,r+1,s),
\label{eqn:appendix_bl7}\\
& \Big(\sqrt{-1}D_t+D_x^2\Big)\, \sigma(\nu+1,r,s)\cdot\sigma(\nu,r,s)
= 0,
\label{eqn:appendix_bl8}\\
& \frac{1}{2}D_x^2\, \sigma(\nu,r,s)\cdot\sigma(\nu,r,s)
= \sigma(\nu+1,r,s)\,\sigma(\nu-1,r,s).
\label{eqn:appendix_bl9} 
 \end{align}
Here, \eqref{eqn:appendix_bl9} is equivalent to 
\begin{equation}
 \frac{1}{2}D_xD_y\, \sigma(\nu,r,s)\cdot\sigma(\nu,r,s)
= \sigma(\nu+1,r,s)\,\sigma(\nu-1,r,s), \label{eqn:appendix_bl10} 
\end{equation}
because of \eqref{eqn:appendix_reduction_y}.

The bilinear equations \eqref{eqn:appendix_bl6}--\eqref{eqn:appendix_bl8} and
\eqref{eqn:appendix_bl10} are obtained by the technique of Laplace expansion by choosing the common
and distinct columns as follows.
\begin{center}
\begin{tabular}{|c|c|c|}
\hline
& common columns & distinct columns \\
\hline
\eqref{eqn:appendix_bl6}
& \scalebox{0.8}{$-r, \cdots, N+\nu-2-r\ ;\ 1-s, \cdots, N-\nu-1-s$}
& \scalebox{0.8}{$-1-r, N+\nu-1-r, N+\nu+1-r\ ;\ -s$}\\
\hline
\eqref{eqn:appendix_bl7}
& \scalebox{0.8}{$1-r, \cdots, N+\nu-2-r\ ;\ -s, \cdots, N-\nu-1-s$}
& \scalebox{0.8}{$-1-r, -r, N+\nu-1-r, N+\nu+1-r\ ;$}\\
\hline
\eqref{eqn:appendix_bl8} 
& \scalebox{0.8}{$-r, \cdots, N+\nu-2-r\ ;\ -s, \cdots, N-\nu-2-s$}
& \scalebox{0.8}{$N+\nu-1-r, N+\nu-r, N+\nu+1-r\ ;\ N-\nu-1-s$}\\
\hline
\eqref{eqn:appendix_bl10}
& \scalebox{0.8}{$-r, \cdots, N+\nu-2-r\ ;\ -s, \cdots, N-\nu-2-s$}
& \scalebox{0.8}{$N+\nu-1-r, N+\nu-r\ ;N-\nu-1-s, N-\nu-s$}\\
\hline
\end{tabular} 
\end{center}

We remark that some of the bilinear equations are not directly obtained from the Laplace expansion
of the determinant specified by the above table. For example, the Laplace expansion 
for \eqref{eqn:appendix_bl6} in the table gives
\begin{displaymath}
\begin{split}
& \Big(\sqrt{-1} \frac{\partial}{\partial t}-\frac{\partial^2}{\partial x^2}\Big)\, 
\sigma(\nu+1,r+1,s-1)\ \sigma(\nu,r,s) 
- \sigma(\nu+1,r+1,s-1)\ \Big(\sqrt{-1} 
\frac{\partial}{\partial t}-\frac{\partial^2}{\partial x^2}\Big)\, \sigma(\nu,r,s) \\
& =-2 \sigma(\nu,r+1,s)\ \frac{\partial}{\partial x}\sigma(\nu+1,r,s-1).
\end{split}
\end{displaymath}
We obtain \eqref{eqn:appendix_bl6} by adding $x$-derivative of \eqref{eqn:appendix_1_8} to the above
equation. Similarly, \eqref{eqn:appendix_bl7} is obtained from the Laplace expansion and
$x$-derivative of \eqref{eqn:appendix_1_9}. This completes the proof of Proposition
\ref{prop:time_evolution_continuous}.\qed

We next prove Proposition \ref{prop:time_evolution_semi-discrete}. First, we comment that the gauge
factor $A$ in \eqref{eqn:sdNLS_tau_identification} does not change the statement of Proposition
\ref{prop:discrete_tau_formula}. We consider the determinant $\sigma_{nn'}$ given in
\eqref{eqn:appendix_2_2} with $\zeta_i$ in \eqref{eqn:phi_semi-discrete}. Then we obtain the
following bilinear equations from the Laplace expansions specified in the table shown below:
\begin{align}
& \Big( \sqrt{-1} \epsilon^2 D_t + 2\Big)\,\sigma_{n,n'+1}(\nu+1,r+1,s)
\cdot \sigma_{nn'}(\nu,r,s+1) \nonumber\\
& = \sigma_{nn'}(\nu,r+1,s+1)\,\sigma_{n,n'+1}(\nu+1,r,s)
- \sigma_{n+1,n'}(\nu,r+1,s+1)\,\sigma_{n-1,n'+1}(\nu+1,r,s),
\label{eqn:appendix_sdbl5}\\
& \frac{\sqrt{-1}}{2}\epsilon^2 D_zD_t
\, \sigma_{nn'}(\nu,r,s)\cdot\sigma_{nn'}(\nu,r,s)\nonumber\\
& = \sigma_{nn'}(\nu,r-1,s)\,\sigma_{nn'}(\nu,r+1,s)
- \sigma_{n-1,n'}(\nu,r-1,s)\,\sigma_{n+1,n'}(\nu,r+1,s),
\label{eqn:appendix_sdbl6}\\
& \sqrt{-1}\epsilon^2 D_t \, \sigma_{nn'}(\nu+1,r,s)\cdot\sigma_{nn'}(\nu,r,s)
\nonumber\\
& + \sigma_{n+1,n'}(\nu+1,r,s)\, \sigma_{n-1,n'}(\nu,r,s) 
+ \sigma_{n-1,n'}(\nu+1,r,s)\, \sigma_{n+1,n'}(\nu,r,s)=0,
\label{eqn:appendix_sdbl7}\\
& \sigma_{n+1,n'}(\nu,r,s)\, \sigma_{n,n'-1}(\nu,r,s)
- \sigma_{n+1,n'-1}(\nu,r,s)\, \sigma_{nn'}(\nu,r,s)\nonumber\\
&= \sigma_{nn'}(\nu+1,r,s)\,\sigma_{n+1,n'-1}(\nu-1,r,s).
\label{eqn:appendix_sdbl8} 
 \end{align}
\begin{center}
\begin{tabular}{|c|c|c|}
\hline
Equation & common columns & distinct columns \\
\hline
\eqref{eqn:appendix_sdbl5} & $\col{n}{r}, \cdots, \col{n+N+\nu-2}{r}\ ;\ \col{n'+1}{s}, \cdots, \col{n'+N-\nu-1}{s}$ &
$\col{n+N+\nu-1}{r},\col{n+N+\nu}{r}\ ,\col{n+N+\nu-1}{r+1}\ ;\  \col{n'}{s+1}$\\
\cline{2-3}
 & $\col{n+1}{r}, \cdots, \col{n+N+\nu-1}{r}\ ;\ \col{n'+1}{s}, \cdots, \col{n'+N-\nu-1}{s}$ &
$\col{n-1}{r},\col{n}{r},\col{n+N+\nu}{r+1}\ ;\  \col{n'}{s+1}$\\
\hline
\eqref{eqn:appendix_sdbl6} & 
$\col{n}{r-N-\nu+2}, \cdots, \col{n}{r-1}\ ;\  \col{n'}{s}, \cdots, \col{n'+N-\nu-1}{s}$ & 
$\col{n}{r-N-\nu}, \col{n}{r-N-\nu+1}, \col{n}{r}, \col{n}{r+1}\ ;$\\
\cline{2-3}
& 
$\col{n+1}{r-N-\nu+2}, \cdots, \col{n+N+\nu-2}{r-1}\ ;\  \col{n'}{s}, \cdots, \col{n'+N-\nu-1}{s}$ & 
$\col{n-1}{r-N-\nu}, \col{n}{r-N-\nu+1}, \col{n+N+\nu-1}{r}, \col{n+N+\nu}{r+1}\ ;$\\
\hline
\eqref{eqn:appendix_sdbl7} & 
$\col{n}{r}, \cdots, \col{n+N+\nu-2}{r}\ ;\ \col{n'+1}{s}, \cdots, \col{n'+N-\nu-1}{s}$
& $\col{n-1}{r}, \col{n+N+\nu-1}{r}, \col{n+N+\nu}{r}\ ;\ \col{n'}{s}$\\
\cline{2-3}
& 
$\col{n+1}{r}, \cdots, \col{n+N+\nu-1}{r}\ ;\ \col{n'}{s}, \cdots, \col{n'+N-\nu-2}{s}$
& $\col{n-1}{r}, \col{n}{r}, \col{n+N+\nu}{r}\ ;\ \col{n'+N-\nu-1}{s}$\\
\hline
\eqref{eqn:appendix_sdbl8} 
& $\col{n}{r}, \cdots, \col{n+N+\nu-1}{r}\ ;\ \col{n'+1}{s}, \cdots, \col{n'+N-\nu-2}{s}$
& $\col{n+N+\nu}{r+1}\ ; \col{n'}{s+1}, \col{n'}{s}, \col{n'+N-\nu-1}{s}$\\
\hline
\end{tabular} 
\end{center}
Note that for each of \eqref{eqn:appendix_sdbl5}--\eqref{eqn:appendix_sdbl7} we need to consider two
Laplace expansions since the time evolution corresponds to superposition of a positive and a
negative KP flows.  By using \eqref{eqn:appendix_2_4} and
\eqref{eqn:appendix_reduction_semidiscrete_tau} with \eqref{eqn:sdNLS_tau_identification}, we obtain
the following bilinear equations from \eqref{eqn:appendix_sdbl5}--\eqref{eqn:appendix_sdbl8},
respectively:
\begin{align}
& \sqrt{-1}\epsilon^2 D_t\,H_n\cdot F_n = f_n^* g_{n-1} - f_{n-1}^* g_{n}, 
\label{eqn:appendix_sdbl1}\\
& \frac{\sqrt{-1}}{2}\epsilon^2 D_zD_t\,F_n\cdot F_n = f_n f_{n-1}^* - f_{n-1} f_{n}^*,
\label{eqn:appendix_sdbl2}\\
& \Big(\sqrt{-1}\epsilon^2 D_t-2\Big)\,G_n\cdot F_n + G_{n+1}F_{n-1} + G_{n-1}F_{n+1}=0,\label{eqn:appendix_sdbl3}\\
& F_{n+1}F_{n-1} - F_n^2 = G_nG_n^*. \label{eqn:appendix_sdbl4}
\end{align}
Then \eqref{eqn:appendix_sdbl1} and \eqref{eqn:appendix_sdbl2} gives
\eqref{eqn:discrete_binormal_flow}, and \eqref{eqn:appendix_sdbl3} and \eqref{eqn:appendix_sdbl4}
gives \eqref{eqn:sdNLS}, respectively. This completes the proof of Proposition
\ref{prop:time_evolution_semi-discrete}. \qed

We finally prove Proposition \ref{prop:time_evolution_discrete}.  First it can be seen that although
the factors $A^m$ and $B^m$ are introduced in \eqref{eqn:dNLS_tau_identification}, Proposition
\ref{prop:discrete_tau_formula} is still valid for the $\tau$ functions
\eqref{eqn:dNLS_tau_identification} with the gauge.  By using \eqref{eqn:u_discrete},
\eqref{eqn:T_tau}--\eqref{eqn:B_tau}, \eqref{eqn:v_continuous} and
\eqref{eqn:kappa_formula_discrete}, we have
\begin{equation}\label{eqn:discrete_alternative_N2}
\left(\tan\frac{\kappa_n^m}{2}\right)N_n^m - T_n^m = -\frac{1}{2(F_n^m)^2}
\begin{bmatrix}
g_n^{m\, *}f_{n-1}^m + f_n^mg_{n-1}^{m\, *} + g_n^mf_{n-1}^{m\,*} + f_n^{m\,*}g_{n-1}^m \\[1mm]
(- g_n^{m\,*}f_{n-1}^m - f_n^mg_{n-1}^{m\,*} + g_n^mf_{n-1}^{m\,*} + f_n^{m\,*}g_{n-1}^m)/\sqrt{-1} \\[1mm]
f_n^{m\,*}f_{n-1}^m + f_n^mf_{n-1}^{m\,*} - g_n^{m\,*}g_{n-1}^m - g_n^mg_{n-1}^{m\,*}
\end{bmatrix},
\end{equation}
\begin{equation}
e^{\sqrt{-1}\Lambda_n^m}(N_n^m + \sqrt{-1}B_n^m) 
= \frac{1}{F_{n+1}^mF_n^m}
\begin{bmatrix}
(f_n^m)^2-(g_n^m)^2 \\[1mm]
-\Big((f_n^m)^2+(g_n^m)^2\Big)/\sqrt{-1}\\[1mm]
-2f_n^mg_n^m
\end{bmatrix},
\end{equation}
and \eqref{eqn:discrete_def_Frenet} is rewritten in terms of $\tau$ functions.  The first and second
entries of \eqref{eqn:discrete_def_Frenet} are equivalent to
\begin{equation}\label{eqn:discrete_1st_2nd_component}
\begin{split}
 \frac{H_n^{m+1}}{F_n^{m+1}} - \frac{H_n^m}{F_n^m} = &\frac{\frac{\delta^2}{\epsilon^4}}{(F_n^m)^2}
\left[\vphantom{\frac{\Gamma_\infty}{F_n^{m+1}}}
  \left({\scriptstyle \sqrt{-1}}\tfrac{\epsilon^2}{\delta} - 1\right) g_n^m f_{n-1}^{m\,*}
- \left({\scriptstyle\sqrt{-1}}\tfrac{\epsilon^2}{\delta} + 1\right) f_n^{m\,*} g_{n-1}^{m}\right.\\
& \left.
+ \frac{\Gamma_\infty}{F_n^{m+1}}\left(2 f_n^{m\,*} g_n^m F_{n-1}^{m+1} - (f_n^{m\,*})^2 G_{n-1}^{m+1} + (g_n^m)^2 G_{n-1}^{m+1\,*}\right)
\right] ,
\end{split}
\end{equation}
and its complex conjugate. From the third entry we get
\begin{align}\label{eqn:discrete_3rd_component}
&\frac{2\delta^2}{\epsilon^4}(\Gamma_\infty - 1) - 2\Big(\log \frac{F_{n}^{m+1}}{F_n^m}\Big)_z \nonumber\\
= & \frac{\frac{\delta^2}{\epsilon^4}}{(F_n^m)^2}
\left[\vphantom{\frac{2\Gamma_\infty}{F_n^{m+1}}}
  \left({\scriptstyle \sqrt{-1}}\tfrac{\epsilon^2}{\delta} - 1\right) (f_n^m f_{n-1}^{m\,*} - g_n^m g_{n-1}^{m\,*})
- \left({\scriptstyle \sqrt{-1}}\tfrac{\epsilon^2}{\delta} + 1\right) (f_n^{m\,*} f_{n-1}^{m} - g_n^{m\,*} g_{n-1}^{m})\right.\nonumber\\
& \left.
+ \frac{2\Gamma_\infty}{F_n^{m+1}}\left((f_n^m f_n^{m\,*} - g_n^m g_n^{m\,*}) F_{n-1}^{m+1} + f_n^{m\,*} g_n^{m\,*} G_{n-1}^{m+1} 
+ f_n^m g_n^m G_{n-1}^{m+1\,*}\right)
\right].
\end{align}
These equations and dNLS \eqref{eqn:dNLS} are bilinearized into the following form:
\begin{align}
& {\textstyle {\scriptstyle \sqrt{-1}}\frac{\epsilon^2}{\delta}}
\Big(H_n^{m+1} F_n^m-H_n^m F_n^{m+1}\Big)
= \sqrt{\Gamma_{\infty}}\,
\Big( f_n^{m\,*} g_{n-1}^{m+1} - f_{n-1}^{m+1\,*} g_{n}^{m}\Big),
 \label{eqn:appendix_dbl1}\\[1mm]
& {\textstyle \left(1 + {\scriptstyle \sqrt{-1}}\frac{\delta}{\epsilon^2}
(\Gamma_{\infty}-1) - {\scriptstyle \sqrt{-1}}\frac{\epsilon^2}{\delta}D_z
\right)}\,F_n^{m+1}\cdot F_n^m
= \sqrt{\Gamma_{\infty}}\Big(f_{n-1}^{m+1} f_n^{m\,*}
+ g_n^m g_{n-1}^{m+1\,*}\Big),\label{eqn:appendix_dbl2}\\[1mm]
& {\textstyle \left({\scriptstyle  \sqrt{-1}}\frac{\epsilon^2}{\delta} - 1\right
)}\, G_{n}^{m+1}F_{n}^{m}
- {\textstyle \left({\scriptstyle\sqrt{-1}}\frac{\epsilon^2}{\delta} + 1\right)}
\, G_{n}^m F_{n}^{m+1}
+ \Gamma_{\infty}\Big(G_{n+1}^{m} F_{n-1}^{m+1} + G_{n-1}^{m+1} F_{n+1}^{m}\Big)
 = 0,\label{eqn:appendix_dbl3}\\[1mm]
& {\textstyle {\scriptstyle \sqrt{-1}}\frac{\epsilon^2}{\delta}}
\sqrt{\Gamma_{\infty}}\,g_{n-1}^{m+1} F_n^m
- {\textstyle \left({\scriptstyle\sqrt{-1}}\frac{\epsilon^2}{\delta} + 1\right)}\,
g_{n-1}^{m} F_{n}^{m+1}
+ \Gamma_{\infty}\,\Big(g_n^m F_{n-1}^{m+1} - f_n^{m\,*} G_{n-1}^{m+1}\Big) = 0,
\label{eqn:appendix_dbl4}\\[1mm]
& {\textstyle {\scriptstyle \sqrt{-1}}\frac{\epsilon^2}{\delta}}
\sqrt{\Gamma_{\infty}}\,f_{n-1}^{m+1} F_n^m
- {\textstyle \left({\scriptstyle\sqrt{-1}}\frac{\epsilon^2}{\delta} + 1\right)}\,
f_{n-1}^{m} F_{n}^{m+1}
+ \Gamma_{\infty}\,\Big(f_n^m F_{n-1}^{m+1} + g_n^{m\,*} G_{n-1}^{m+1}\Big) = 0,
\label{eqn:appendix_dbl5}\\[1mm]
& F_{n+1}^{m}F_{n-1}^{m} - (F_{n}^{m})^2 = G_{n}^{m}G_{n}^{m\,*}, \label{eqn:appendix_dbl6}
\end{align}
where $\Gamma_\infty$ is either $1$ or $1+\epsilon^4/\delta^2$.  Actually by using
\eqref{eqn:appendix_dbl4} and complex conjugate of \eqref{eqn:appendix_dbl5}, we can eliminate
$g_{n-1}^{m+1}$ and $f_{n-1}^{m+1\,*}$ from \eqref{eqn:appendix_dbl1} and obtain
\eqref{eqn:discrete_1st_2nd_component}.  Similarly we take imaginary part of
\eqref{eqn:appendix_dbl2} and eliminate $f_{n-1}^{m+1}$ and $g_{n-1}^{m+1}$ by using
\eqref{eqn:appendix_dbl4} and \eqref{eqn:appendix_dbl5}.  Then \eqref{eqn:discrete_3rd_component} is
obtained in a straightforward way.  Finally dNLS \eqref{eqn:dNLS} follows from
\eqref{eqn:appendix_dbl3} and \eqref{eqn:appendix_dbl6} through the variable transformation
\eqref{eqn:variable_transformation_dNLS}.  Here we comment that the equations
\eqref{eqn:appendix_dbl1}--\eqref{eqn:appendix_dbl6} seem over-determining, however they are indeed
compatible.

In the following we prove that the $\tau$ functions in Proposition
\ref{prop:time_evolution_discrete} actually satisfy the bilinear equations
\eqref{eqn:appendix_dbl1}--\eqref{eqn:appendix_dbl6}.  We note that \eqref{eqn:appendix_dbl6} is the
same as \eqref{eqn:appendix_sdbl4} which holds irrespective of the time evolutions.  Now let us
define a $2N\times 2N$ determinant $\sigma_{nn'}^{mm'}(\nu,r,s)$ by
\begin{equation}
\setlength{\arraycolsep}{4pt}
\begin{split}
&\sigma_{nn'}^{mm'}(\nu,r,s)=\\
& \left|\begin{matrix}
\phi_1^{(n,m)}(r) &\phi_1^{(n+1,m)}(r)
 &\cdots &\phi_1^{(n+N+\nu-1,m)}(r) &\vbl{5}
 &\psi_1^{(n',m')}(s) &\psi_1^{(n'+1,m')}(s) &\cdots &\psi_1^{(n'+N-\nu-1,m')}(s)
\\
\phi_2^{(n,m)}(r) &\phi_2^{(n+1,m)}(r)
 &\cdots &\phi_2^{(n+N+\nu-1,m)}(r) &\vbl{5}
 &\psi_2^{(n',m')}(s) &\psi_2^{(n'+1,m')}(s) &\cdots &\psi_2^{(n'+N-\nu-1,m')}(s)
\\
\vdots &\vdots &&\vdots &\vbl{6} &\vdots &\vdots &&\vdots
\\
\phi_{2N}^{(n,m)}(r) &\phi_{2N}^{(n+1,m)}(r)
 &\cdots &\phi_{2N}^{(n+N+\nu-1,m)}(r) &\vbl{5}
 &\psi_{2N}^{(n',m')}(s) &\psi_{2N}^{(n'+1,m')}(s) &\cdots &\psi_{2N}^{(n'+N-\nu-1,m')}(s)
\end{matrix}\right|, 
\end{split}
\end{equation}
where
\begin{equation}
\begin{split}
&\phi_i^{(n,m)}(r)=p_i^n\,\Big(1-\frac{a^*}{p_i}\Big)^{-m}\,(1-p_i)^{-r}\,e^{\zeta_i},\quad  
\zeta_i=\frac{p_i+1}{p_i-1}\frac{z}{2}+\zeta_{i0}, \\
& \psi_i^{(n',m')}(s)=\bigg(\frac{1}{p_i}\bigg)^{n'}\,(1-ap_i)^{-m'}\,(1-p_i)^{-s}\,e^{\eta_i},
\end{split}
\end{equation}
and $\eta_i$ are constants. Similarly to \eqref{eqn:sd_FN}, we simply denote 
\begin{equation}
\sigma_{nn'}^{mm'}(\nu,r,s)=\left|
\ \colu{n}{m}{r} \colu{n+1}{m}{r} \cdots\ \colu{n+N+\nu-1}{m}{r}\ ;
\ \colu{n'}{m'}{s} \colu{n'+1}{m'}{s} \cdots \colu{n'+N-\nu-1}{m'}{s} \right|.
\end{equation}
Then we have the following difference and differential formulas:
\begin{align}
&(1-a^*)\,\sigma_{n,n'+1}^{m+1,m'+1}(\nu+1,r+1,s)
=\left|
\colu{n}{m+1}{r} \colu{n+1}{m}{r} \cdots \colu{n+N+\nu-1}{m}{r} \colu{n+N+\nu}{m}{r+1}\ ;
\ \colu{n'+1}{m'+1}{s} \colu{n'+2}{m'}{s} \cdots \colu{n'+N-\nu-1}{m'}{s} \right|,\\
& (1-a)\,\sigma_{nn'}^{m+1,m'+1}(\nu,r,s+1)=
\left|
\colu{n}{m+1}{r}\colu{n+1}{m}{r}\cdots \colu{n+N+\nu-1}{m}{r}\ ;
\ \colu{n'+1}{m'}{s+1}\colu{n'+1}{m'+1}{s} \colu{n'+2}{m'}{s} \cdots \colu{n'+N-\nu-1}{m'}{s} \right|,\\
&\scalebox{0.9}{$\displaystyle \left((a^*-1)\Big(\tfrac{\partial }{\partial z}-\tfrac{N+\nu}{2}\Big) - 1\right)\,
\sigma_{nn'}^{m+1,m'+1}(\nu,r,s)$}
=
\left|
\colu{n}{m+1}{r-1}\colu{n+1}{m}{r-N-\nu+2} \cdots \colu{n+1}{m}{r-1}\colu{n+1}{m}{r+1}\ ;
\ \colu{n'}{m'+1}{s}\colu{n'+1}{m'}{s}\cdots \colu{n'+N-\nu-1}{m'}{s} \right|,
\end{align}
and so on. By applying the Laplace expansion to the identities,
\begin{equation}
\setlength{\arraycolsep}{2.5pt}
\left|\begin{matrix}
\colu{n}{m+1}{r} \colu{n+1}{m}{r} \cdots \colu{n+N+\nu-1}{m}{r} & \vbl{5}  &\colu{n+N+\nu}{m}{r+1}
& ; 
& \colu{n'+1}{m'+1}{s} \colu{n'+2}{m'}{s} \cdots \colu{n'+N-\nu-1}{m'}{s} &\vbl{5} &\colu{n}{m}{r} &\vbl{5} & \hbox{ \O}
&\vbl{5} & \colu{n'+1}{m'}{s+1} \colu{n'+1}{m'}{s} &\vbl{5} & \hbox{ \O} \\[2mm]
\multispan{13}\hblfil\
\\
\hbox{\O} &\vbl{5} &\colu{n+N+\nu}{m}{r+1} & \vbl{5} & \hbox{\O} 
&\vbl{5} & \colu{n}{m}{r} &\vbl{5} & \colu{n+1}{m}{r}\cdots\colu{n+N+\nu-1}{m}{r}  & ;
&\colu{n'+1}{m'}{s+1} \colu{n'+1}{m'}{s} &\vbl{5} & \colu{n'+2}{m'}{s} \cdots \colu{n'+N-\nu-1}{m'}{s}
\end{matrix}\right|=0,
\end{equation}

\begin{equation}
\setlength{\arraycolsep}{2.75pt}
\left|\begin{matrix}
\colu{n}{m+1}{r-1} \colu{n+1}{m}{r-N-\nu+2} \cdots \colu{n+1}{m}{r-1} & \vbl{5}  &\colu{n+1}{m}{r+1}
& ; 
& \colu{n'}{m'+1}{s} \colu{n'+1}{m'}{s} \cdots \colu{n'+N-\nu-1}{m'}{s} &\vbl{5} &\colu{n}{m}{r-1} &\vbl{5} & \hbox{ \O}
&\vbl{5} & \colu{n+1}{m}{r} &;& \colu{n'}{m'}{s} &\vbl{5} & \hbox{ \O} \\[2mm]
\multispan{15}\hblfil\
\\
\hbox{\O} &\vbl{5} &\colu{n+1}{m}{r+1} & \vbl{5} & \hbox{\O} 
&\vbl{5} & \colu{n}{m}{r-1} &\vbl{5} & \colu{n+1}{m}{r-N-\nu+2}\cdots
\colu{n+1}{m}{r-1} &\vbl{5}
&\colu{n+1}{m}{r}
& ; &\colu{n'}{m'}{s} &\vbl{5} & \colu{n'+1}{m'}{s} \cdots
\colu{n'+N-\nu-1}{m'}{s}
\end{matrix}\right|=0,
\end{equation}

\begin{equation}\label{eqn:Laplace_expansion}
\setlength{\arraycolsep}{2.5pt}
\left|\begin{matrix}
\colu{n}{m+1}{r} \colu{n+1}{m}{r} \cdots \colu{n+N+\nu-1}{m}{r}
& ;
&\colu{n'}{m'+1}{s}
\colu{n'+1}{m'}{s} \cdots
\colu{n'+N-\nu-2}{m'}{s}
& \vbl{5}
& \colu{n'+N-\nu-1}{m'}{s}
&\vbl{5}
&\colu{n}{m}{r}
&\vbl{5}
& \hbox{ \O}
&\vbl{5}
& \colu{n+N+\nu}{m}{r}
& ;
& \colu{n'}{m'}{s}
&\vbl{5}
& \hbox{ \O} \\[2mm]
\multispan{15}\hblfil\
\\
\multicolumn{3}{c}{\hbox{\O}}
& \vbl{5}
&\colu{n'+N-\nu-1}{m'}{s}
& \vbl{5}
& \colu{n}{m}{r}
&\vbl{5}
& \colu{n+1}{m}{r}\cdots\colu{n+N+\nu-1}{m}{r} 
& \vbl{5}
&\colu{n+N+\nu}{m}{r}
& ;
& \colu{n'}{m'}{s} 
&\vbl{5} 
& \colu{n'+1}{m'}{s} \cdots \colu{n'+N-\nu-2}{m'}{s}
\end{matrix}\right|=0,
\end{equation}
we get the bilinear equations for $\sigma_{nn'}^{mm'}$, 
\begin{align}
& \scalebox{0.82}{$\displaystyle (1-a^*)\,\sigma_{n,n'+1}^{m+1,m'+1}(\nu+1,r+1,s)\,\sigma_{nn'}^{mm'}(\nu,r,s+1)
 -(1-a)\,\sigma_{n,n'+1}^{mm'}(\nu+1,r+1,s)\,\sigma_{nn'}^{m+1,m'+1}(\nu,r,s+1)$} \nonumber\\
& \scalebox{0.85}{$\displaystyle = a^*\sigma_{n+1,n'}^{mm'}(\nu,r+1,s+1)\,\sigma_{n-1,n'+1}^{m+1,m'+1}(\nu+1,r,s)
+ a\,\sigma_{nn'}^{m+1,m'+1}(\nu,r,s)\,\sigma_{n,n'+1}^{mm'}(\nu+1,r+1,s+1),$}
\label{eqn:appendix_dbl7}
\\[2mm]
& \scalebox{0.82}{$\displaystyle \Big((1-a^*)D_z+a^*\Big)
\, \sigma_{nn'}^{m+1,m'+1}(\nu,r,s)\cdot\sigma_{nn'}^{mm'}(\nu,r,s)$}\nonumber\\
& \scalebox{0.82}{$\displaystyle
= a^* \sigma_{n-1,n'}^{m+1,m'+1}(\nu,r-1,s)\,\sigma_{n+1,n'}^{mm'}(\nu,r+1,s)
+ a\, \sigma_{n,n'+1}^{mm'}(\nu+1,r+1,s)\,\sigma_{n,n'-1}^{m+1,m'+1}(\nu-1,r-1,s),$}
\label{eqn:appendix_dbl8}
\\[2mm]
& \scalebox{0.82}{$\displaystyle \sigma_{nn'}^{m+1,m'+1}(\nu+1,r,s)\,\sigma_{nn'}^{mm'}(\nu,r,s)
- \sigma_{nn'}^{mm'}(\nu+1,r,s)\,\sigma_{nn'}^{m+1,m'+1}(\nu,r,s)$}\nonumber\\
& \scalebox{0.85}{$\displaystyle + a\, \sigma_{n,n'+1}^{mm'}(\nu+1,r,s)\, \sigma_{n,n'-1}^{m+1,m'+1}(\nu,r,s)
- a^*\sigma_{n-1,n'}^{m+1,m'+1}(\nu+1,r,s)\,\sigma_{n+1,n'}^{mm'}(\nu,r,s) 
= 0,$}
\label{eqn:appendix_dbl9}
\end{align}
respectively. In \eqref{eqn:Laplace_expansion}, if we replace the column
$\colu{n+N+\nu}{m}{r}$ by $\colu{n+N+\nu}{m}{r+1}$,
we get instead of \eqref{eqn:appendix_dbl9},
\begin{align}
& \scalebox{0.82}{$\displaystyle 
(1-a^*) \sigma_{nn'}^{m+1,m'+1}(\nu+1,r+1,s)\,\sigma_{nn'}^{mm'}(\nu,r,s)
- \sigma_{nn'}^{mm'}(\nu+1,r+1,s)\,\sigma_{nn'}^{m+1,m'+1}(\nu,r,s)$}
\nonumber\\
& \scalebox{0.85}{$\displaystyle
+ a\, \sigma_{n,n'+1}^{mm'}(\nu+1,r+1,s)\, \sigma_{n,n'-1}^{m+1,m'+1}(\nu,r,s)
- a^*\sigma_{n-1,n'}^{m+1,m'+1}(\nu+1,r,s)\,\sigma_{n+1,n'}^{mm'}(\nu,r+1,s) 
= 0.$}
\label{eqn:appendix_dbl10}
\end{align}
Similarly the replacement of column $\colu{n+N+\nu}{m}{r}$ in
\eqref{eqn:Laplace_expansion} by $\colu{n'}{m'}{s+1}$ gives
\begin{align}
& \scalebox{0.82}{$\displaystyle
(1-a) \sigma_{n,n'-1}^{m+1,m'+1}(\nu,r,s+1)\,\sigma_{nn'}^{mm'}(\nu,r,s)
- \sigma_{n,n'-1}^{mm'}(\nu,r,s+1)\,\sigma_{nn'}^{m+1,m'+1}(\nu,r,s)$}
\nonumber\\
& \scalebox{0.85}{$\displaystyle
+ a\, \sigma_{nn'}^{mm'}(\nu,r,s+1)\, \sigma_{n,n'-1}^{m+1,m'+1}(\nu,r,s)
- a^*\sigma_{n-1,n'}^{m+1,m'+1}(\nu+1,r,s)\,\sigma_{n+1,n'-1}^{mm'}(\nu-1,r,s+1) 
= 0.$}
\label{eqn:appendix_dbl11}
\end{align}
Under the specialization \eqref{eqn:sd_specialization}, $\tau_n^m$ in Proposition
\ref{prop:time_evolution_discrete} is obtained by
\begin{equation}\label{eqn:appendix_reduction_discrete_tau2}
\tau_n^m(\nu,r)=\sigma_{n0}^{mm}(\nu,r,0)\prod_{i=1}^N
(p_i^*)^n\,(1-ap_i)^m\,(1-a^*p_i^*)^m\,\bigg(1-\frac{1}{p_i^*}\bigg)^r\,e^{\zeta_i^*},
\end{equation}
and \eqref{eqn:appendix_dbl1}--\eqref{eqn:appendix_dbl5} are derived from \eqref{eqn:appendix_dbl7}--\eqref{eqn:appendix_dbl11},
respectively, through the reduction condition \eqref{eqn:appendix_2_4} and the identification given in \eqref{eqn:dNLS_tau_identification}.
This completes the proof of Proposition \ref{prop:time_evolution_discrete}. \qed


\begin{thebibliography}{9}
\bibitem{Ablowitz-Ladik:SIAM1976}
M.J. Ablowitz and  J.F. Ladik, A nonlinear difference scheme and inverse scattering, Stud. in Appl. Math. {\bf 55} (1976), pp. 213--229.
\bibitem{Ablowitz-Ladik:SIAM1977}
M.J. Ablowitz and  J.F. Ladik, On the solution of a class of nonlinear partial difference equations, Stud. in Appl. Math. {\bf 57} (1977), pp. 1--12.
\bibitem{Ablowitz-Prinari-Trubach:NLS_book} M.J. Ablowitz, B. Prinari and A.D. Trubach, {\it Discrete and continuous nonlinear Schr\"odinger systems}, London Mathematical Society Lecture Note Series Vol. 302 (Cambridge University Press, Cambridge, 2004).
\bibitem{DJKM:LL} E. Date, M. Jimbo, M. Kashiwara and T. Miwa, On Landau-Lifschitz equation and infinite-dimentional groups, Infinite Dimentional Groups with Applications, ed. by V. Kac  (Springer-Verlag, New York, 1985).
\bibitem{Doliwa-Santini:JMP} A. Doliwa and P.M. Santini, Integrable dynamics of a discrete curve and the Ablowitz-Ladik hierarchy, J. Math. Phys. {\bf 36} (1995), pp.1259--1273.
\bibitem{Doliwa-Santini:dsG} A. Doliwa and P. M. Santini, The integrable dynamics of a discrete	curve,
in {\it Symmetries and Integrability of Difference Equations},  eds. by D. Levi, L. Vinet and P. Winternitz,
CRM Proceedings \& Lecture Notes Vol.9 (American Mathematical Society, Providence, RI, 1996), pp. 91--102.
\bibitem{Fukumoto_Miyazaki:JPSJ1986}
Y. Fukumoto and T. Miyazaki, N-Solitons on a curved vortex filament, J. Phys. Soc. Jpn. {\bf 55} (1986), pp. 4152--4155.
\bibitem{Hasimoto:JFM1972}
H. Hasimoto, Soliton on a vortex filament, J. Fluid Mech. {\bf 51} (1972), pp. 477--485.
\bibitem{Hirota-Ohta:PSJ_91Spring} R. Hirota and Y. Ohta, Discrete nonlinear Schr\"odinger equation, talk delivered at spring meeting of the Physical Society of Japan (1991). Abstract available online: http://ci.nii.ac.jp/naid/110001908012 (in Japanese).
\bibitem{Hisakado-Nakayama-Wadati} M. Hisakado, K. Nakayama and M. Wadati, Motion of discrete curves in the plane, J. Phys. Soc. Jpn.  {\bf 64} (1995), pp.2390--2393. 
\bibitem{Hisakado-Wadati} M. Hisakado and M. Wadati, Moving discrete curve and geometric phase, Phys. Lett. {\bf A214} (1996), pp.252--258.
\bibitem{Hoffmann:LN} T. Hoffmann, Discrete Differential Geometry of Curves and Surfaces,
COE lecture Notes Vol. 18 (Kyushu University, Fukuoka, 2009).
\bibitem{Hoffmann-Kutz} T.~Hoffmann and N.~Kutz, Discrete curves in $\mathbb{C}P^1$ and the Toda lattice, Stud. Appl. Math.  {\bf 113} (2004) 31--55.
\bibitem{Hoffmann:dNLS_HM} T. Hoffmann, On the equivalence of the discrete nonlinear Schr\"odinger equation and the 
discrete isotropic Heisenberg magnet, Phys. Lett. A {\bf 265} (2000), pp. 62--67.
\bibitem{Hoffmann:dNLS} T. Hoffmann, Discrete Hashimoto surfaces and a doubly discrete smoke-ring flow, in {\it Discrete Differential Geometry}, eds. by A.I. Bobenko, P. Schr\"oder, J.M. Sullivan and	G.M. Ziegler, Oberwolfach Seminars Vol.38 (Birkh\"auser, Basel, 2008), pp. 95--115.
\bibitem{IKMO:KJM} J. Inoguchi, K. Kajiwara, N. Matsuura and Y. Ohta, Motion and B\"acklund transformations of discrete plane curves, Kyushu J. Math. {\bf 66} (2012), pp. 303--324.
\bibitem{IKMO:JPA} J. Inoguchi, K. Kajiwara, N. Matsuura and Y. Ohta, Explicit solutions to the semi-discrete modified KdV equation and motion of discrete plane curves, J. Phys. A: Math. Theor. {\bf 45} (2012), 045206.
\bibitem{IKMO:dmKdV_space_curve} J. Inoguchi, K. Kajiwara, N. Matsuura and Y. Ohta, Discrete mKdV and discrete sine-Gordon flows on discrete space curves, J. Phys. A: Math. Theor. {\bf 47} (2014), 235202.
\bibitem{IKMO:MEIS2013}
J. Inoguchi, K. Kajiwara, N. Matsuura and Y. Ohta, Discrete models of isoperimetric deformation of plane curves, in {\it Mathematical progress in expressive image synthesis}, ed. by Ken Anjyo, Mathematics for Industry {\bf 4} (Springer, Tokyo, 2014) pp. 111--122.
\bibitem{JM:RIMS1983} M. Jimbo and T. Miwa, Solitons and infinite dimensional Lie algebras, Publ. RIMS {\bf 19} (1983), pp.943-1001.
\bibitem{Lamb:JMP1977}
G.L. Lamb, Solitons on moving space curves, moving discrete curve and geometric phase, J. Math. Phys. {\bf 18} (1977), pp. 1654--1659. 
\bibitem{Matsuura:IMRN} N. Matsuura, Discrete KdV and discrete modified KdV equations arising from motions of planar discrete curves, Int. Math. Res. Not. {\bf 2012} (2012), pp.1681--1698.
\bibitem{Nakayama:JPSJ2007} K. Nakayama, Elementary vortex filament model of the discrete nonlinear Schr\"odinger equation, J. Phys. Soc. Jpn. {\bf 76} (2007), 074003.
\bibitem{Nakayama_Segur_Wadati:PRL} K. Nakayama, H. Segur and M. Wadati, Integrability and the motions of curves, Phys. Rev. Lett. {\bf 69} (1992), pp. 2603--2606.
\bibitem{Nijhoff:discrete_LL} F.W. Nijhoff and V. Papageorgiou, Lattice equations associated with the Landau-Lifschitz equations, 
Phys. Lett. {\bf 141A} (1989) 269--274.

\bibitem{Nishinari} K. Nishinari, A discrete model of an extensible string in three-dimensional space, J. Appl. Mech. {\bf 66} (1999), pp.695--701.
\bibitem{Ohta:RIMS1989} Y. Ohta, Wronskian solutions for soliton equations, RIMS Kokyuroku {\bf 684} (1984) 1--17 (in Japanese).
\bibitem{Pinkall:dNLS} U. Pinkall, B. Springborn and S. Wei{\ss}mann, A new doubly discrete analogue of smoke ring flow and the real time simulation of fluid flow, J. Phys. A: Math. Theor. {\bf 40} (2007), pp. 12563--12576.
\bibitem{Rogers-Schief:book} C. Rogers and W.K. Schief, {\it B\"acklund and Darboux transformations:
	geometry and modern applications in soliton theory}
(Cambridge University Press, Cambridge, 2002).
\bibitem{Sadakane:sdNLS} T. Sadakane, Ablowitz-Ladik hierarchy and two-component Toda lattice
	hierarchy, J. Phys. A: Math. Gen. {\bf 36} (2003), pp.87--97.
\bibitem{Sym:Sym-Tafel} A. Sym, Soliton surfaces, Lett. Nuovo Cimento (2) {\bf 33} (1982), pp.394--400.
\bibitem{Hirota-Tsujimoto-Ohta:JSIAM1993} S. Tsujimoto, Y. Ohta and R. Hirota, Difference scheme of nonlinear Schr\"odinger equation, Proceedings of the annual meeting of Japan Society of Industrial and Applied Mathematics (1993), pp.203--204 (in Japanese).
\bibitem{Tsujimoto:App_book} S. Tsujimoto, Discretization of integrable systems, in {\sl Applied integrable systems}, ed. by Y. Nakamura (Shokabo, Tokyo, 2000), pp. 1--52 (in Japanese).
\bibitem{Ueno-Takasaki:book} K. Ueno and K. Takasaki, Toda lattice hierarchy, in {\sl Group representations and systems of differential equations ({T}okyo, 1982)},
Adv. Stud. Pure Math. {\bf 4} (North-Holland, Amsterdam, 1984), pp.1--95.
\bibitem{Weissmann-Pinkall:dNLS_simulation} S. Wei{\ss}mann and U. Pinkall, Real-time interactive simulation of smoke using discrete integrable vortex filaments, in {\it vriphys: Workshop on virtual reality interaction and physical simulations}, eds. by H. Prautzsch et al (The Eurographics Association, 2009), pp. 1--10.
\end{thebibliography}
\end{document}